\documentclass[12pt,preprint]{aastex}
\shortauthors{Spergel et al.}

\newcommand{\WMAP}{\textsl{WMAP}}

\newcommand{\bi}{\begin{itemize}}
\newcommand{\ei}{\end{itemize}}
\newcommand{\be}{\begin{equation}}
\newcommand{\ee}{\end{equation}}
\newcommand{\bea}{\begin{eqnarray}}
\newcommand{\eea}{\end{eqnarray}}

\renewcommand{\l}{{\ell}}

\newcommand{\G}{{\cal G}}

\renewcommand{\arcmin}{'}

\newcommand{\fg}{\bf}

\newcommand{ \eg }{{\it e.g. }}
\newcommand{ \ie }{{\it i.e. }}

\newcommand{\hatn}{{\bf \hat n}}

\begin{document}

\title{Wilkinson Microwave Anisotropy Probe (WMAP) Three Year Observations:\ Implications for Cosmology}
\author{
D. N. Spergel\altaffilmark{1,2},
R. Bean\altaffilmark{1,3},
O. Dor\'{e}\altaffilmark{1,4},
M. R. Nolta \altaffilmark{4,5},
C. L. Bennett\altaffilmark{6,7},
J. Dunkley\altaffilmark{1,5},
G. Hinshaw\altaffilmark{6},
N. Jarosik \altaffilmark{5},
E. Komatsu \altaffilmark{1,8},
L. Page\altaffilmark{5},
H. V. Peiris \altaffilmark{1,9,10},
L. Verde \altaffilmark{1,11},
M. Halpern \altaffilmark{12},
R. S. Hill\altaffilmark{6,15},
A. Kogut \altaffilmark{6},
M. Limon \altaffilmark{6},
S. S. Meyer \altaffilmark{9},
N. Odegard \altaffilmark{6,15},
G. S. Tucker \altaffilmark{13},
J. L. Weiland\altaffilmark{6,15},
E. Wollack \altaffilmark{6},
E. L. Wright \altaffilmark{14}
}
\altaffiltext{1}{Dept. of Astrophysical Sciences, 
Peyton Hall, Princeton University, Princeton, NJ 08544-1001}
\altaffiltext{2}{Visiting Scientist, Cerro-Tololo Inter-American Observatory}
\altaffiltext{3}{612 Space Sciences Building, 
Cornell University, Ithaca, NY  14853}
\altaffiltext{4}{Canadian Institute for Theoretical Astrophysics, 
60 St. George St, University of Toronto, 
Toronto, ON  Canada M5S 3H8}
\altaffiltext{5}{Dept. of Physics, Jadwin Hall, 
Princeton University, Princeton, NJ 08544-0708}
\altaffiltext{6}{Code 665, NASA/Goddard Space Flight Center, 
Greenbelt, MD 20771}
\altaffiltext{7}{Dept. of Physics \& Astronomy, 
The Johns Hopkins University, 3400 N. Charles St., 
Baltimore, MD  21218-2686}
\altaffiltext{8}{Department of Astronomy, University of Texas, Austin, TX}
\altaffiltext{9}{Depts. of Astrophysics and Physics, KICP and EFI, 
University of Chicago, Chicago, IL 60637}
\altaffiltext{10}{Hubble Fellow}
\altaffiltext{11}{Department of Physics, University of Pennsylvania, Philadelphia, PA}
\altaffiltext{12}{Dept. of Physics and Astronomy, University of 
British Columbia, Vancouver, BC  Canada V6T 1Z1}
\altaffiltext{13}{Dept. of Physics, Brown University, 
182 Hope St., Providence, RI 02912-1843}
\altaffiltext{14}{UCLA Astronomy, PO Box 951562, Los Angeles, CA 90095-1562}
\altaffiltext{15}{Science Systems and Applications, Inc. (SSAI), 
10210 Greenbelt Road, Suite 600 Lanham, Maryland 20706}
\email{dns@astro.princeton.edu}

\slugcomment{ApJ, in press, January 5, 2007}
\begin{abstract}
A simple cosmological model with only six parameters 
(matter density, $\Omega_m h^2$, baryon density, $\Omega_b h^2$, Hubble  Constant, $H_0$,
amplitude of fluctuations, $\sigma_8$, optical depth, $\tau$,
and a slope for the scalar perturbation spectrum, $n_s$)
fits not only the three year WMAP temperature and polarization data,
 but also small scale CMB data, light element abundances, large-scale
 structure observations, and the supernova
 luminosity/distance relationship. Using WMAP data only, the best fit values for cosmological
parameters for the power-law flat $\Lambda$CDM model are
$(\Omega_m h^2, \Omega_b h^2, h, n_s, \tau, \sigma_8) = 
\ensuremath{0.1277^{+ 0.0080}_{- 0.0079}},
\ensuremath{0.02229\pm 0.00073},
\ensuremath{0.732^{+ 0.031}_{- 0.032}},
\ensuremath{0.958\pm 0.016},
\ensuremath{0.089\pm 0.030},
\ensuremath{0.761^{+ 0.049}_{- 0.048}})$.
The three year data dramatically shrinks 
the allowed volume in this six dimensional parameter space.

Assuming that the primordial fluctuations are adiabatic with
a power law spectrum, the WMAP data {\it alone} 
require dark matter, 
and  favor a spectral index that is significantly less
than the Harrison-Zel'dovich-Peebles scale-invariant spectrum
$(n_s=1,r=0)$.
Adding additional data sets improves the
constraints on these components and the spectral slope.
For power-law models, WMAP data alone  puts an improved upper limit on
the tensor to scalar ratio, 
\ensuremath{r_{0.002} < 0.65\ \mbox{(95\% CL)}} 
and the combination of WMAP
and the lensing-normalized SDSS galaxy survey implies
\ensuremath{r_{0.002} < 0.30\ \mbox{(95\% CL)}}.

Models that suppress large-scale
power through a running spectral index or a large-scale cut-off in
the power spectrum are a better fit to the WMAP and small
scale CMB data than the power-law $\Lambda$CDM model; 
however,  the improvement in
the fit to the WMAP data is 
only  $\Delta \chi^2 = 3$ for 1 extra degree of freedom.  Models with a
running-spectral
index are consistent with a higher amplitude of gravity waves.

In a flat universe, the combination of WMAP and the Supernova
Legacy Survey (SNLS)  data yields  a significant constraint
on the equation of state of the dark energy,
\ensuremath{w = -0.967^{+ 0.073}_{- 0.072}}.  
If we assume $w=-1$, then 
the deviations from the critical density, $\Omega_K$,
are small: 
the combination
of WMAP and the SNLS data imply
\ensuremath{\Omega_k = -0.011\pm 0.012}.
The combination of WMAP three year data plus the HST key project
constraint on $H_0$ implies 
\ensuremath{\Omega_k = -0.014\pm 0.017}
and 
\ensuremath{\Omega_\Lambda = 0.716\pm 0.055}.
Even if we do not include the prior that the universe is flat,
by combining WMAP, large-scale structure and supernova data,
we can still put a strong constraint on the
dark energy equation of state,
\ensuremath{w = -1.08\pm 0.12}.

For a flat universe, the combination of WMAP and other
astronomical  data
yield a constraint on the sum of the neutrino masses, $\sum m_\nu < $
\ensuremath{0.66\ \mbox{eV}\ \mbox{(95\% CL)}}.
Consistent with the predictions of simple inflationary theories,
 we detect no significant deviations from Gaussianity in the CMB maps
 using Minkowski functionals, the bispectrum, trispectrum, and a
 new statistic designed to detect large-scale anisotropies
in the fluctuations. 

\end{abstract}
\maketitle

\section{Introduction\label{sec:intro}}
The  power-law {$\Lambda$CDM}
model  fits not only the 
Wilkinson Microwave Anisotropy Probe (WMAP) first year data,
but also a wide range of astronomical data \citep{bennett/etal:2003b,spergel/etal:2003}. In this model, the
universe is spatially flat, homogeneous and isotropic on large scales.
It is composed of ordinary matter, radiation, and dark matter and has a cosmological
constant.  The primordial fluctuations in this model are adiabatic, nearly
scale-invariant Gaussian random fluctuations \citep{komatsu/etal:2003}.   Six
cosmological parameters (the density of matter, the density of atoms,
the expansion rate of the universe, the amplitude of the primordial
fluctuations, their scale dependence and the optical depth of the
universe) are enough to predict not only the statistical  
properties of the microwave sky, measured by WMAP at several hundred
thousand points on the sky, but also the large-scale distribution of
matter and galaxies, mapped by the Sloan Digital Sky Survey (SDSS) and
the 2dF Galaxy Redshift Survey (2dFGRS).  

With three years of integration, improved beam models, better
understanding of systematic errors \citep{jarosik/etal:prep}, temperature data \citep{hinshaw/etal:prep}, and
polarization data \citep{page/etal:prep}, the WMAP data has
significantly improved.  There have also been significant improvements
in other astronomical data sets: analysis of galaxy
clustering in the SDSS \citep{tegmark/etal:2004b,eisenstein/etal:2005} and the completion
of the 2dFGRS  \citep{cole/etal:2005};  improvements in small-scale
CMB measurements  \citep{kuo/etal:2004,readhead/etal:2004,readhead/etal:2004c,
  grainge/etal:2003,leitch/etal:2005,piacentini/etal:2005,montroy/etal:2005,odwyer/etal:2005}, 
much larger samples of high redshift
supernova \citep{riess/etal:2004,astier/etal:2005,nobili/etal:2005,clocchiatti/etal:2005,krisciunas/etal:2005}; and significant improvements in the
lensing data \citep{refregier:2003,heymans/etal:2005,semboloni/etal:2005,hoekstra/etal:2005}. 

In \S \ref{sec:method}, we describe the basic analysis methodology used, with an emphasis on changes since the first year.
In \S \ref{sec:lcdm}, we fit the $\Lambda$CDM model to the WMAP temperature
and polarization data.  With its basic parameters fixed at
$z \sim 1100$, this model
predicts the properties of the low redshift universe: the galaxy power
spectrum, the gravitational lensing power spectrum, the Hubble
constant, and the luminosity-distance relationship.  In \S \ref{sec:wmap_astro}, we compare the
predictions of this model to a host of astronomical observations.  We
then discuss the results of combined analysis of WMAP data, other
astronomical data, and other CMB data sets. In \S \ref{sec:power}, we use the WMAP
data to constrain the shape of the power spectrum. 
In \S \ref{sec:inf}, we consider the implications of
the WMAP data for our understanding of inflation.  In \S \ref{sec:constraints}, we
use these data sets to constrain the composition of the universe: the
equation of state of the dark energy, the neutrino masses and 
the effective number of  neutrino species.  In \S \ref{sec:ng}, we search for
non-Gaussian features in the microwave background data.  The
conclusions of our analysis are described in \S \ref{sec:conclusions}.

\section{Methodology\label{sec:method}}
The basic approach of this paper is similar to that of the first-year WMAP
analysis: our goal is to find the simplest model that fits the CMB and large-scale structure
data.    Unless explicitly noted in \S 2.1, we use the
methodology described in \citet{verde/etal:2003} and applied in
\citet{spergel/etal:2003}. We use Bayesian statistical techniques to
explore the shape of the likelihood function, we use Monte Carlo Markov chain
methods to explore the likelihood surface and we quote both our maximum
likelihood parameters and the marginalized expectation value for each
parameter in a given model: 
\begin{equation} 
\label{eq:mean}
<\alpha_i> = \int d^N \alpha\ {\cal L}(d|\alpha) p(\alpha) \alpha_i  =\frac{1}{M} \sum_{j=1}^M \alpha_i^{j}
\end{equation}
where $\alpha_i^j$ is the value of the $i-$th parameter in the chain and 
$j$ indexes the chain  element. The number
of elements ($M$) in the  typical merged Markov chain is at
least 50,000 and is always long enough to satisfy the 
\citet{gelman/rubin:1992} convergence test with $R < 1.1$.  In addition,
we use the spectral convergence test described in \citet{dunkley/etal:2005} to
confirm convergence.  Most
merged chains have over 100,000 elements.
We use a uniform  prior on cosmological parameters, $p(\alpha)$
unless otherwise  specified.   We refer to $<\alpha_i>$ as
the best fit value for the parameter and the peak of the likelihood
function as the best fit model.

The Markov chain outputs and the marginalized values of the cosmological parameters listed
in Table \ref{tab_def} are available at http://lambda.gsfc.nasa.gov for all of the models discussed in the paper.

\subsection{Changes in analysis techniques}

We now use not only the measurements of the temperature power spectrum
(TT) and the temperature polarization power spectrum (TE), but also
measurements of the polarization power spectra (EE) and (BB).

At the lowest multipoles, a number of the approximations used in the
first year analysis were suboptimal. \citet{efstathiou:2004a} notes that a
maximum likelihood analysis is significantly better than a quadratic
estimator analysis at $\ell=2$.  \citet{slosar/seljak/makarov:2004} note
that the shape of the likelihood function at $\ell=2$ 
is not well approximated  by the fitting function used in the first
year analysis \citep{verde/etal:2003}.  More accurate treatments of the low $\ell$ likelihoods decrease the significance of the evidence for a running spectral index \citep{efstathiou:2004a,slosar/seljak/makarov:2004,odwyer/etal:2004}.
\citet{hinshaw/etal:prep} and \citet{page/etal:prep} describe our
approach to addressing this concern:  
for low multipoles, we explicitly compute the likelihood function for the WMAP
temperature and polarization maps.  For the  analysis of the
polarization maps, we use
the resolution $N_{side} = 8$ $N^{-1}$ matrices.
This pixel-based method is
used for $C_{\ell}^{TT}$ for $2\le \ell\le 30$ and polarization for $2\le \ell \le 23$.
For most of the analyses in the paper, we use a $N_{side}=8$ version of the temperature
map for the analysis of the low $l$ likelihood that uses a pixel-based version
for $\ell \le 12$.  For the WMAP $\Lambda$CDM only
case, we use the more time-consuming $N_{side}=16$ version of the code.
\citet{hinshaw/etal:prep} compares various approaches towards computing the low $l$ likelihood.
In Appendix \ref{appendix:sz_marg}, we discuss various choices made in the maximum likelihood code. 
For the $\Lambda$CDM model, we have computed the best fit parameters using
a range of assumptions for the amplitude of point source contamination and 
different treatments of the low $l$ likelihood.  

There are several improvements in our analysis of high $\l$ temperature data
\citep{hinshaw/etal:prep}: better beam models,
improved foreground models, and the use of maps with smaller pixels ($N_{side} = 1024$).
The improved foreground model is significant at $\l < 200$.  The $N_{side}=1024$ maps significantly reduce the effects
of sub-pixel CMB fluctuations and other pixelization effects. 
We found that  $N_{side} = 512$ maps had higher $\chi^2$ than
$N_{side}=1024$ maps, particularly for 
$\l = 600 - 700$, where there is significant signal-to-noise and
pixelization effects are significant.

We now marginalize over  the amplitude of Sunyaev-Zel'dovich (SZ)
fluctuations.
The expected level of SZ fluctuations
\citep{refregier/etal:2000, komatsu/seljak:2001, bond/etal:2005} is
$\ell(\ell+1)C_\ell/(2\pi)=19 \pm 3 \mu{\rm K}^2$ at $\ell=450-800$
for $\Omega_m=0.26$, $\Omega_b=0.044$,
$h=0.72$, $n_s=0.97$ and $\sigma_8=0.80$.
The amplitude of SZ fluctuations is very sensitive to $\sigma_8$
\citep{komatsu/kitayama:1999, komatsu/seljak:2001}.
For example at 60 GHz, $\ell(\ell+1)C_\ell/(2\pi)=65 \pm 15(\mu{\rm K})^2$ at $\ell=450-800$
for $\sigma_8=0.91$, which is comparable to the WMAP statistical errors
at the same multipole range.
Since the WMAP spectral coverage is not sufficient
to be able to distinguish CMB fluctuations from SZ fluctuations (see
discussion in \citet{hinshaw/etal:prep}),  we marginalize over its
amplitude using the \citet{komatsu/seljak:2002} analytical model for
the shape of the SZ fluctuations.  We impose the
prior that the SZ signal is between $0$ and $2$ times the 
\citet{komatsu/seljak:2002} value.
Consistent with the analysis of
\citet{huffenberger/seljak/makarov:2004}, we find that the SZ
contribution is not a  significant contaminant to the CMB signal on
the scales probed by the WMAP experiment.  We
report the amplitude of the SZ signal normalized to the
\citet{komatsu/seljak:2002} predictions for the cosmological parameters
listed above with $\sigma_8=0.80$.  
$A_{SZ} =1$ implies that the SZ contribution is
8.4, 18.7 and 25.2 $(\mu{\rm K})^2$ at $\ell = 220, 600$
and $1000$ respectively.
We discuss the effects
of this marginalization in Appendix \ref{appendix:sz_marg}.
We have checked that gravitational lensing of the microwave background, the next most significant 
secondary effect after the thermal SZ effect \citep{lewis/challinor:2006} does not have a significant effect on
parameters.

We now use the CAMB code 
\citep{lewis/challinor/lasenby:2000} for our
analysis of the WMAP power spectrum.  The CAMB code is derived from
CMBFAST \citep{zaldarriaga/seljak:2000}, but has the advantage of
running a factor of 2 faster on the Silicon Graphics, Inc.  (SGI)  machines used for the analysis
in this paper.   For the multipole range probed by WMAP, the numerical uncertainties and
physical uncertainties in theoretical calculations of multipoles are about 1 part in 10$^3$
\citep{seljak/etal:2003}, significantly smaller than the experimental uncertainties.  When we compare the results to large-scale structure and lensing calculations, the analytical treatments of the growth
of structure in the non-linear regime is accurate to better than 10\% on the smallest scales considered 
in this paper \citep{smith/etal:2003}.
 
\subsection{Parameter choices}

We consider constraints on the hot big bang cosmological scenario
with Gaussian, adiabatic primordial fluctuations as would arise from
single field, slow-roll inflation. We do not consider the influence of
isocurvature modes nor the subsequent production of fluctuations from
topological defects or unstable particle decay. 

We parameterize our cosmological model in terms of 15 parameters:
\begin{equation} 
{\bf p} = \{ \omega_{b},\omega_{c},\tau,\Omega_{\Lambda},
w,\Omega_{k},f_{\nu},N_{\nu},\Delta^{2}_{\mathcal
  R},n_{s},r,dn_{s}/d\ln k,A_{SZ},b_{SDSS},z_s\}
\end{equation}
where these parameters are defined in Table \ref{tab_def}.
For the basic power-law $\Lambda$CDM model, we use
$\omega_b$, $\omega_c$, $\exp(-2\tau)$, $\Theta_s$, $n_s,$  and $C^{TT}_{\ell=220}$,
as the cosmological parameters in the chain, $A_{SZ}$ as a nuisance
parameter,
and unless otherwise noted, we assume a flat prior on these parameters.  Note that
$\tau$ is the optical depth since reionization.  Prior to reionization,
$x_e$ is set to the standard value for the residual ionization computed in RECFAST 
\citep{seager/sasselov/scott:2000}.   For other models, we use
these same basic seven parameters plus the additional parameters noted in the text.
Other standard cosmological
parameters (also defined in Table \ref{tab_def}), such as $\sigma_8$ and $h$, are functions of these six parameters.
Appendix \ref{appendix:sz_marg} discusses the dependence of results on
the choice of priors.

With only one year of WMAP data, there were significant degeneracies even
in the $\Lambda$CDM model: there was a long degenerate valley in $n_s-\tau$ space
and there was also a significant degeneracy between $n_s$ and $\Omega_b h^2$
(see Figure 5 in \citet{spergel/etal:2003}).  With the measurements of
the rise to the third peak \citep{hinshaw/etal:prep} and the EE power spectrum \citep{page/etal:prep},
these degeneracies are now broken (see \S \ref{sec:lcdm}).
However,
more general models, most notably those with non-flat
cosmologies and with richer dark energy or matter content, have 
strong parameter degeneracies.
For models with adiabatic fluctuations, the WMAP data constrains the ratio of
the matter density/radiation density, effectively, $\Omega_m h^2$, the baryon density, $\Omega_b h^2$,
the slope of the primordial power spectrum and the distance to the surface of last scatter.
In a flat vacuum energy dominated universe, this distance is a function only of $\Omega$
and $h$, so the matter density and Hubble constant are well constrained.  On the other hand,
in non-flat models, there is a degeneracy between $\Omega_m$, $h$ and the curvature
(see \S \ref{sec:geom}).  Similarly, in models with more complicated dark energy properties
($w \ne -1$), there is a degeneracy between $\Omega_m$, $h$ and $w$. In models
where the number of neutrino species is not fixed, the energy density in radiation is no
longer known so that the WMAP data only constrains
a combination of $\Omega_m h^2$ and the number of neutrino species. 
These degeneracies slow convergence as the Markov chains  need to
explore degenerate valleys in the
likelihood surface.  

\begin{table}[htbp!] 
\linespread{1.25}
\begin{center}{\footnotesize\begin{tabular}{|cll|}
\hline
Parameter & Description & Definition \\
\hline
$H_{0}$ & Hubble expansion factor &   $H_{0} = 100h$ Mpc$^{-1}$km\ s$^{-1}$ \\
$\omega_{b}$ &Baryon density & $\omega_b = \Omega_{b}h^{2}= \rho_{b}/1.88\times 10^{-26}$ kg\ m$^{-3}$ \\
$\omega_{c}$ &Cold dark matter density & $\omega_c = \Omega_{c}h^{2} =\rho_{c}/18.8$ yoctograms/ m$^{-3}$ \\
$f_{\nu}$ &Massive neutrino fraction & $f_{\nu} =\Omega_{\nu}/\Omega_{c} $ \\
$\sum m_{\nu}$ & Total neutrino mass (eV)  &  $\sum m_{\nu}=94 \Omega_{\nu}h^2$ \\
$N_{\nu}$ & Effective number of relativistic neutrino species 
& \\
$\Omega_{k}$ & Spatial curvature & \\ 
$\Omega_{DE}$ & Dark energy density  &
For $w=-1$, $\Omega_{\Lambda} = \Omega_{DE}$ \\
 $\Omega_{m}$ & Matter
energy density & $\Omega_{m} = \Omega_{b}+\Omega_{c}+\Omega_{\nu}$\\ 
$w$ & Dark energy equation of state & $w= p_{DE}/\rho_{DE}
$\\
$\Delta_{\mathcal R}^{2}$ & Amplitude of curvature perturbations
$\mathcal{R}$ & $\Delta_{\mathcal R}^{2} (k=0.002/Mpc)\approx
29.5\times 10^{-10} A$ \\ 
$A$ & Amplitude of density fluctuations ($k=0.002$/Mpc)
 & See \citet{spergel/etal:2003} \\
$n_{s}$ &Scalar spectral index at 0.002/Mpc &  \\
$\alpha$ & Running in scalar spectral index  & $\alpha = d n_{s} /d ln k$ (assume constant)\\
$r$ & Ratio of the amplitude of tensor fluctuations &\\
&to scalar potential fluctuations at k=0.002/Mpc & \\
$n_{t}$ & Tensor spectral index &  Assume $n_{t} = -r/8$ \\
$\tau$  & Reionization optical depth & \\
$\sigma_8$ & Linear theory amplitude of matter&  \\ & fluctuations on
8 $h^{-1}$ Mpc & \\	
$\Theta_{s}$ & Acoustic peak scale (degrees) & see 
\citet{kosowsky/milosavljevic/jimenez:2002}\\
$A_{SZ}$ & SZ marginalization factor & see appendix \ref{appendix:sz_marg} \\
$b_{sdss}$ & Galaxy bias factor for SDSS sample & $b= [P_{sdss}(k,z=0)/P(k)]^{1/2}$  (constant)\\
$C_{220}^{ \ \ \ TT}$ & Amplitude of the TT temperature power spectrum at $\l=220$ & \\
$z_{s}$ & Weak lensing source redshift&\\
\hline
\end{tabular}}\end{center}
\caption{\footnotesize Cosmological parameters used in the analysis.  http://lambda.gsfc.nasa.gov lists the marginalized values for these parameters for all of the models discussed in this paper. \label{tab_def}}
\linespread{1}
\end{table}

\section{$\Lambda$CDM\ Model:\ Does it still fit the data?
\label{sec:lcdm}}

\subsection{WMAP only}
\begin{figure}[htbp!]  
\centering
\includegraphics[width=6in]{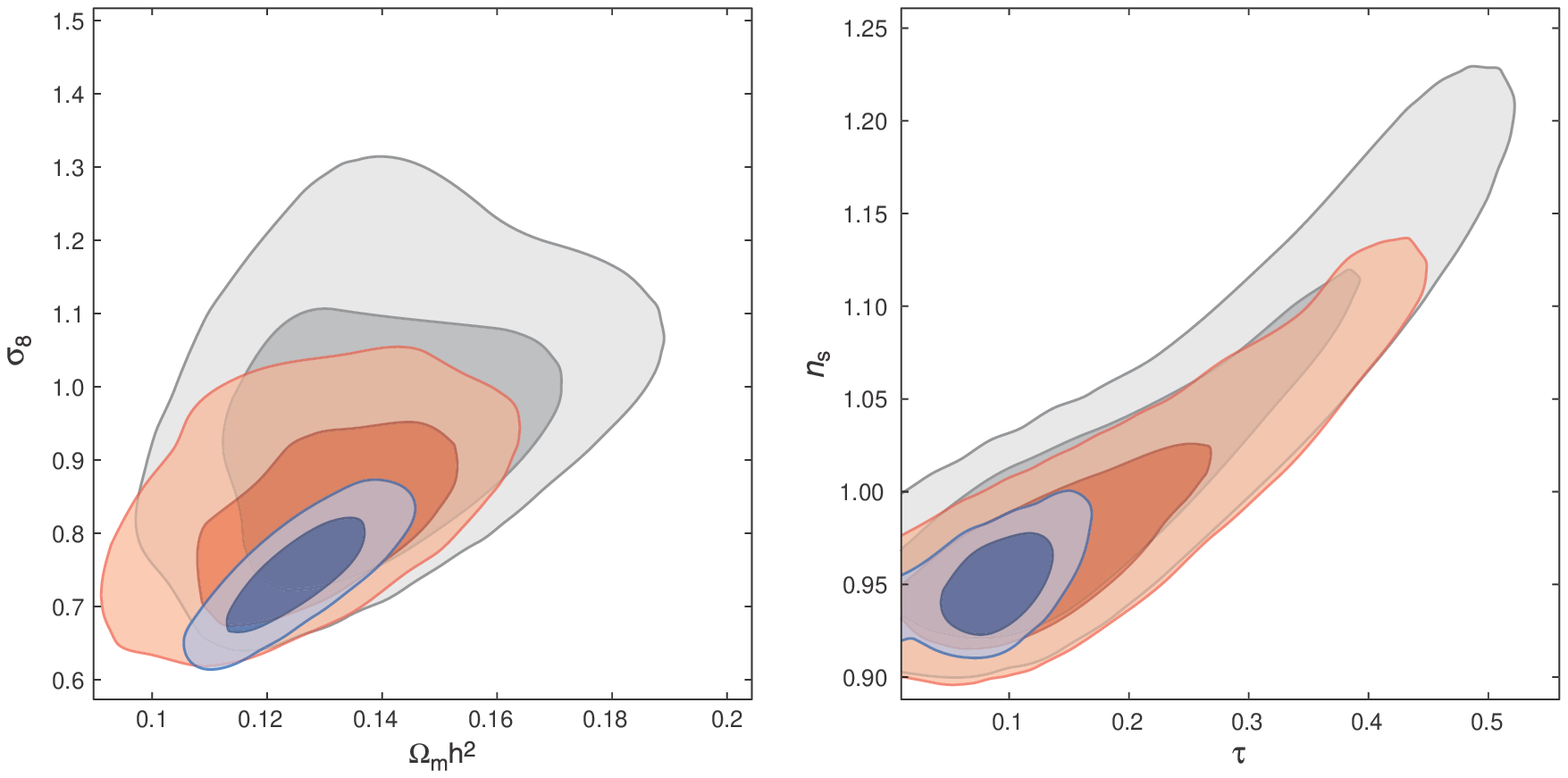}
\caption{\fg The improvement in
parameter constraints for the power-law $\Lambda$CDM
model (Model M5 in Table~\ref{tab:how_many_parameters}).  
The contours show the 68\% and 95\% joint 2-d marginalized
contours for the $(\Omega_m h^2,\sigma_8)$ plane (left)
and the $(n_s,\tau)$ plane (right).
The black contours are for the first year WMAP data (with no prior
on $\tau$).  The red contours are for the first year WMAP data combined
with CBI and ACBAR (WMAPext in \citet{spergel/etal:2003}).  The blue
contours are for the 
three year WMAP data only with the SZ contribution set to 0 to 
maintain consistency with the first year analysis.
The WMAP measurements of EE power spectrum
provide a strong constraint on the value of $\tau$. 
The models with no reionization ($\tau=0$) or a scale-invariant 
spectrum ($n_s=1$) are both disfavored at 
$\Delta\chi^2_{eff}>6$ for 5 parameters (see Table~\ref{tab:how_many_parameters}).
Improvements in
the measurement of the amplitude of the third peak  yield better
constraints on $\Omega_m h^2$.  
\label{fig:nstauwmaponly}}
\end{figure}

\begin{figure}[htbp!] 
\centering
\includegraphics[width=5in]{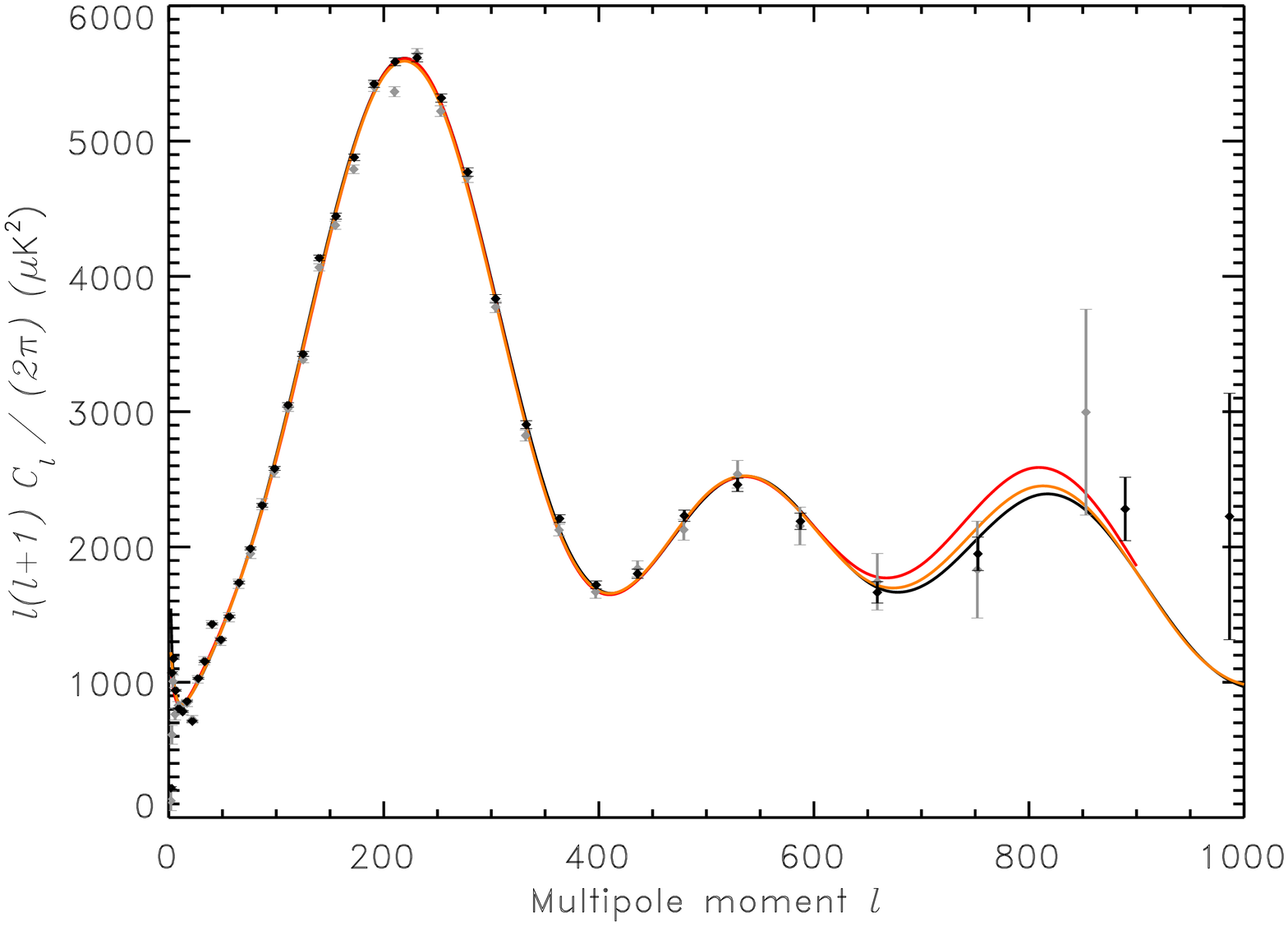}
\caption{\fg Comparison of the predictions of the different best fit models
to the data.  The  
{\it black} line is the angular power spectrum predicted for the best fit
three-year WMAP only $\Lambda$CDM  model.  The {\it red} line is the best fit to
the 1-year WMAP data.  The {\it orange} line is the best fit to the combination
of the 1-year WMAP data, CBI and ACBAR (WMAPext in \citet{spergel/etal:2003}).
The solid data points are for the 3 year data and the light gray data points
are for the first year data.
\label{fig:compare_years}}
\end{figure}

{\it The $\Lambda$CDM model is still an excellent fit to the WMAP data}.
With longer integration times and smaller pixels, the errors 
in the high $l$ temperature multipoles have shrunk by more
than 
a factor of three.
As the data have improved, the likelihood function 
remains peaked around the maximum likelihood peak of the first year WMAP value.  With longer integration, the most discrepant
high $\ell$ points from the year-one data are now much closer to the best fit model (see Figure \ref{fig:compare_years}).   
For the first year  WMAP TT and TE data \citep{spergel/etal:2003},
the reduced $\chi^2_{eff}$ 
was 1.09 for 893 degrees of freedom (D.O.F.)  for the TT
data and
was 1.066
for  the combined TT and TE data (893+449=1342 D.O.F.).
For the three year data, which
has much smaller errors for $\ell > 350$,
the reduced $\chi^2_{eff}$
for  
982 D.O.F. ($\ell=13-1000$- 7 parameters)
is now 1.068 for the TT data 
and 1.041 for the combined TT and TE data (
1410 D.O.F., including TE $\ell=24-450$),
where the TE data contribution is evaluated from $\ell=24-500$.
For the T, Q, and U maps using the pixel based likelihood we obtain a reduced $\chi^2_{eff} = 0.981$ for 1838 pixels
(corresponding to $C_\ell^{TT}$ for   $\ell =2-12$ and $C_\ell^{TE}$ for $\ell=2-23$).  The combined reduced $\chi^2_{eff} =1.037 $ for 3162 degrees of freedom
for the combined fit to the $TT$ and $TE$ power spectrum at high $\ell$
and the $T, Q$ and $U$ maps at low $\ell$.

While many of the maximum likelihood parameter values (Table 
\ref{tab:lcdmwmaponly}, columns 3 and 7 and
Figure \ref{fig:nstauwmaponly}) have not changed significantly, there
has been a noticeable reduction in the marginalized 
value for the optical depth, $\tau$, and a shift in
the best fit value of $\Omega_m h^2$. (Each shift is slightly larger than
 $1 \sigma$).    
The addition of the EE data now
eliminates a large region of parameter space with large $\tau$ and
$n_s$ that was consistent with the first year data.    With only 
the first year data set, the likelihood surface was very flat. It covered
a ridge in $\tau-n_s$ over a region that extended from $\tau \simeq 0.07$
to nearly $\tau = 0.3$. If the optical depth of the universe were as large as $\tau=0.3$ (a value 
consistent with the first year data), then the measured EE signal would have
been 10 times larger than the value reported in
\citet{page/etal:prep}. On the other hand, an optical depth of $\tau =0.05$ would produce one quarter of the detected EE signal.  As discussed in \citet{page/etal:prep}, the reionization signal is now based primarily on the EE signal rather than the TE signal.  See figure 26 in \citet{page/etal:prep}
for the likelihood plot for $\tau$: note that the form of this likelihood function is relatively
insensitive to the cosmological model (over the range considered in this paper).

There has also been a significant reduction in the uncertainties in the matter density,
$\Omega_m h^2$.  With the first year of WMAP data, the third peak was poorly constrained
(see the light gray data points in Figure \ref{fig:compare_years}).  With three years of
integration, the WMAP data better constrain the height of the third peak:
WMAP is now cosmic variance limited up to $\ell = 400$  and the signal-to-noise
ratio exceeds unity up to $\ell=850$.
The new best fit
WMAP-only model is close to the WMAP (first year)+CBI+ACBAR model in the third peak
region.  As a result, the preferred value of $\Omega_m h^2$ now shifts closer to the
``WMAPext" value reported in \citet{spergel/etal:2003}.  Figure \ref{fig:nstauwmaponly}
shows the $\Omega_m h^2-\sigma_8$ likelihood surfaces for the first year WMAP data, the
first year WMAPext data and the three year WMAP data.  The accurately determined peak position
constrains $\Omega_m^{0.275} h$ \citep{page/etal:2003b},
fixes the cosmological age,  and determines the direction of the
degeneracy surface.  With 1 year data, the best fit value is
$\Omega_m^{0.275} h = 0.498$.  With three years of data, the best fit shifts to
$0.492^{+0.008}_{-0.017}$. The lower third peak implies a smaller value
of $\Omega_m h^2$ and because of the peak constraint, a lower value of $\Omega_m$.  

The best fit value of $\sigma_8$  (marginalized over the other parameters) is now 
noticeably smaller for the three year data alone, 
$\ensuremath{0.761^{+ 0.049}_{- 0.048}}$
than for the first-year WMAP data alone, $0.92\pm0.10$.  
This lower value is due to a smaller third peak height, which 
leads to a lower value of
$\Omega_m$, and less structure growth and a lower best fit value for $\tau$. 
The height of the third peak was very uncertain with the first-year data alone.  In \citet{spergel/etal:2003},
we used external CMB data sets to constrain the third peak and with these data, the
Maximum Likelihood value was 0.84.  With three years of data, the third peak is
better determined and its height is close to the value estimated from the ground based data.
With the EE measurements, we have eliminated most of the high $\tau$ region of parameter
space.  Since higher values of $\tau$ imply a higher amplitude of primordial fluctuations,
the best fit value of $\sigma$ is proportional to $\exp(\tau)$.  For a model with all
other parameters fixed, $\tau = 0.10$ implies a 7\% lower value of $\sigma_8$.
As discussed in \S \ref{sec:wmap_pred}, this lower value of $\sigma_8$ is more consistent with the X-ray measurements but lower than the best fit value from recent lensing surveys.  Lower $\sigma_8$ implies later growth of structure.

In the first year data, we assumed that the SZ contribution
to the WMAP data was negligible.   Appendix \ref{appendix:sz_marg}  discusses
the change in priors and the change in the SZ treatment and their
effects on parameters: marginalizing over SZ most significantly shifts $n_s$
and $\sigma_8$ by 1\% and 3\% respectively.  In Table \ref{tab:lcdmwmaponly}
in the column labeled No SZ and
Figure \ref{fig:nstauwmaponly}, we assume $A_{SZ}= 0$ and use
the first  year likelihood code to make a consistent comparison 
between the first-year and three-year results.   As in the first year analysis, we use
a flat prior on the logarithm of the amplitude and a flat prior on $\Theta_s$ and 
a flat prior on $\tau$. 
The first column of Table \ref{tab:lcdm_low} list the parameters
fit to the WMAP three-year data with $A_{SZ}$ allowed to vary
between 0 and 2.
In the tables, the ``mean" value is calculated according to
equation (\ref{eq:mean}) and the ``Maximum Likelihood (ML)" value is the value
at the peak of the likelihood function.  In the last two columns, we provide our current best
estimate of parameters including SZ marginalization and using the full $N_{side}=16$ likelihood
code to compute the TT likelihoods.
In subsequent tables and figures, we will allow the 
SZ contribution to vary and quote the appropriate
marginalized values.  Allowing for an SZ contribution lowers the best
fit primordial contribution at high $\l$, thus, the best fit models
with an SZ contribution have lower $n_s$ and $\sigma_8$ values.
However, in other tables, we use the faster $N_{side}=8$ likelihood code unless specifically noted.
In all of the Tables, we quote the 68\% confidence intervals
on parameters and the 95\%  confidence limits on bounded parameters.

\begin{table} [htbp!] 
\begin{center}
\caption{Power Law $\Lambda$CDM Model Parameters and 68\% Confidence
Intervals.  The Three Year fits in the columns labeled ``No SZ"  use the 
likelihood formalism of the first year paper and assume no SZ contribution,
$A_{SZ} = 0$, to allow direct comparison with the First Year results.
Fits that include SZ marginalization are given in the last two
columns of the upper and lower tables and represent our best estimate
of these parameters.
The last column includes all data sets.
\label{tab:lcdmwmaponly}}
\begin{tabular}{|c|cc|cc|c||}
\hline
\hline
{Parameter}  &  {First Year} &{WMAPext}&{Three Year} &
{Three Year} & {Three Year+ALL}  \\
 & Mean   & Mean & Mean (No SZ)& Mean & Mean\\
\hline
\hline
$100 \Omega_b h^2$ &
$2.38^{+0.13}_{-0.12}$ &
$2.32^{+0.12}_{-0.11}$&
$2.23\pm 0.08$ &
\ensuremath{2.229\pm 0.073} &
\ensuremath{2.186\pm 0.068} \\
$\Omega_m h^2$ &
$0.144^{+0.016}_{-0.016}$ & 
$0.134^{+0.006}_{-0.006}$ & 
$0.126 \pm 0.009$ &
\ensuremath{0.1277^{+ 0.0080}_{- 0.0079}}&
\ensuremath{0.1324^{+ 0.0042}_{- 0.0041}}\\
$H_0$ &
$72^{+5}_{-5}$&
$73^{+3}_{-3}$ & 
\ensuremath{73.5\pm 3.2}&
\ensuremath{73.2^{+ 3.1}_{- 3.2}}&
\ensuremath{70.4^{+ 1.5}_{- 1.6}}\\
$\tau$  &
$0.17^{+0.08}_{-0.07}$ & 
$0.15^{+0.07}_{-0.07}$  &
\ensuremath{0.088^{+ 0.029}_{- 0.030}}&
\ensuremath{0.089\pm 0.030}&
\ensuremath{0.073^{+ 0.027}_{- 0.028}}\\
$n_s$ &
$0.99^{+0.04}_{-0.04}$ & 
$0.98^{+0.03}_{-0.03}$ & 
$0.961 \pm 0.017$  &
\ensuremath{0.958\pm 0.016}&
\ensuremath{0.947\pm 0.015}\\
$\Omega_m$ &
$0.29^{+0.07}_{-0.07}$ &  
$0.25^{+0.03}_{-0.03}$  & 
$0.234 \pm 0.035$ &
\ensuremath{0.241\pm 0.034}&
\ensuremath{0.268\pm 0.018}\\
$\sigma_8$ &
$0.92^{+0.1}_{-0.1}$ &
$0.84^{+0.06}_{-0.06}$ & 
$0.76 \pm 0.05$ &
\ensuremath{0.761^{+ 0.049}_{- 0.048}}&
\ensuremath{0.776^{+ 0.031}_{- 0.032}}\\
\hline
\hline
\hline
{Parameter}  &  
{First Year} &{WMAPext}&{Three Year} &{Three Year}  & {Three Year + ALL} \\
 & ML & ML & ML (No SZ) & ML & ML \\
\hline
\hline
$100 \Omega_b h^2$ &
$2.30$ &
$2.21$&
\ensuremath{2.23} &
\ensuremath{2.22} &
\ensuremath{2.19} \\
$\Omega_m h^2$ &
$0.145$ & 
$0.138$ & 
\ensuremath{0.125}&
\ensuremath{0.127}&
\ensuremath{0.131}\\
$H_0$ &
$68$&
$71$ & 
\ensuremath{73.4}&
\ensuremath{73.2}&
\ensuremath{73.2}\\
$\tau$  &
$0.10$ & 
$0.10$  &
\ensuremath{0.0904}&
\ensuremath{0.091}&
\ensuremath{0.0867}\\
$n_s$ &
$0.97$ & 
$0.96$ & 
\ensuremath{0.95}&
\ensuremath{0.954}&
\ensuremath{0.949}\\
$\Omega_m$ &
$0.32$ &  
$0.27$  & 
\ensuremath{0.232}&
\ensuremath{0.236}&
\ensuremath{0.259}\\
$\sigma_8$ &
$0.88$ &
$0.82$ & 
\ensuremath{0.737}&
\ensuremath{0.756}&
\ensuremath{0.783}\\
\hline
\hline
\end{tabular}
\end{center}
\end{table}
\subsection{Reionization History \label{sec:reion}}
\begin{figure}[htbp!] 
\plotone{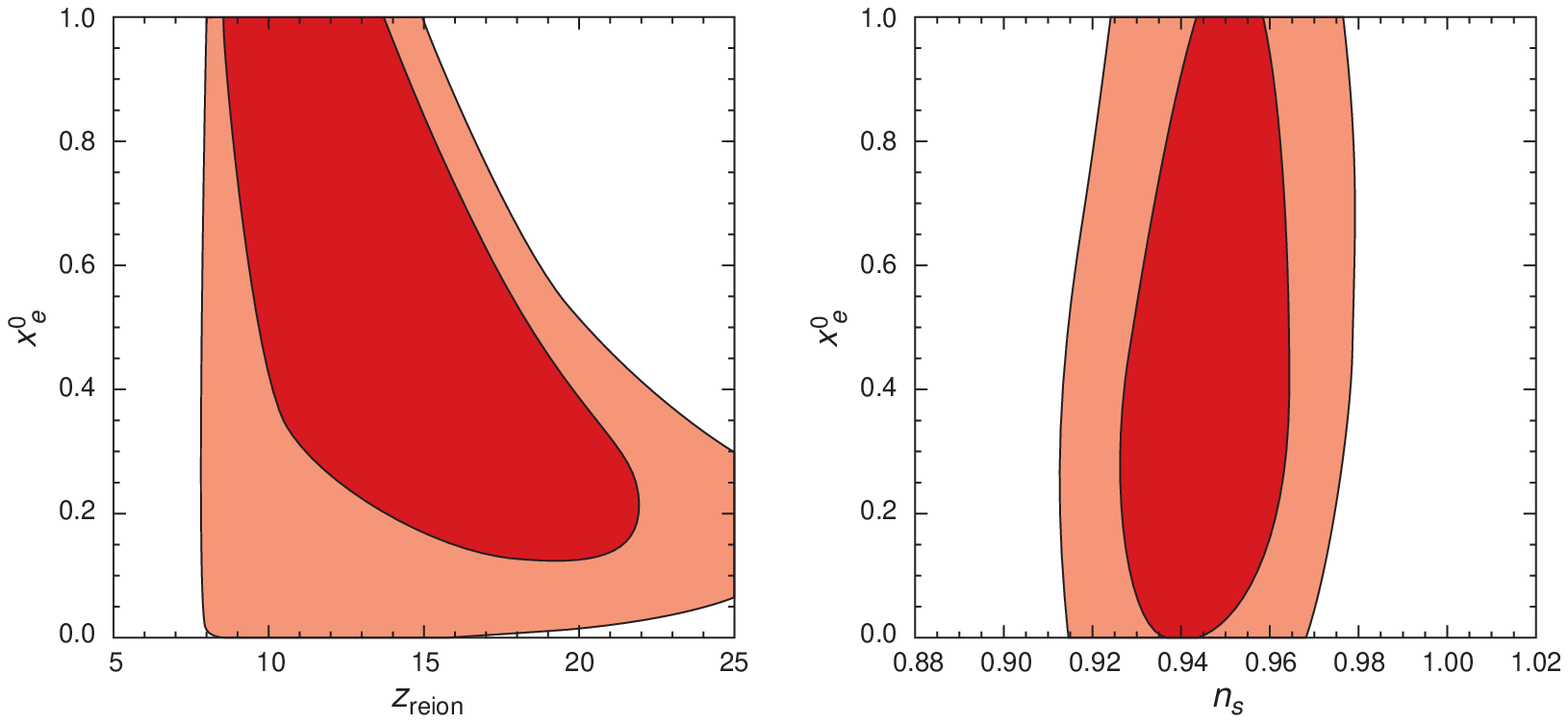}
\caption{\fg WMAP constraints on the reionization history. 
({\it Left}) The 68\% and 95\% joint 2-d marginalized confidence level
 contours for $x_e^0-z_{reion}$
for a power law $\Lambda$ Cold Dark Matter ($\Lambda$CDM) model with the reionization history described
by equation \ref{eq:reion} and fit to the WMAP three year data.
In equation~\ref{eq:reion} we assume that the universe was
 partially reionized at $z_{\rm reion}$ to an ionization fraction of
 $x_e^0$, and then became fully ionized at $z=7$.
({\it Right}) The 68\% and 95\% joint 2-d marginalized 
confidence level contours for $x_e^0-n_s$. This figure shows that $x_e^0$ and $n_s$ are nearly
independent for a given value of $\tau$, indicating that WMAP
determinations of cosmological parameters are not affected by details
of the reionization history.
Note that we assume a uniform prior on
$z_{reion}$  in this calculation, which favors models with lower
$x_e^0$ values in the right panel.
\label{fig:reion}
}
\end{figure}

Since the \citet{kogut/etal:2003} detection of $\tau$,
the physics of reionization has been a subject of extensive theoretical
study
\citep{cen:2003,ciardi/ferrara/white:2003,haiman/holder:2003,madau/etal:2004,oh/haiman:2003,
venkatesan/tumlinson/shull:2003,
ricotti/ostriker:2004,sokasian/etal:2004,somerville/livio:2003,wyithe/loeb:2003a,iliev/etal:prep}.
\citet{page/etal:prep} provides a detailed discussion of the new polarization data:   while the best fit value for $\tau$ has not changed significantly,
the new EE data, combined with an improved treatment of the TE data, implies
 smaller marginalized maximum likelihood value.
The three year data favors 
$\tau \simeq 0.1$, consistent with the
predictions of a number of simulations of $\Lambda$CDM models.
For example, \citet{ciardi/ferrara/white:2003}
$\Lambda$CDM simulations predict $\tau = 0.104$ for parameters
consistent with the WMAP primordial power spectrum.
\citet{tumlinson/venkatesan/shull:2004} use the nucleosynthetic data to derive
and construct an initial mass function (IMF) for reionization and find $\tau \sim 0.1$.
\citet{chiu/fan/ostriker:2003} found that their joint analysis
of the WMAP and SDSS quasar data favored a model with $\tau = 0.11$,
$\sigma_8 = 0.83$ and $n = 0.96$, very close to our new best fit values.
\citet{wyithe/cen:prep} predict that if the product of star formation 
efficiency and escape fraction for Pop-III stars is comparable to that for 
Pop-II stars, $\tau=0.09-0.12$ with reionization histories characterized by an extended ionization plateau from $z=7-12.$
They argue that  this result holds regardless of the redshift 
where the intergalactic medium (IGM)  becomes enriched with metals.

Measurements of the EE and TE power spectrum  are a powerful probe
of early star formation and an important complement to other astronomical
measurements.  Observations of galaxies \citep{malhotra/rhoads:2004},
quasars \citep{fan/etal:2005} and gamma ray bursts \citep{totani/etal:2005}
imply that the universe was mostly ionized by $z = 6$.
The detection of large-scale TE and EE signal \citep{page/etal:prep} implies
that the universe was mostly reionized at even higher redshift.
CMB observations have the potential to constrain some of the details
of reionization, as the shape of the CMB EE power spectrum is
sensitive to reionization history \citep{kaplinghat/etal:2003,hu/holder:2003}.  Here,
we explore the ability of the current EE data to constrain
reionization by postulating a two stage process
as a toy model.  During the first stage, the universe is partially
reionized at redshift $z_{reion}$ and complete reionization occurs at 
$z = 7$:
\begin{eqnarray}
\label{eq:reion}
x_e &=& 0 \qquad \qquad z > z_{reion} \nonumber \\
    &=& x_e^0  \qquad \qquad z_{reion} > z > 7 \nonumber \\
    &=& 1 \qquad \qquad z < 7
\end{eqnarray}
We have modified CAMB to include this reionization history.

Figure \ref{fig:reion} shows the likelihood surface for $x_e^0$ and
$z_{reion}$.  The plot shows that the data does not yet constrain
$x_e^0$ and that the characteristic redshift of reionization is sensitive
to our assumptions about
reionization.  If we assume that the universe
is fully reionized, $x_e^0 = 1$, then  the maximum likelihood peak is 
$z_{reion} = 
\ensuremath{11.3}$. The maximum
  likelihood peak value of the cosmic age at the reionization epoch is 
  $t_{reion} =  365 Myr.$

Reionization alters the TT power spectrum by suppressing fluctuations
on scales smaller than the horizon size at the epoch of reionization.
Without strong
constraints from polarization data on $\tau$, there is a strong degeneracy between spectral
index and $\tau$ in likelihood fits \citep{spergel/etal:2003}.  
The polarization measurements now strongly constrain $\tau$; however,
there is still significant uncertainty in $x_e$ and the details
of the reionization history.  Fortunately, 
the temperature power spectrum mostly depends on  the amplitude
of the optical depth signal, $\tau$, so that
the other fit parameters (e.g., $n_s$) are  insensitive 
to the details of the reionization history (see
Figure \ref{fig:reion}).
Because of this weak correlation,
we will assume a simple reionization  history ($x_e^0 = 1$)
in all of the other analysis in this paper.  Allowing for 
a more complex history is not likely to alter any of the conclusions
of the other sections.

\subsection{How Many Parameters Do We Need to Fit the WMAP Data?}
\begin{table} 
\begin{center}
\caption{Goodness of Fit, $\Delta \chi_{eff}^2 \equiv
-2 \ln {\cal L}$, for WMAP data only 
relative to a Power-Law $\Lambda$CDM model. $\Delta \chi_{eff}^2 >0$ is a worse fit to the data.
\label{tab:how_many_parameters}}
\begin{tabular}{|l|c|c|c|}
\hline
 & Model & $-\Delta (2 \ln {\cal L})$ & $N_{par}$ \\
\hline
\hline
M1 &Scale Invariant Fluctuations ($n_s=1$)& 6 & 5\\
M2 &No Reionization ($\tau=0$) & 7.4& 5 \\
M3 &No Dark Matter ($\Omega_c =0, \Omega_\Lambda \ne 0$)& 248 & 6 \\
M4 &No Cosmological Constant ($\Omega_c\ne 0, \Omega_\Lambda = 0)$& 0 & 6 \\
\hline
M5 &{\bf Power Law $\Lambda$CDM} & 0 & 6 \\
\hline
M6 &Quintessence ($w\neq -1$) & 0 & 7\\
M7 &Massive Neutrino ($m_{\nu}>0$) &-1 & 7\\
M8&Tensor Modes ($r>0$) &0  & 7 \\
M9&Running Spectral Index ($dn_{s}/d\ln k \ne 0 $) & $-4$  & 7 \\
M10 &Non-flat Universe ($\Omega_k\neq 0$) &  $-2$ & 7\\
M11&Running Spectral Index \& Tensor Modes& $-4$  & 8 \\
M12&Sharp cutoff & $-1$ & 7 \\
M13 &Binned $\Delta_{\mathcal R}^2(k)$ & $-22$  & 20 \\
\hline
\hline
\end{tabular}
\end{center}
\end{table}
In this subsection, 
we compare the power-law $\Lambda$CDM to other cosmological models.
We consider both simpler models with fewer parameters and models with
additional physics, characterized by
additional parameters.  
We quantify the relative goodness of fit of the models,
\begin{equation}
\Delta \chi_{eff}^2 \equiv -\Delta(2 \ln {\cal L}) = 2 \ln {\cal L}(\Lambda {\rm CDM}) -
			2 \ln {\cal L}({\rm model})
\end{equation}
A positive value for $\Delta \chi_{eff}^2$ implies 
the model is disfavored.  A negative value means that the model
is a better fit.  We also characterize each model
by the number of free parameters, $N_{par}$.   There are 3162
degrees of freedom in the combination of T, Q, and U maps 
and high $\ell$ TT and TE power spectra used in the fits and 1448
independent $C_l$'s, so that the effective number of data degrees of freedom is between 1448 and 3162.

Table \ref{tab:how_many_parameters} shows that the power-law $\Lambda$CDM
is a significantly better fit than the simpler models.   For consistency, all of
the models are computed with the $N_{side}=8$ likelihood code and the higher
value of the point source amplitude. If we reduce the
number of parameters in the model, the cosmological fits significantly worsen:
\begin{itemize}
\item 
Cold dark matter serves as a significant forcing term that changes the acoustic
peak structure.
Alternative gravity models (e.g., MOND), and all
baryons-only models, lack this forcing
term so they predict a much lower third peak than is observed by WMAP and small scale CMB 
experiments \citep{mcgaugh:2004,skordis/etal:2006}.  Models
without dark matter (even if we allow for a cosmological constant)
are very poor fits to the data.
\item Positively curved models without a cosmological constant  are consistent 
with the WMAP data alone: a model with the same six parameters
and the prior that there is no dark energy, $\Omega_\Lambda = 0$, fits as well as the standard model
with the flat universe prior, $\Omega_{m} + \Omega_\Lambda = 1$.
However, if
we imposed a prior that $H_0 > 40$ km s$^{-1}$ Mpc$^{-1}$, then the WMAP data would
not be consistent with $\Omega_\Lambda = 0$.  
Moreover, the parameters fit to the no-cosmological-constant model, $(H_0 = 30$ km s$^{-1}$ Mpc$^{-1}$ and
$\Omega_m = 1.3$)
are terrible fits to a host of astronomical data:
large-scale structure observations,
supernova data and measurements of local dynamics. 
As discussed in \S \ref{sec:geom},
the combination
of WMAP data and other astronomical data solidifies the evidence against these models.
The detected cross-correlation between
CMB fluctuations and large-scale structure 
provides further evidence for the existence of dark energy (see \S
\ref{sec:isw}).
\item The simple scale invariant ($n_s =1.0$) model is no longer a good fit to the WMAP data.
As discussed in the previous subsection,  combining the WMAP data with other astronomical data
sets further strengthens the case for $n_s < 1$.
\end{itemize}
The conclusion that the WMAP data demands the existence of dark matter and dark energy is
based on the assumption that the primordial power spectrum is a power-law spectrum.  By
adding additional features in the primordial perturbation spectrum, these alternative models
may be able to better mimic the $\Lambda$CDM model.  This possibility requires further study.

The bottom half of Table 
\ref{tab:how_many_parameters}
lists the relative improvement of the generalized models over the
power-law $\Lambda$CDM.  As the Table shows, the WMAP data alone does not require the existence
of tensor modes, quintessence, or modifications in neutrino properties.   Adding these parameters does not improve
the fit significantly.  For the WMAP data, the region in likelihood space where 
($r=0$, $w=-1$, and $m_\nu = 0$  lies  within the $1 \sigma$ contour.
In the \S \ref{sec:constraints}, we consider the limits on these parameters based
on WMAP data and other astronomical data sets.  

If we allow for a non-flat universe, then models with
small negative $\Omega_k$ are a better fit than the power-law
$\Lambda$CDM model.  
These models have a lower intervening Sachs-Wolfe
(ISW) signal at low $l$ and are a better fit to the low $\ell$ multipoles.
The best fit closed  universe
 model has  $\Omega_m = 0.415$, $\Omega_\Lambda = 0.630$
and $H_0 = 55$ km$s^{-1}$Mpc$^{-1}$ and is a better fit to
the WMAP data alone than the flat universe model($\Delta \chi^2_{eff} = 2$)
However,
as discussed in \S \ref{sec:geom},
the combination of WMAP data with either SNe data, large-scale structure
data or measurements of $H_0$  favors models
with $\Omega_K$ close to 0.

In section 5, we consider several different modifications to the shape
of the power spectrum.  As noted in Table \ref{tab:how_many_parameters}
, none
of the modifications lead to significant improvements in the fit.
Allowing the spectral index to vary as a function of scale improves the goodness-of-fit.
The low $\ell$ multipoles, particularly $\ell=2$, are lower than predicted in the $\Lambda$CDM model.
However, the relative improvement in fit is not large, $\Delta \chi^{2}_{eff} = 3$, so the WMAP
data alone do not require a running spectral index.   

Measurement of the goodness of fit is a simple approach to
test the needed number of parameters.
These results should be confirmed by  Bayesian model
comparison techniques
\citep{beltran/etal:2005,trotta:prep,mukherjee/parkinson/liddle:2006,bridges/lasenby/hobson:prep}.
Bayesian methods, however, require an estimate of the number of data points in the fit.  It is not
clear whether we should use the $\sim 10^{10}$ points in the TOD, the $10^6$ points in the
temperature maps, the $3 \times 10^3$ points in the TT, TE, or EE power spectrum,
or the $\sim 10-20$ numbers needed to fit the peaks and valleys in the TT data in evaluating
the significance of new parameters.
\section{WMAP $\Lambda$CDM Model and Other Astronomical Data
\label{sec:wmap_astro}}

In this paper, our approach is to 
show first that a wide range of astronomical data sets
are consistent with the predictions of the $\Lambda$CDM model with
its parameters fitted to the WMAP data
(see section \S \ref{sec:wmap_pred}).  We then use the
external data sets to constrain extensions of the standard model.

In our analyses, we consider several different types of data sets.
We consider the SDSS LRGs, the SDSS full
sample and the 2dFGRS data separately, this allows a check of systematic
effects.  We divide the small scale CMB data sets into low frequency 
experiments (CBI, VSA) 
and high frequency 
experiments (BOOMERanG,
ACBAR).  We divide the supernova data sets into two groups as described below.
The details of the data sets are also described in \S \ref{sec:wmap_pred}.

When we consider models with more
parameters, there are significant degeneracies, and external data sets
are essential for parameter constraints.  We use this approach in
\S \ref{sec:joint} and subsequent sections.

\subsection{Predictions from the WMAP Best Fit $\Lambda$CDM Model
\label{sec:wmap_pred}}

The WMAP data alone is now able to accurately constrain the basic six parameters
of the $\Lambda$CDM model.  
In this section, we focus on this model
and begin by using  only the WMAP data to fix the cosmological
parameters.  We then use the Markov chains (and linear theory) to
predict the amplitude of fluctuations in the local universe and
compare to other astronomical observations.  These comparisons test
the basic physical assumptions of the $\Lambda$CDM model. 

\subsubsection{Age of the Universe and $H_0$}

The
CMB data do not directly measure $H_0$; however, by measuring
$\Omega_m H_0^2$ through the height of the peaks and the conformal
distance to the surface of last scatter through the peak positions
\citep{page/etal:2003c}, the CMB data produces a determination of
$H_0$ {\it if we assume the simple flat $\Lambda$CDM model.}  Within
the context of the basic model of adiabatic fluctuations, the CMB
data provides a relatively
robust determination of the
age of the universe as the degeneracy in other cosmological parameters
 is  nearly orthogonal to measurements of the age
 \citep{knox/christensen/skordis:2001,hu/etal:2001}. 

\begin{figure} 
\centering
\includegraphics[width=5in]{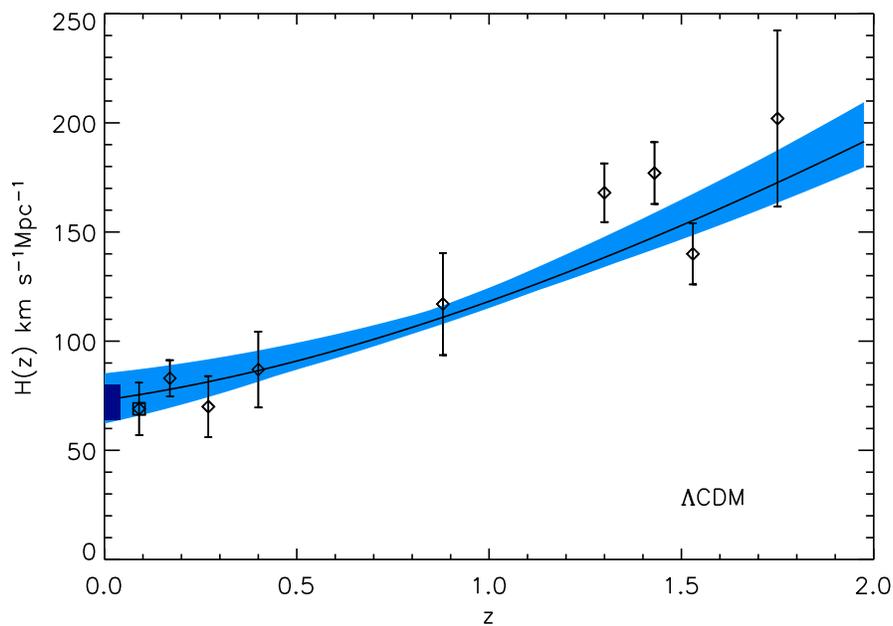}
\caption{\fg
The 
$\Lambda$CDM model fit to the WMAP data 
predicts the Hubble parameter redshift relation.  The blue band
shows the 68\% confidence interval for the Hubble parameter, $H$.
The dark blue rectangle shows the HST key project estimate for $H_0$
and its uncertainties \citep{freedman/etal:2001}.  The other points are from measurements
of the differential ages of galaxies, based on fits of
synthetic stellar population models to galaxy spectroscopy.  The squares
show values  from \citet{jimenez/etal:2003} analyses of SDSS galaxies.  The
diamonds show values from \citet{simon/verde/jimenez:2005} analysis of a high
redshift sample of red galaxies.
\label{fig:ages}}
\end{figure}

The WMAP $\Lambda$CDM best fit value for the age:
$\ensuremath{t_0 = 13.73^{+ 0.16}_{- 0.15}\ \mbox{Gyr}}$,
agrees with estimates of ages based on globular clusters
 \citep{chaboyer/krauss:2002} and white dwarfs \citep{hansen/etal:2004,richer/etal:2004}.  
Figure \ref{fig:ages} compares the predicted evolution of $H(z)$ to
the HST key project value \citep{freedman/etal:2001}
and to values from 
analysis of differential ages as a function
of redshift \citep{jimenez/etal:2003, simon/verde/jimenez:2005}. 
The WMAP best fit value, $H_0=$\ensuremath{73.2^{+ 3.1}_{- 3.2}}
km/s/Mpc, 
is also consistent with HST measurements
\citep{freedman/etal:2001}, $H_0=72\pm 8$ km/s/Mpc,  where
the error includes random and systematic uncertainties and the
estimate is based on several different methods (Type Ia supernovae,
Type II supernovae, surface brightness fluctuations and fundamental plane).
It also agrees with 
detailed studies
of gravitationally lensed systems such as B1608+656
\citep{koopmans/etal:2003}, which yields $75^{+7}_{-6}$ km/s/Mpc,
measurements of the Hubble constant from SZ and X-ray observations of clusters
\citep{bonamente/etal:2005}
that find $H_0 = 76^{+3.9}_{-3.4}\ ^{+10.0}_{-8.0}$ km/s/Mpc,
and recent measurements of the Cepheid distances to nearby galaxies that host type Ia supernova
\citep{riess/etal:2005}, $H_0 =73 \pm 4 \pm 5$  km/s/Mpc.

\subsubsection{Big Bang Nucleosynthesis}

Measurements of the light element abundances are among the most
important tests of the standard big bang model.   The WMAP estimate of the baryon abundance depends on our
understanding of acoustic oscillations 300,000 years 
after the big bang.  The BBN abundance
predictions depend on our understanding of physics in the first minutes after the big bang.

Table \ref{tab:bbn} lists
the primordial deuterium abundance,
$y_{D}^{FIT} $, the primordial $^3$He abundance, $y_{3}$, the primordial helium abundance, $Y_P$,
and the primordial $^7$Li abundance, $y_{\rm Li}$, based on
analytical fits to the predicted BBN abundances \citep{kneller/steigman:2004}  and the power-law $\Lambda$CDM 68\% confidence range for the
baryon/photon ratio, $\eta_{10} = (273.9\pm 0.3) \Omega_bh^2$. 
The lithium abundance is often expressed as a logarithmic abundance, [Li]$_P = 12 +
{\rm log_{10}}$(Li/H).
\begin{table}[h!]  
\begin{center}
\caption{Primordial abundances  based
on using  the \citet{steigman:2005} fitting formula
for the $\Lambda$CDM 3-year WMAP only value
for the baryon/photon ratio,  $\eta_{10} = 6.116^{+0.197}_{-0.249}$.
\label{tab:bbn}}
\begin{tabular}{|c|c|c|}
\hline
  & CMB-based BBN prediction & Observed Value \\
\hline
$10^{5} y_{D}^{FIT}$ & $2.57_{-0.13}^{+0.17}$&  1.6 -  4.0 \\
$10^{5} y_{3}$ & $1.05 \pm 0.03 \pm 0.03$ (syst.) &  $< 1.1 \pm 0.2$ \\
$Y_P$ & $0.24819^{+0.00029}_{-0.00040}  \pm 0.0006 ({\rm syst}.)$& 0.232 - 0.258\\ 
$[{\rm Li}]$& $2.64 \pm 0.03$ &2.2 - 2.4 \\
\hline
\hline
\end{tabular}
\end{center}
\end{table}

The systematic uncertainties in the helium abundances are due to the uncertainties in nuclear parameters, particularly neutron lifetime \citep{steigman:2005}. Prior to the measurements of
the CMB power spectrum, uncertainties in the baryon abundance were the biggest
source of uncertainty in CMB predictions.
 \citet{serebrov/etal:2005} argues that
the currently accepted value, $\tau_n = 887.5$\ s, should be reduced by $7.2$\ s, a shift of several times the reported errors in the Particle Data Book.  This (controversial) shorter lifetime lowers
the predicted best fit helium abundance to $Y_P = 0.24675$ \citep{mathews/kajino/shima:2005,steigman:2005}.

The deuterium abundance measurements provide the strongest test of the 
predicted baryon abundance.  \citet{kirkman/etal:2003} estimate
a primordial deuterium abundance, [D]/[H]$=2.78^{+0.44}_{-0.38} \times 10^{-5}$,
based on five QSO absorption systems.   The six systems used in
the \citet{kirkman/etal:2003} analysis show a significant range in abundances:
$1.65 - 3.98 \times 10^{-5}$ and have a scatter much larger than the quoted observational
errors.  Recently,
\citet{crighton/etal:2004} report a deuterium abundance of 
$1.6^{+0.5}_{-0.4} \times 10^{-5}$ for PKS 1937-1009.  Because of the large scatter,
we quote the range in [D]/[H] abundances in Table \ref{tab:bbn}; however, note that the mean
abundance is in good   agreement with the CMB prediction.

It is difficult to directly measure
the primordial $^3$He abundance.  \citet{bania/rood/balser:2002} quote an upper limit on the primordial $^3$He abundance of 
$y_3 < 1.1 \pm 0.2 \times 10^{-5}$.  This limit is compatible with the BBN predictions.

\citet{olive/skillman:2004} have reanalyzed the estimates of primordial helium abundance
based on observations of metal-poor HII regions.  They conclude that the errors in
the abundance are dominated by systematic errors and argue that a number of these
systematic effects have not been properly included in earlier analysis.  In Table \ref{tab:bbn},
we quote their estimate of the allowed range of $Y_P$.  \citet{olive/skillman:2004} find
a representative value of 
 $0.249 \pm 0.009$ for a linear fit of [O]/[H] to the helium abundance, significantly
 above earlier estimates and consistent with WMAP-normalized BBN.

While the measured abundances of the other light elements appear to be consistent
with BBN predictions, measurements of neutral  lithium abundance in low metallicity stars imply
values that are a factor of 2 below the BBN predictions:  most recent 
measurements \citep{charbonnel/primas:2005, boesgaard/stephens/deliyannis:2005} find an abundance of [{\rm Li}]$_P
\simeq 2.2 - 2.25$.  While  \citet{melendez/ramirez:2004} find a higher value,
[Li]$_P \simeq 2.37\pm0.05$, even this value is  still significantly below the BB-predicted value, $2.64 \pm 0.03$.  These discrepancies could be
due to systematics in the inferred lithium abundance \citep{steigman:2005}, 
uncertainties in the stellar temperature scale \citep{fields/olive/vangioni-flam:2005},
destruction of lithium in an early generation of stars or the signature of new early universe physics 
\citep{coc/etal:2004, jedamzik:2004, richard/michaud/richer:2005, ellis/olive/vangioni:2005, larena/alimi/serna:2005, jedamzik/etal:2005}.   The recent detection
\citep{asplund/etal:proceedings} of $^{6}$Li in several
low metallicity stars further constrains chemical evolution models and exacerbates the tensions between the BBN predictions and observations.

\subsubsection{Small Scale CMB Measurements}

\begin{figure}[h] 
\centering
\includegraphics[width=5in]{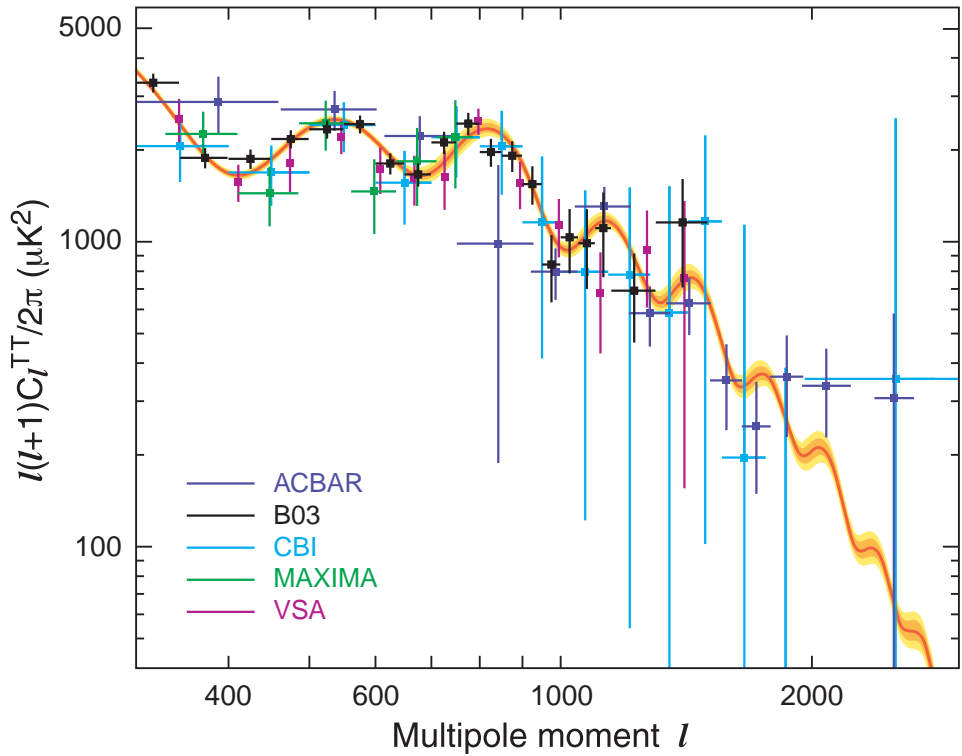}
\caption{\fg 
The prediction for the small-scale angular power spectrum seen by ground-based
and balloon CMB experiments from the $\Lambda$CDM model fit to the WMAP data
 only.
The colored lines 
show the best fit (red)  and the 68\%
 (dark orange) and 95\% confidence levels (light orange)
based on fits of the $\Lambda$CDM models to the WMAP data.
The points in the figure show 
small scale CMB measurements \citep{ruhl/etal:2003,abroe/etal:2004,
kuo/etal:2004,readhead/etal:2004,dickinson/etal:2004}.  
The plot shows that the $\Lambda$CDM model (fit to the
WMAP data alone) can accurately
predict the amplitude of fluctuations on the small scales measured by
ground and balloon-based experiments.
\label{fig:predict_small_scale_cmb}}
\end{figure}

With the  parameters of the $\Lambda${\it CDM} model fixed by the measurements of the first three  acoustic peaks, the basic properties
of the small scale CMB fluctuations are determined by the assumption
of a power law for the amplitude of potential fluctuations and by the
physics of Silk damping.  We test these assumptions by comparing the
WMAP best fit power law $\Lambda$CDM model to data from several recent
small scale CMB experiments (BOOMERanG,  MAXIMA, ACBAR, CBI, VSA).
These experiments probe smaller angular scales than the WMAP
experiment and are more sensitive to the details of recombination and
the physics of acoustic oscillations.  The good agreement seen in
Figure ({\ref{fig:predict_small_scale_cmb}}) suggests that the standard
cosmological model is accurately characterizing the basic physics at
$z \simeq 1100$. 
In subsequent sections,
we combine WMAP with small scale experiments.
We include four external CMB datasets which complement the WMAP
observations at smaller scales: the Cosmic Background Imager (CBI:
\citet{mason/etal:2003,sievers/etal:2003,pearson/etal:2003,readhead/etal:2004}),
the Very Small Array (VSA: \cite{dickinson/etal:2004}), the
Arcminute Cosmology Bolometer Array Receiver (ACBAR: \cite{kuo/etal:2004}) and BOOMERanG \citep{ruhl/etal:2003,montroy/etal:2005,piacentini/etal:2005}
We do not include results from a
number of experiments that overlap in $\ell$ range coverage with WMAP as these
experiments have non-trivial cross-correlations with WMAP that would have to be
included in the analysis.  

We do not use the small-scale polarization results for parameter determination as they do not yet noticeably improve constraints.  These polarization measurements, however,
already provide important tests on the basic assumptions of the model (e.g., adiabatic fluctuations and standard recombination history).

The measurements beyond the third peak improve constraints on the cosmological
parameters. These observations constrict the $\{ \tau,\omega_b, A_{s},n_{s}\}$
degeneracy and provide an improved probe of a running tilt in
the primordial power spectrum. In each case we only use bandpowers
that do not overlap with signal-dominated WMAP observations,
so that they can be treated as independent measurements.   

In the subsequent sections,
we perform likelihood analysis for two combinations of WMAP
data with other  CMB data sets:  
WMAP + high frequency bolometer experiments (ACBAR + BOOMERanG) and
WMAP + low frequency radiometer experiments (CBI+VSA).
The CBI data set is described in
\cite{readhead/etal:2004}.  We use 7 band powers, with mean $\ell$ values
of 948, 1066, 1211, 1355, 1482, 1692 and 1739,  from the even binning
of observations rescaled to a 32 GHz Jupiter temperature of 147.3
$\pm$ 1.8K. The rescaling reduces the calibration uncertainty to 2.6\%
from 10\% assumed in the first year  analyses; CBI
beam uncertainties scale the entire power spectrum and, thus,  act like
a calibration error.  We use a log-normal
form of the likelihood as in \cite{pearson/etal:2003}. The VSA data
\citep{dickinson/etal:2004} uses 5 band powers with mean $\ell$-values of
894, 995, 1117, 1269 and 1407, which are calibrated to the WMAP 32 GHz
Jupiter temperature measurement. The calibration uncertainty is
assumed to be 3\% and again we use a log-normal form of the likelihood.  
For ACBAR \citep{kuo/etal:2004}, we use the same band powers as in the first year analysis, with 
central $\ell$ values 842, 986, 1128, 1279, 1426, 1580, and 1716, and
errors from the ACBAR web site\footnote{See
  http://cosmology.berkeley.edu/group/swlh/acbar/data}. We assume a calibration uncertainty of 20\% in $C_\ell$, and the quoted 3\% beam uncertainty in Full Width Half Maximum.
We use the temperature data from the 2003 flight of BOOMERanG, based on the ``NA pipeline" \citep{jones/etal:2005} considering the 7 datapoints and covariance matrix for bins with mean $\ell$ values, 924, 974, 1025, 1076, 1117, 1211 and 1370.  While there is overlap between some
of the small scale power spectrum measurements and WMAP measurements
at $800 < l < 1000$, the WMAP measurements are noise-dominated so that there
is little covariance between the measurements.

\subsubsection{Large-Scale Structure}
\label{sec:lss}

With the WMAP polarization measurements constraining the  suppression
 of temperature anisotropy by reionization, we now  have an accurate measure of 
the amplitude of fluctuations at the redshift of decoupling. If 
the power-law $\Lambda$CDM model is an
accurate description of the large-scale properties of the universe,
then we can extrapolate forward the roughly 1000-fold growth in the
amplitude of fluctuations due to gravitational clustering and compare
the predicted amplitude of fluctuations to the large-scale structure
observations.  This is
a powerful test of the theory as some alternative models fit the CMB data 
but predict significantly different galaxy
power spectra (e.g.,
\citet{blanchard/etal:2003}).

\begin{figure}[h] 
\plottwo{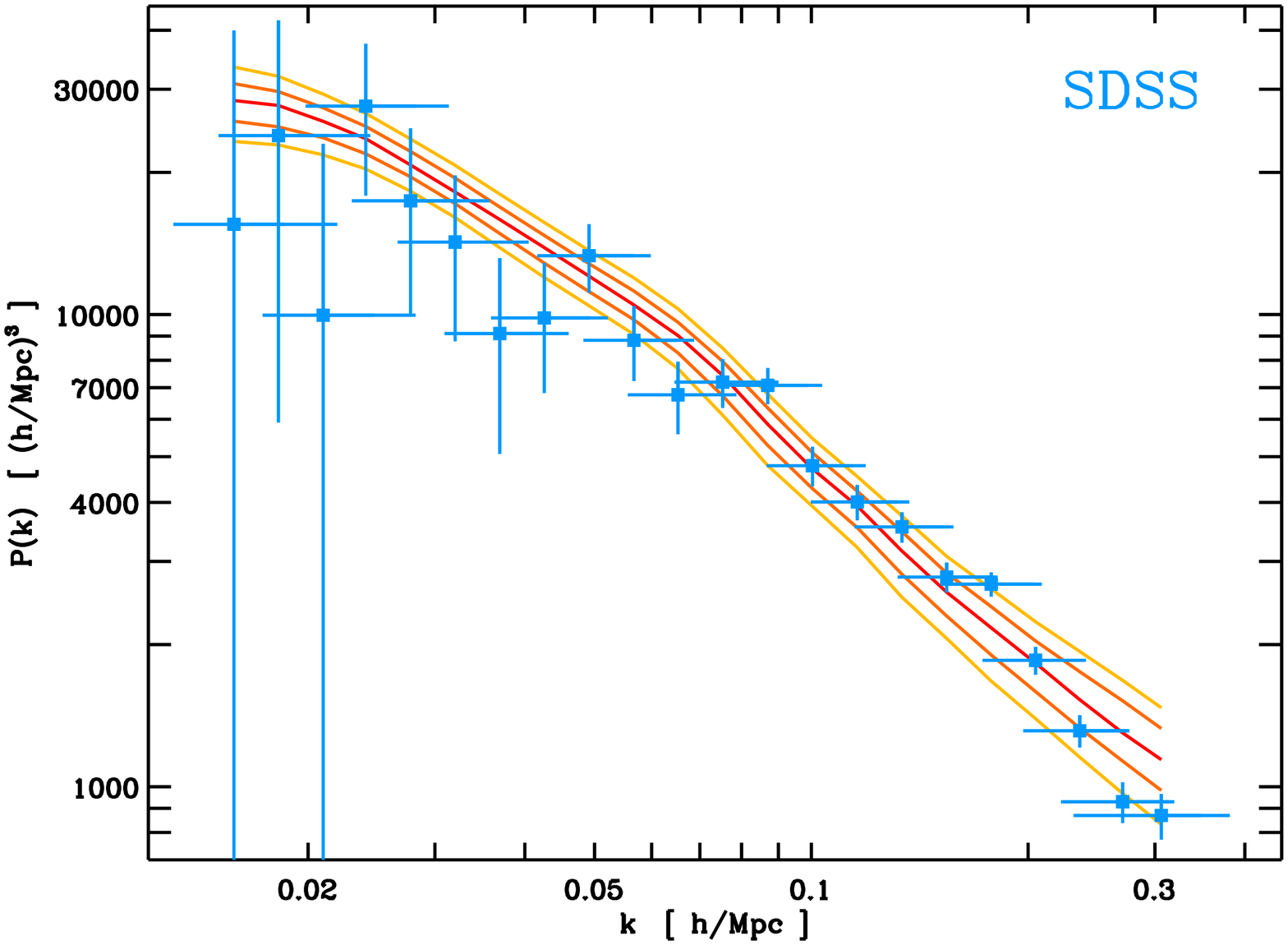}{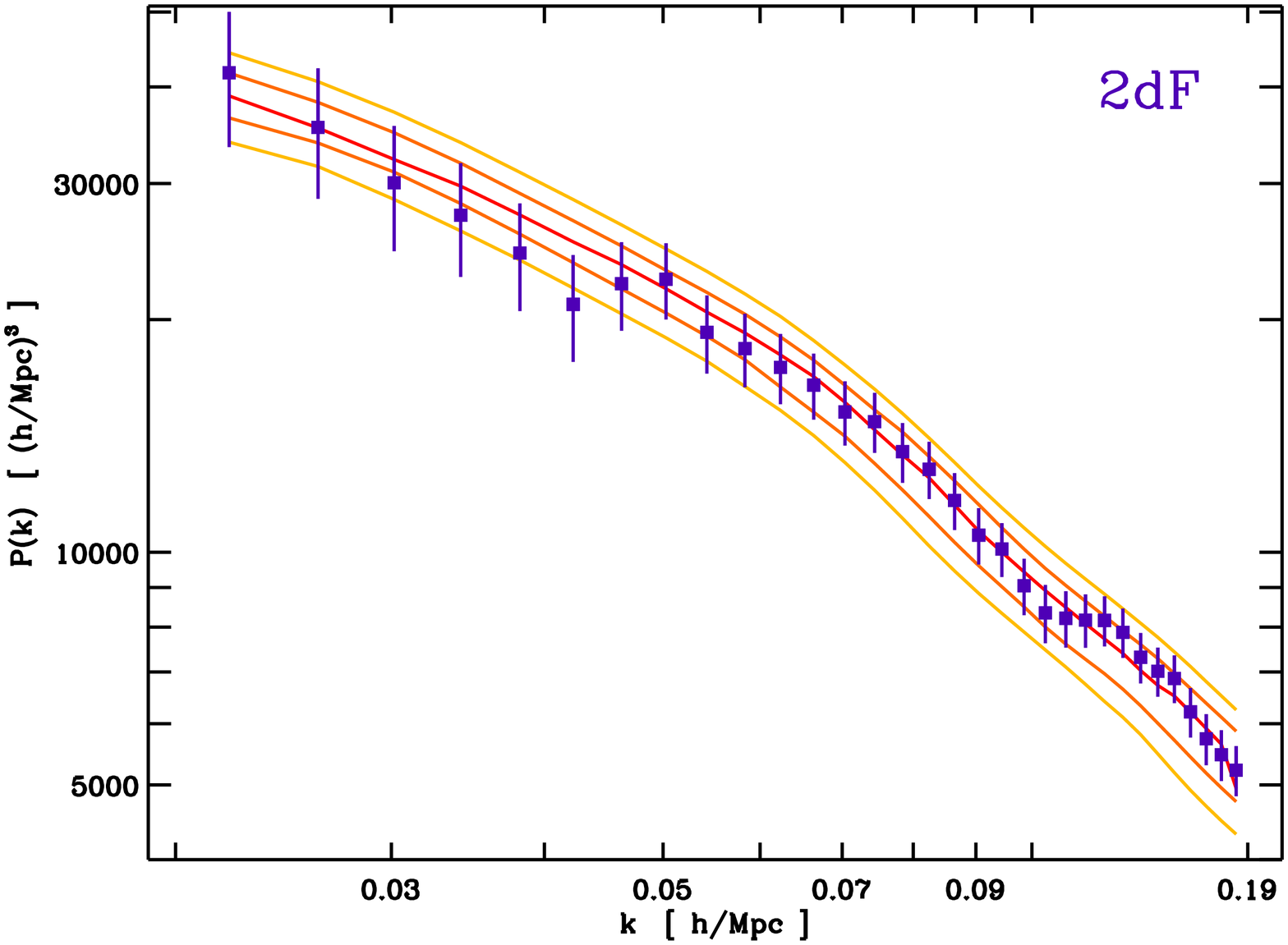}
\caption{\fg 
({\it Left}) The predicted
power spectrum (based on the range of parameters consistent with the
WMAP-only  parameters) is compared to the mass power spectrum inferred from the
SDSS galaxy power spectrum \citep{tegmark/etal:2004} as normalized
by weak lensing measurements \citep{seljak/etal:2005b}.
({\it Right})
The predicted
power spectrum is compared 
to the mass power spectrum inferred from
the 2dFGRS galaxy power spectrum \citep{cole/etal:2005} with the best
fit value for $b_{2dFGRS}$ based on the fit to the WMAP model.
Note that the 2dFGRS data points shown are correlated.
\label{fig:predictsdss}}
\end{figure}

Using {\it only} the WMAP data, together with linear theory, we can predict
the amplitude and shape of the matter power spectrum. The  band in
Figure \ref{fig:predictsdss} shows the 68\% confidence interval for the
matter power spectrum.  Most of the uncertainty in the figure is due
to the uncertainties in $\Omega_{m}h$.  The points in the figure show
the SDSS Galaxy power spectrum \citep{tegmark/etal:2004} with the
amplitude of the fluctuations normalized by the galaxy lensing
measurements and the 2dFGRS data \citep{cole/etal:2005}.  The figure shows that the $\Lambda$CDM model, when
normalized to observations at $z \sim 1100$, accurately predicts the
large-scale properties of the matter distribution in the nearby
universe. It also shows that adding the large-scale structure
measurements will reduce uncertainties in cosmological parameters. 

When we combine WMAP with large-scale structure observations
in subsequent sections,  
we consider the measurements of the power
spectrum from the two large-scale structure surveys. 
Since
the galaxy power spectrum does not suffer the optical
depth-driven suppression in power seen in the CMB, large scale
structure data gives an independent measure of the normalization of
the initial power spectrum (to within the uncertainty of the galaxy
biasing and redshift space distortions)  and significantly truncates
the $\{ \tau,\omega_{b},A_{s},n_{s}\}$ degeneracy.  In addition the
galaxy power spectrum shape is determined by $\Omega_{m}h$ as opposed
to the $\Omega_{m}h^{2}$ dependency of the CMB. Its inclusion
therefore further helps to break the $\{ \omega_{m},\Omega_{\Lambda},
w$ or $\Omega_{k}\}$ degeneracy.  

The 2dFGRS survey probes the universe at redshift $z\sim$0.1 (we
assume $z_{eff}=0.17$ for the effective
redshift for the survey) and probes the power spectrum on scales $0.022\
 h$Mpc$^{-1}<k<0.19\ h$Mpc$^{-1}$. 
From the data and covariance described in \cite{cole/etal:2005} we use 32 of the 36 bandpowers in the range $0.022\
 h$Mpc$^{-1}<k<0.19\ h$Mpc$^{-1}$.
We correct for non-linearities and non-linear redshift space distortions using the prescription employed by the 2dF team,
\begin{equation}
P_{gal}^{redsh}(k)=\frac{1+Qk^2}{1+Ak}P_{gal}^{theory}(k)
\end{equation}
where $P_{gal}^{redshift}$ and $P_{gal}^{theory}$ are the redshift space
and theoretical real space galaxy power spectra
with $Q=4.$ (Mpc/h)$^2$ and $A=1.4$ Mpc/h. We analytically marginalize over the power spectrum amplitude, effectively applying no prior on the linear bias and on linear redshift space distortions, in contrast to our first year analyses.

The SDSS  main galaxy survey  measures the galaxy distribution
at redshift of  $z\sim0.1$; however, as
in the analysis of the SDSS team \citep{tegmark/etal:2004} we assume
$z_{eff}=0.07$, and we use 14
power spectrum bandpowers between $0.016 h\ {\rm Mpc}^{-1}<k<0.11 h$\ Mpc$^{-1}$. We follow the approach  used in the 
SDSS analysis  in \citet{tegmark/etal:2004b}:  we  use
the nonlinear evolution of clustering as described in
\citet{smith/etal:2003} and include a linear bias
factor, $b_{sdss}$, and the linear redshift space distortion parameter, $\beta$.
\begin{eqnarray}
P_{gal}^{redsh}(k)& =& (1+\frac{2}{3}\beta+\frac{1}{5}\beta^2) P_{gal}^{theory}(k)
\end{eqnarray}
Following \citet{lahav/etal:1991}, we use
$\beta b = d\ln \delta/d\ln a$
where
$\beta \approx [\Omega_{m}^{4/7}+(1+\Omega_{m}/2)(\Omega_{\Lambda}/70)]/b$.
We impose a Gaussian prior on the bias of b = 1.03 +/- 0.15, based on an estimate from weak
lensing of the same SDSS galaxies used to derive the matter power spectrum. This value  includes a 4\% calibration uncertainty in quadrature with the reported bias error  \footnote{M. Tegmark private communication.}
and  is a symmetrized form of the 
bias constraint in Table 2 of \citet{seljak/etal:2005b}.
While the WMAP first year data was used in the \citet{seljak/etal:2005b}
analysis, the covariance between the data sets are small.
We restrict our analysis to
scales where the bias of a given galaxy population does not show significant scale
dependence \citep{zehavi/etal:2005}.  Analyses that use galaxy clustering data 
on smaller scales require detailed modeling of non-linear dynamics
and the relationship between galaxy halos and galaxy properties (see,
e.g., \citet{abazajian/etal:2005}). 

The SDSS luminous red galaxy (LRG) survey uses the brightest class of galaxies in the
SDSS survey \citep{eisenstein/etal:2005}.   While a much smaller galaxy sample than the main SDSS galaxy sample, it has
the advantage of  covering 0.72 $h^{-3}$ Gpc$^3$ between $0.16 < z < 0.47$.   Because
of its large volume, this survey was able to detect the acoustic peak in the galaxy correlation,
one of the distinctive predictions of the standard adiabatic cosmological model \citep{peebles/yu:1970,sunyaev/zeldovich:1970,silk/wilson:1981,bond/efstathiou:1984,vittorio/silk:1984,bond/efstathiou:1987}. We use the SDSS acoustic peak results to constrain the balance of the matter content, using the well measured combination,
\begin{equation}
A(z=0.35) \equiv  \left[\frac{d_A(z=0.35)^{2}}
{H(z=0.35)}\right]\sqrt{\Omega_{m}H_{0}^{2}}
\end{equation}  
where $d_A$ is the comoving angular diameter distance and $c$
is the speed of light.   We  impose a Gaussian prior of 
$A = 0.469\left(\frac{n_{s}}{0.98}\right)^{-0.35}\pm 0.017$ based
on the analysis of \citet{eisenstein/etal:2005} .

\subsubsection{Lyman $\alpha$ Forest}
Absorption features in high redshift quasars (QSO) at around the
frequency of the Lyman-$\alpha$ emission line are thought to be
produced by regions of low-density gas at
redshifts $2<z<4$ \citep{croft/etal:1998,gnedin/hamilton:2002}. These features allow the
matter distribution to be characterized on scales of $0.2 < k < 5\
h{\ \rm Mpc}^{-1}$ and as such extend the lever arm provided by 
combining large-scale structure data and CMB. These observations also
probe a higher redshift range ($z \sim 2- 3$). Thus, these
observations nicely complement CMB measurements and large scale
structure observations.  
While there has been
significant progress in understanding systematics in the past few years \citep{mcdonald/etal:2005,meiksin/white:2004}, time constraints limit our ability to consider all relevant data sets.

Recent fits to the Lyman-$\alpha$ forest imply a higher
amplitude of density fluctuations:
\cite{jena/etal:2005}
 find that  $\sigma_8 = 0.9,
\Omega_m = 0.27, h = 0.71$ provides a good fit to the Lyman $\alpha$ data.
\citet{seljak/etal:2005}
combines first year WMAP data, other CMB experiments, 
large-scale structure and Lyman $\alpha$
to find: $n_s = 0.98 \pm 0.02, \sigma_8 = 0.90 \pm 0.03, h = 0.71 \pm 0.021,$ 
and
$\Omega_m = 0.281_{-0.021}^{+0.023}$.
Note that if they assume $\tau = 0.09$, the best fit value drops to $\sigma_8 = 0.84$.
While these models have somewhat higher amplitudes than the new best fit WMAP values, a recent analysis by \citet{desjacques/nusser:2005}
find that the Lyman $\alpha$ data is consistent with $\sigma_8$ between
$0.7 -0.9$.  This suggests that the Lyman $\alpha$ data is consistent with the 
new WMAP best
fit values; however, further analysis is needed.

\subsubsection{Galaxy Motions and Properties}
Observations of galaxy peculiar velocities probe the growth rate of
structure and are sensitive to the matter density and the amplitude of
mass fluctuations. The \citet{feldman/etal:2003} analysis of peculiar
velocities of nearby ellipticals and spirals finds $\Omega_m =  0.30^{+0.17}_{-0.07}$ and $\sigma_8 = 1.13^{+0.22}_{-0.23}$, within $1 \sigma$ of
the WMAP best fit value for $\Omega_m$ and 1.5$\sigma$ higher
than the WMAP value for $\sigma_8$. These estimates are based on dynamics
and are not sensitive to the shape of the power spectrum.  
\citet{mohayaee/tully:2005} apply orbit retracing methods to motions
in the local supercluster and obtain $\Omega_m = 0.22 \pm0.02$,
consistent with the WMAP values.

Modeled galaxy properties can be compared to the clustering properties of galaxies on smaller scales.
The best fit parameters for WMAP only 
are consistent with the recent \citet{abazajian/etal:2005}
analysis of the pre-three year release CMB data combined with the SDSS data.
In their analysis, they fit a Halo Occupation Distribution model to
the galaxy distribution so as to use the galaxy clustering data at
smaller scales. Their best fit parameters ($H_0 = 70 \pm 2.6\ {\rm km/s/Mpc}, \Omega_m = 0.271 \pm 0.026$) are
consistent with the results found here.  \citet{vale/ostriker:prep} fit
the observed galaxy luminosity functions with $\sigma_8 = 0.8$ and 
$\Omega_m = 0.25$.  \citet{vandenbosch/etal:2003} 
used the conditional luminosity function to fit the 2dFGRS luminosity function and the
correlation length as a function of luminosity.  Combining with the first-year WMAP data,
they found $\Omega_m = 0.25^{+0.10}_{-0.07}$ and $\sigma_8 = 0.78\pm0.12$  (95\% CL),
again in remarkable agreement with the three year WMAP best fit values. 

\subsubsection{Weak Lensing \label{sec:lensing}}
\begin{figure}[htbp!] 
\centering
\includegraphics[width=4in]{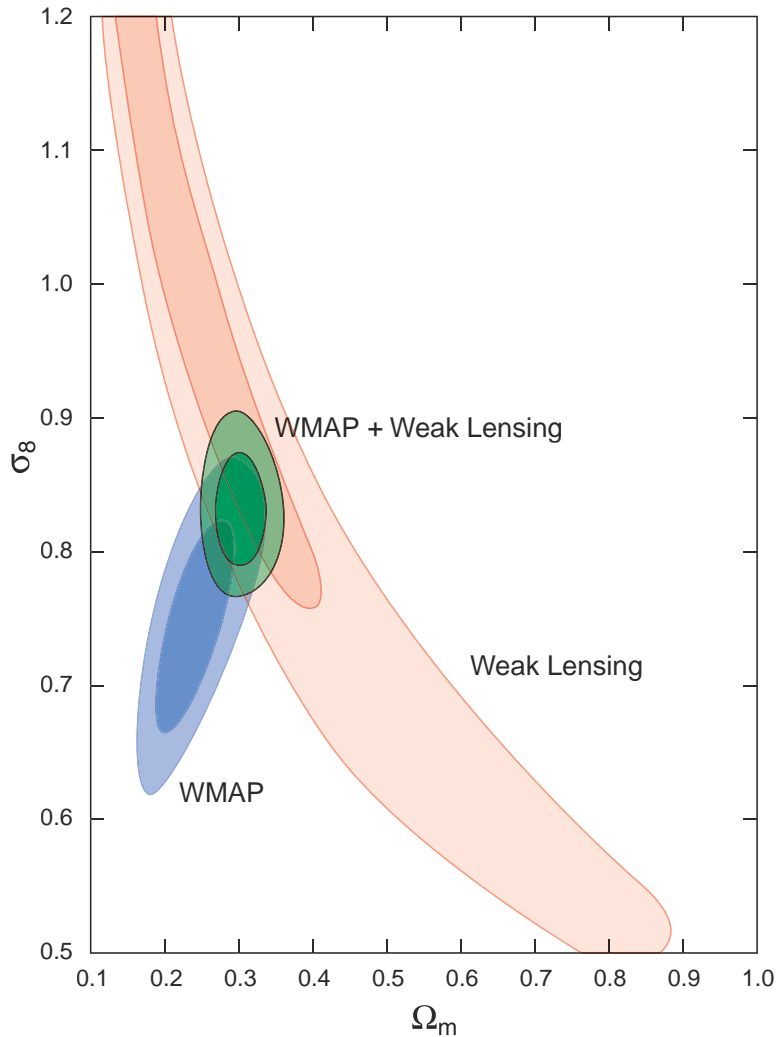}
\caption{\fg
Prediction for the
mass fluctuations measured by the CFTHLS weak-lensing survey
from the $\Lambda$CDM model fit to the WMAP data only.
The blue, red and green contours show the joint 2-d marginalized
68\% and 95\% confidence limits in the ($\sigma_8$, $\Omega_m$) plane for
for WMAP only, CFHTLS only and WMAP + CFHTLS, respectively, 
for the power law $\Lambda$CDM models.  All constraints come from assuming the same priors on input parameters, with the additional marginalization over $z_{s}$ in the weak lensing analysis, using a top hat prior of $0.613 <z_s< 0.721$ .
While lensing data favors higher values of $\sigma_8 \simeq 0.8 - 1.0$
(see \S \ref{sec:lensing}), X-ray cluster studies favor lower
values of $\sigma_8 \simeq 0.7-0.8$ (see \S \ref{sec:cluster}).
\label{fig:lensing}}
\end{figure}
Over the past few years, there has been dramatic progress in using weak lensing data as a probe of mass
fluctuations in the nearby universe.  Lensing surveys complement CMB measurements \citep{contaldi/hoekstra/lewis:2003,tereno/etal:2005},
and their dominant systematic uncertainties differ from the large-scale
structure surveys.

Measurements of weak gravitational lensing, the distortion of galaxy
images by the distribution of mass along the line of sight, directly
probe the distribution of mass fluctuations along the line of sight
(see \citet{refregier:2003} for a recent review).  Figure~\ref{fig:lensing}
shows that the WMAP 
$\Lambda$CDM model predictions for $\sigma_8$ and $\Omega_m$ 
are lower than the amplitude found in most recent lensing surveys:
\citet{hoekstra/yee/gladders:2002} calculate $\sigma_8 = 
0.94^{+0.10}_{-0.14} (\Omega_m/0.25)^{-0.52}$ (95\% confidence) from the
RCS survey and
\citet{vanwaerbeke/mellier/hoekstra:2005} determine
$\sigma_8 = 0.91 \pm 0.08 (\Omega_m/0.25)^{-0.49}$ from the VIRMOS-DESCART survey; 
however, \citet{jarvis/etal:2003} find $\sigma_8 = 0.79^{+0.13}_{-0.16}
(\Omega_m/0.25)^{-0.57}$ (95\% confidence level) from the 
75 Degree CTIO survey.

In \S \ref{sec:joint}, we use the data set provided by the first weak
gravitational lensing analysis of the Canada-France-Hawaii Telescope Legacy Survey (CFHTLS)
\footnote{http://www.cfht.hawaii.edu/Science/CFHTLS} as conducted by
\citet{hoekstra/etal:2005} (Ho05) and \citet{semboloni/etal:2005}.
Following Ho05, we use only the wide fields W1 and W3, hence a total
area of 22 deg$^2$ observed in the $i^\prime$ band limited to
a magnitude of $i^\prime=24.5$. We follow the same methodology as Ho05
and \citet{tereno/etal:2005}. For each given model and set of
parameters,  we compute the predicted shear variance at various
smoothing scales, $\langle\gamma^2 \rangle$, and then evaluate its likelihood (see Ho05 equation 13).

Since  the lensing data
 are in a noise dominated regime, we neglect
the cosmological dependence of the covariance matrix. To account
conservatively for a possible residual systematic contamination, we
use $\langle \gamma^2_{B} \rangle$ as a  monitor and add it in quadrature to the
diagonal of the noise covariance matrix,  as in Ho05. We
furthermore marginalize over the mean  source redshift, $z_s$ (defined in equation (16) of HoO5)
assuming a uniform prior between 0.613 and 0.721. This marginalization is
performed by including these extra parameters in the Monte Carlo Markov Chain. 
Our analysis differs however from the likelihood analysis of Ho05 in
the choice of the transfer function. We use the 
\citet{novosyadlyj/durrer/lukash:1999}(NDL) CDM transfer function
(with the assumptions of  \citet{tegmark/zaldarriaga/hamilton:2001})
rather than the \citet{bardeen/etal:1986} (BBKS) CDM transfer
function. The NDL transfer function includes more accurately
baryon oscillations and neutrino effects.  This modification alters
the shape of the likelihood surface in
the 2-dimensional $(\sigma_8,\Omega_m)$ likelihood space. 

\subsubsection{Strong  Lensing}
Strong lensing provides another potentially 
powerful probe of cosmology.  The number
of multiply-lensed arcs and quasars is very sensitive to the underlying
cosmology.  The cross-section for lensing depends 
on the number of systems with surface densities 
above the critical density, which in turn
is sensitive to the angular diameter distance relation \citep{turner:1990}.
The CLASS lensing survey \citep{chae/etal:2002} finds that
the number of lenses detected in the radio survey is consistent with
a flat universe with a cosmological constant and $\Omega_m =0.31_{-0.14}^{+0.27}$.  The statistics of strong lenses in the SDSS is
also consistent with the standard $\Lambda$CDM cosmology \citep{oguri:2004}.
The number and the properties of lensed arcs are also quite
sensitive to cosmological parameters (but also to the details
of the data analysis).  \citet{wambsganss/bode/ostriker:2004} conclude
that arc statistics are consistent with the concordance $\Lambda$CDM
model.

\citet{soucail/kneib/golse:2004} has used multiple lenses 
in Abell 2218 to provide another 
geometrical tests of cosmological parameters.  They find that $0 < \Omega_m < 0.33$ and $w < -0.85$ for a flat universe with dark energy.  This method
is another independent test of the standard  cosmology.

\subsubsection{Clusters and the Growth of Structure
\label{sec:cluster}}

The numbers and properties of rich clusters are another
tool for testing the emerging standard model.  Since clusters are rare, the number of clusters as a function of redshift is a
sensitive probe of cosmological parameters.  Recent analyses of both
optical and X-ray cluster samples yield cosmological parameters
consistent with the best fit WMAP $\Lambda$CDM model
\citep{borgani/etal:2001,bahcall/bode:2003, allen/etal:2003,vikhlinin/etal:2003,henry:2004}. 
The parameters are, however, sensitive to
uncertainties in the conversion between observed properties and cluster
mass \citep{pierpaoli/etal:2003,rasia/etal:2005}.  

Clusters can also be used to infer cosmological parameters through
measurements of the baryon/dark matter ratio as a function of redshift
\citep{pen:1997, ettori/tozzi/rosati:2003, allen/etal:2004}. Under the
assumption that the baryon/dark matter ratio is constant with
redshift, the Universe is flat, and standard baryon densities,
\citet{allen/etal:2004} find $\Omega_m = 0.24\pm 0.04$ and $w =-1.20_{-0.28}^{+0.24}.$   \citet{voevodkin/vikhlinin:2004} determine $\sigma_8 = 0.72 \pm 0.04$
and $\Omega_m h^2 = 0.13 \pm 0.07$ from measurements of the baryon fraction.
These parameters are consistent with the 
values found here and in \S \ref{sec:w}. 

\subsubsection{Integrated Sachs-Wolfe (ISW) effect}
\label{sec:isw}

The $\Lambda$CDM model
predicts a statistical
correlation between CMB temperature fluctuations and 
the large-scale distribution of matter \citep{crittenden/turok:1996}.
Several groups have detected correlations between the WMAP
measurements and various tracers of large-scale structure at levels
consistent with the concordance $\Lambda$CDM model
\citep{boughn/crittenden:2004, nolta/etal:2004,
  afshordi/loh/strauss:2004, scranton/etal:prep,
  fosalba/gaztanaga:2004, padmanabhan/etal:2005,corasaniti/giannantonio/melchiorri:2005,boughn/crittenden:2005,vielva/martinez-gonzalez/tucci:2006}.  These detections provide an important
independent test of the effects of dark energy on the growth of 
structure.  However, the first 
year WMAP data is already signal dominated on the scales probed by the
ISW effect, thus, improved large-scale structure surveys are needed to
improve the statistical significance of this detection
\citep{afshordi:2004,bean/dore:2004,pogosian/etal:2005}. 

\subsubsection{Supernova}
\begin{figure}[h!] 
\centering
\includegraphics[width=6in]{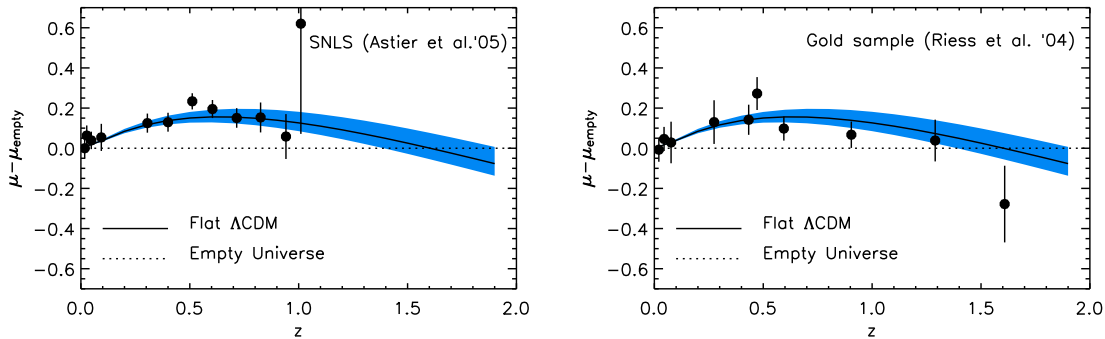}
\caption{\fg
Using the WMAP $\Lambda$ CDM parameters, we predict  
luminosity distance-redshift relationship  and compare it to measurements
from supernova surveys. 
The plots show the deviations of the distance measure ($DM$)
from the empty universe model.  The solid lines are for
the best WMAP $\Lambda$CDM parameters and the blue band shows
the 68\% confidence range. ({\it Left})  SNLS data\citep{astier/etal:2005}.
({\it Right})``gold'' supernova data \citep{riess/etal:2004}.
\label{fig:sn}}
\end{figure}

With the realization that their light curve shapes could be used to
make SN Ia into standard candles, supernovae have
become an important cosmological probe
\citep{phillips:1993,hamuy/etal:1996,riess/press/kirshner:1996}.
They can be used to measure the luminosity distance as a function of
redshift. The dimness of $z \approx 0.5$ supernova provide direct
evidence for the accelerating universe
\citep{riess/etal:1998,schmidt/etal:1998,perlmutter/etal:1999,tonry/etal:2003,knop/etal:2003,nobili/etal:2005,clocchiatti/etal:2005,krisciunas/etal:2005,astier/etal:2005}.
Recent HST measurements \citep{riess/etal:2004} trace the
luminosity distance/redshift relation out to higher redshift and
provide additional evidence for presence of dark energy. Assuming a
flat Universe, the \citet{riess/etal:2004a} analysis of the supernova data
alone finds that $\Omega_m = 0.29_{-0.03}^{+0.05}$ consistent
with the fits to WMAP data alone (see Table
\ref{tab:lcdmwmaponly}) and to various combinations of CMB and LSS
data sets (see Tables \ref{tab:lcdm_low} and \ref{tab:lcdm_high}). \citet{astier/etal:2005} find that
$\Omega_m = 0.263^{+0.042}_{-0.042}(stat.)_{-0.032}^{+0.032} (sys.)$ from the first year supernova legacy survey.   

Within the $\Lambda$CDM model, the supernovae data serve as
a test of our cosmological model.
Figure \ref{fig:sn}
shows the consistency between the supernova and CMB data. Using just
the WMAP data and the $\Lambda$CDM model, we can predict the
distance/luminosity relationship and test it with the supernova data. 

In \S \ref{sec:joint} and subsequent sections,
we consider two recently published high-$z$ supernovae datasets in combination with the WMAP CMB data,   the first sample is 157 supernova in the ``Gold Sample"
as described in \citet{riess/etal:2004} with $0.015 < z < 1.6$ based on a combination of  ground-based data and the  GOODS ACS Treasury program using the Hubble Space Telescope (HST) and the second sample is 115 supernova in
the range $0.015 < z < 1 $  from the Supernova Legacy Survey (SNLS) \citep{astier/etal:2005} .  

Measurements of the apparent magnitude, $m$, and inferred absolute magnitude, $M_0$, of each SN have been  used to derive the distance modulus $\mu_{obs}= m-M_0$, from which a luminosity distance is inferred,
$\mu_{obs} = 5\log [d_{L}(z)/{\rm Mpc}]+25$. The luminosity distance predicted from theory,
$\mu_{th}$,
is compared to observations using a $\chi^{2}$ analysis summing over the SN sample. 
\begin{equation}
\chi^{2} = \sum_{i}\frac{(\mu_{obs,i}(z_{i})-\mu_{th}(z_{i},M_0))^{2}}{\sigma_{obs,i}^{2}}
\end{equation}
where the absolute magnitude, $M_0$, is a ``nuisance parameter", analytically marginalized over in the likelihood analysis \citep{lewis/bridle:2002}, and $\sigma_{obs}$ contains systematic errors related to the light curve stretch factor, K-correction, extinction and the intrinsic redshift dispersion due to SNe peculiar velocities (assumed 400 and 300 km s$^{-1}$ for HST/GOODS and SNLS data sets respectively). 

\subsection{Joint Constraints on $\Lambda$CDM Model Parameters
\label{sec:joint}}
\begin{table}[h!] 
\begin{center}
\caption{$\Lambda$CDM Model: Joint Likelihoods 
\footnotesize These values are calculated using the  $N_{side}=8$ likelihood
code with $A_{PS}=0.017$ \label{tab:lcdm_low}}
\begin{tabular}{|c||c|c|c|c|}
\hline
\hline
&WMAP   & WMAP& WMAP+ACBAR  & WMAP + \\
&Only & +CBI+VSA & +BOOMERanG  &2dFGRS   \\
 &   & & &   \\
\hline
\hline
100$\Omega_b h^2$ & 
\ensuremath{2.230^{+ 0.075}_{- 0.073}} &
\ensuremath{2.208\pm 0.071} &
\ensuremath{2.232\pm 0.074} &
\ensuremath{2.223^{+ 0.069}_{- 0.068}}  \\
$\Omega_m h^2 $ & 
\ensuremath{0.1265^{+ 0.0081}_{- 0.0080}} &
\ensuremath{0.1233^{+ 0.0075}_{- 0.0074}} &
\ensuremath{0.1260\pm 0.0081} &
\ensuremath{0.1261\pm 0.0050}  \\
$h$ & 
\ensuremath{0.735\pm 0.032} &
\ensuremath{0.742\pm 0.031} &
\ensuremath{0.739^{+ 0.033}_{- 0.032}} &
\ensuremath{0.733^{+ 0.020}_{- 0.021}}  \\
$\tau$ & 
\ensuremath{0.088^{+ 0.029}_{- 0.030}} &
\ensuremath{0.087\pm 0.029} &
\ensuremath{0.088^{+ 0.031}_{- 0.032}} &
\ensuremath{0.083\pm 0.028}  \\
$n_s$ & 
\ensuremath{0.951\pm 0.016} &
\ensuremath{0.947\pm 0.015} &
\ensuremath{0.951\pm 0.016} &
\ensuremath{0.948\pm 0.015}  \\
\hline
\hline
$\sigma_8$ & 
\ensuremath{0.742\pm 0.051} &
\ensuremath{0.721^{+ 0.047}_{- 0.046}} &
\ensuremath{0.739^{+ 0.050}_{- 0.051}} &
\ensuremath{0.737\pm 0.036} 
\\
$\Omega_m $ & 
\ensuremath{0.237\pm 0.034} &
\ensuremath{0.226\pm 0.031} &
\ensuremath{0.233^{+ 0.033}_{- 0.034}} &
\ensuremath{0.236\pm 0.020} 
\\
\hline
\end{tabular}
\end{center}
\end{table}

\begin{table}[h!]  
\begin{center}
\caption{$\Lambda$CDM Model \label{tab:lcdm_high}}
\begin{tabular}{|c||c|c|c|c|c|}
\hline
\hline
&WMAP+   & WMAP+& WMAP+  & WMAP + & WMAP+ \\
&SDSS & LRG & SNLS&  SN Gold   & CFHTLS \\
& & & & &   \\
\hline
\hline
$100\Omega_b h^2$ & 
\ensuremath{2.230^{+ 0.071}_{- 0.070}} &
\ensuremath{2.242^{+ 0.069}_{- 0.070}} &
\ensuremath{2.234^{+ 0.075}_{- 0.074}} &
\ensuremath{2.230^{+ 0.069}_{- 0.072}} &
\ensuremath{2.255\pm 0.067} \\
$\Omega_m h^2 $ & 
\ensuremath{0.1327^{+ 0.0063}_{- 0.0064}} &
\ensuremath{0.1336\pm 0.0049} &
\ensuremath{0.1293\pm 0.0059} &
\ensuremath{0.1349^{+ 0.0061}_{- 0.0060}} &
\ensuremath{0.1409\pm 0.0038} \\
$h$ & 
\ensuremath{0.710\pm 0.026} &
\ensuremath{0.709^{+ 0.019}_{- 0.018}} &
\ensuremath{0.724\pm 0.023} &
\ensuremath{0.701\pm 0.021} &
\ensuremath{0.687\pm 0.018} \\
$\tau$ & 
\ensuremath{0.080^{+ 0.029}_{- 0.030}} &
\ensuremath{0.082\pm 0.029} &
\ensuremath{0.085\pm 0.030} &
\ensuremath{0.079^{+ 0.030}_{- 0.029}} &
\ensuremath{0.088^{+ 0.028}_{- 0.027}} \\
$n_s$ & 
\ensuremath{0.948^{+ 0.016}_{- 0.015}} &
\ensuremath{0.951\pm 0.016} &
\ensuremath{0.950^{+ 0.016}_{- 0.017}} &
\ensuremath{0.946\pm 0.016} &
\ensuremath{0.953\pm 0.016} \\
\hline
\hline
$\sigma_8$ & 
\ensuremath{0.772^{+ 0.040}_{- 0.041}} &
\ensuremath{0.780\pm 0.036} &
\ensuremath{0.758\pm 0.041} &
\ensuremath{0.784^{+ 0.042}_{- 0.041}} &
\ensuremath{0.827^{+ 0.026}_{- 0.025}} \\
$\Omega_m $ & 
\ensuremath{0.265\pm 0.030} &
\ensuremath{0.266^{+ 0.020}_{- 0.021}} &
\ensuremath{0.248^{+ 0.024}_{- 0.025}} &
\ensuremath{0.276\pm 0.026} &
\ensuremath{0.300\pm 0.021}  \\
\hline
\end{tabular}
\end{center}
\end{table}

In the previous section we showed that the power-law LCDM model fit to WMAP data only is consistent with other astronomical data.  Motivated by this stringent series of cosmological tests, we combine the WMAP data with other astronomical observations to further constrain cosmological parameters.

\begin{figure} 
\centering
\includegraphics[width=5in]{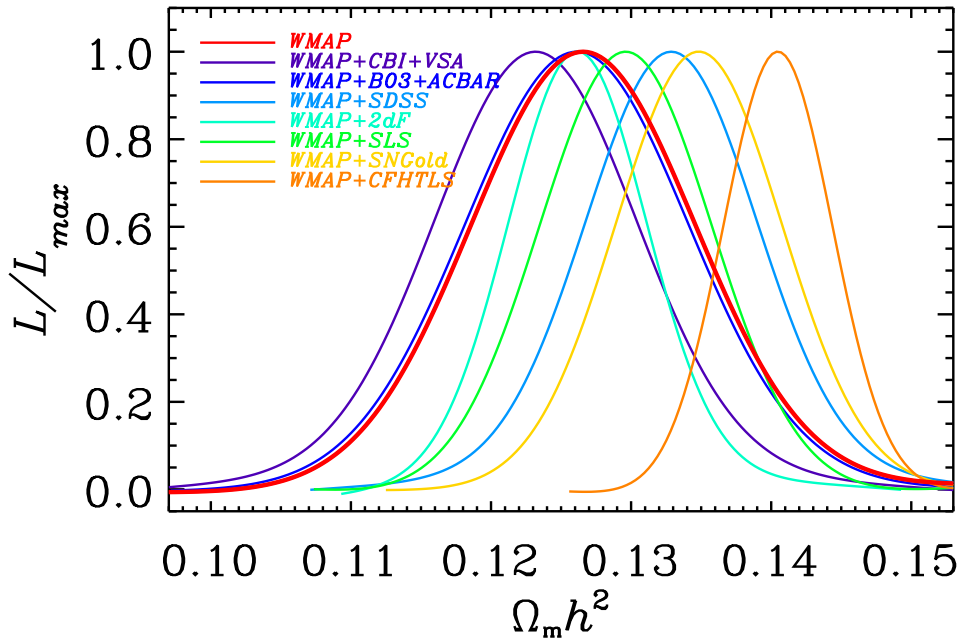}
\caption{\fg 
One-dimensional marginalized distribution of
$\Omega_m h^2$ for WMAP, WMAP+CBI+VSA, WMAP+BOOM+ACBAR,
WMAP+SDSS, WMAP+SN(SNLS), WMAP+SN(HST/GOODS), WMAP+2dFGRS and WMAP+CFHTLS for
the power-law $\Lambda$CDM model.
\label{fig:omegamh2}}
\end{figure}

\begin{figure}[htbp!] 
\centering
\includegraphics[width=6in]{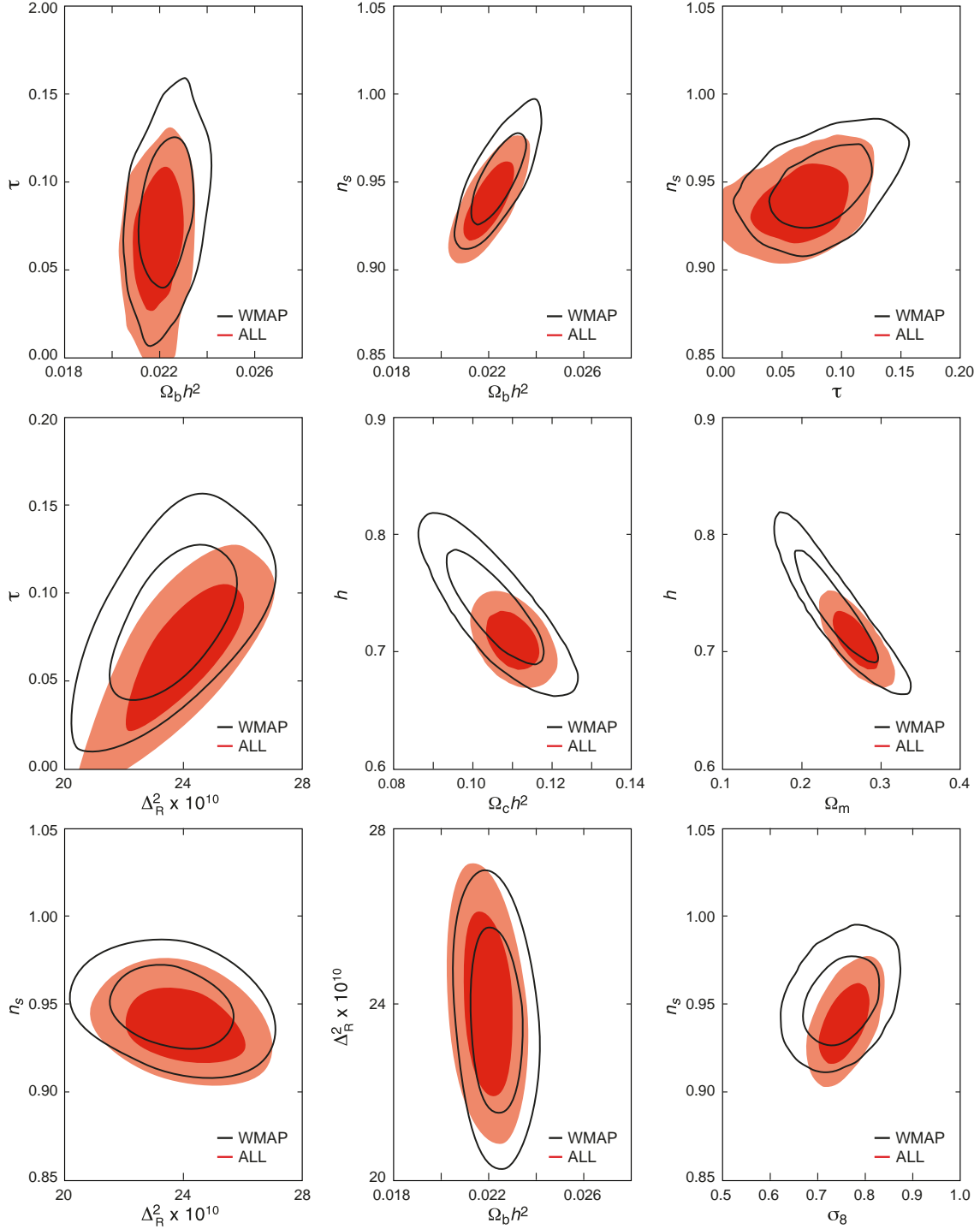}
\caption{\fg 
Joint two-dimensional marginalized contours (68\%, and 95\% confidence levels)
for various combination of parameters for WMAP only (solid lines)
and WMAP+2dFGRS+SDSS+ACBAR+BOOMERanG+CBI+VSA+
SN(HST/GOODS)+SN(SNLS) (filled red) for the 
power-law $\Lambda$CDM model.
\label{fig:CMBall}}
\end{figure}

Tables \ref{tab:lcdm_low} and \ref{tab:lcdm_high} show that adding
external data sets has little effect on several parameters: $\Omega_b h^2$, $n_s$
and $\tau$.   However, the various combinations do reduce the uncertainties
on $\Omega_m$ and the amplitude of fluctuations.  The data sets used
in Table \ref{tab:lcdm_low} favor smaller values of the matter density, higher Hubble
constant values, and lower values of $\sigma_8$.  The data sets used in 
Table \ref{tab:lcdm_high} favor higher values of  $\Omega_m$,  lower Hubble constants
and higher values of $\sigma_8$.  The lensing data is most
discrepant and it most strongly
pulls the combined results towards higher amplitudes and higher $\Omega_m$
(see Figure \ref{fig:lensing}
and \ref{fig:omegamh2}).  The overall effect of combining the data sets is shown
in Figure \ref{fig:CMBall}.

The best fits shown in
Table \ref{tab:lcdm_high}  differ by about 1$\sigma$ 
from the best fits shown
in Table \ref{tab:lcdm_low} in their predictions
for the total matter density, $\Omega_m h^2$
 (See Tables \ref{tab:lcdm_low} and
\ref{tab:lcdm_high} and Figure \ref{fig:omegamh2}).  
More accurate measurements of the third peak will help resolve
these discrepancies.

The differences between the two sets of data may be due to statistical fluctuations.
For example,
the SDSS main galaxy sample power spectrum differs from the power spectrum
measured
from the 2dfGRS: this leads to a lower value
for the Hubble constant for WMAP+SDSS data combination,
$\ensuremath{h = 0.710\pm 0.026}$, than for WMAP+2dFGRS,
$\ensuremath{h = 0.733^{+ 0.020}_{- 0.021}}$.
Note that while the SDSS LRG data parameters values are close to those from
the main SDSS catalog, they are independent determinations with mostly
different systematics.

Lensing measurements are sensitive to the amplitude
 of the local
potential fluctuations, which scale roughly as
$\sigma_8 \Omega_{m}^{0.6}$, so that lensing parameter constraints are nearly
orthogonal to the CMB degeneracies \citep{tereno/etal:2005}.  The CFHTLS lensing data best fit 
value for $\sigma_8 \Omega_{m}^{0.6}$ is $1-2 \sigma$ higher than the 
best fit three year WMAP value.
As a result, the combination of CFHT and WMAP data favors a higher value of $\sigma_8$ and
$\Omega_m$ and a lower value of $H_0$ than WMAP data alone.
Appendix \ref{appendix:sz_marg} shows that
the amplitude of this discrepancy is somewhat sensitive
to our choice of priors.
Because of the small error bars in the CFHT
data set and the relatively small overlap region in parameter space, the CFHT data set
has a strong influence on cosmological parameters.  
Because of the small errors in the CFHT data and the relatively small overlap region in parameter space, the CFHT data has a strong influence on cosmological parameters.  This effect if exacerbated when additional cosmological data sets are included in the analysis.  Because of this, we do not include the lensing data in the full combined data set (WMAP...WMAP+ALL), rather we quote WMAP+CFHT results separately.
The combined data sets place the strongest limits on cosmological parameters.  Because
they are based on the overlap between many likelihood functions, limits based on the WMAP+ALL
data set should be treated with some caution.  
Figure \ref{fig:CMBall} shows the 2-dimensional marginalized
likelihood surface for both WMAP only and for the combination of WMAP+ALL.

\section{Constraining the Shape of the Primordial Power Spectrum
\label{sec:power}}

While the simplest inflationary models predict that the spectral index
varies slowly with scale, inflationary models can produce strong scale
dependent fluctuations (see 
e.g., \citet{kawasaki/etal:2003,hall/moss/berera:2004,yamaguchi/yokoyama:2004}).  The first year WMAP observations provided
some motivation for considering these models as the data, particularly
when combined with the Lyman $\alpha$ forest measurements, were better
fit by models with a running spectral index \citep{spergel/etal:2003}.
Small scale CMB measurements \citep{readhead/etal:2004} also favor
running spectral index models over power law models.

Here, we consider whether a more general function 
for the primordial power spectrum could better fit
the new WMAP data.  We consider three forms for the power spectrum:
\begin{itemize}
\item $\Delta_{\mathcal R}^2(k)$ 
	with a running spectral index: $1+d\ln \Delta^2_R(k)/d \ln k = n(k_0) +
        dn_s/d\ln(k) \ln(k/k_0)$
\item $\Delta_{\mathcal R}^2(k)$ allowed to freely vary in 15  bins in $k$-space, with $k_{1} = 0,\ k_{2}= 0.001/{\rm Mpc},\ k_{15}=0.15/{\rm Mpc},\ k_{i+1}=1.47 k_{i}$ for $ 3\le i \le 14$.  $\Delta_{\mathcal R}^2(k)$ is given by linear interpolation within the bins and $\Delta_{\mathcal R}^2(k) = \Delta_{\mathcal R}^2(0.15/{\rm Mpc})$ for $k> 0.15/{\rm Mpc}$.
\item $\Delta_{\mathcal R}^2(k)$ with a sharp $k$ cut off at $k=k_{c}$,
\begin{equation}\begin{array}{ccll}\Delta_{\mathcal R}^2(k) &=& 0, \ \ \ \  \ \ \ \ & k\le k_{c} \\
&\propto& \left(\frac{k}{ k_{0}}\right)^{(n_{s}-1)}, &  k> k_{c}
\end{array}\label{eq:kcut}\end{equation}
\end{itemize}

\begin{figure}[htbp!] 
\centering
\includegraphics[width=3in]{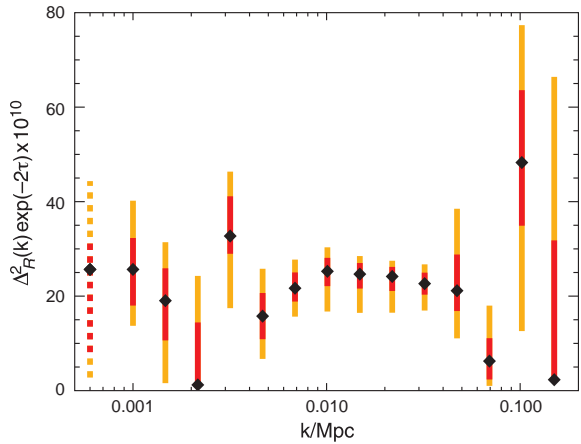}
\caption{\fg The reconstructed primordial curvature fluctuation
spectrum, $\Delta_R^2(k)$, for a $\Lambda$CDM cosmology,
in logarithmically spaced $k$ bins, where $k$ is in Mpc$^{-1}$.  The errors show the  68\% (red) and 95\% (orange) constraints and the black diamonds the peak likelihood value. The dashed line show the values for $k=0$. Consistent with the predictions of simple inflationary theories,
there are no significant features in the spectrum.
The data are consistent with a nearly scale-invariant spectrum.
\label{fig:pkrec}}
\end{figure}
Figure \ref{fig:pkrec} shows how WMAP data alone
can be used to
reconstruct the primordial power spectrum as a function of scale,
parameterized by logarithmically spaced bins out to $k=0.15$ Mpc$^{-1}$.
Even allowing for these additional degrees of freedom, 
the data prefer a nearly featureless power-law  power spectrum.

The deviation of the primordial power spectrum from a simple power law
can be most simply characterized by a
sharp cut-off in the primordial
spectrum. 
Analysis of this model finds that putting in a cut off
of  $k_{c}\sim 3\times10^{-4}/$Mpc  improves the fit by $\Delta \chi^2 =1.2$,
not enough to justify a radical change in the primordial spectrum.

Table \ref{tab:how_many_parameters}
demonstrates that, while models with reduced large scale power provide  slightly improved fits
to the CMB data, the improvements do not warrant additional parameters.

\subsection{External Data Sets and the Running Spectral Index
\label{sec:running_ext}}

\begin{figure}[h] 
\centering
\includegraphics[width=5in]{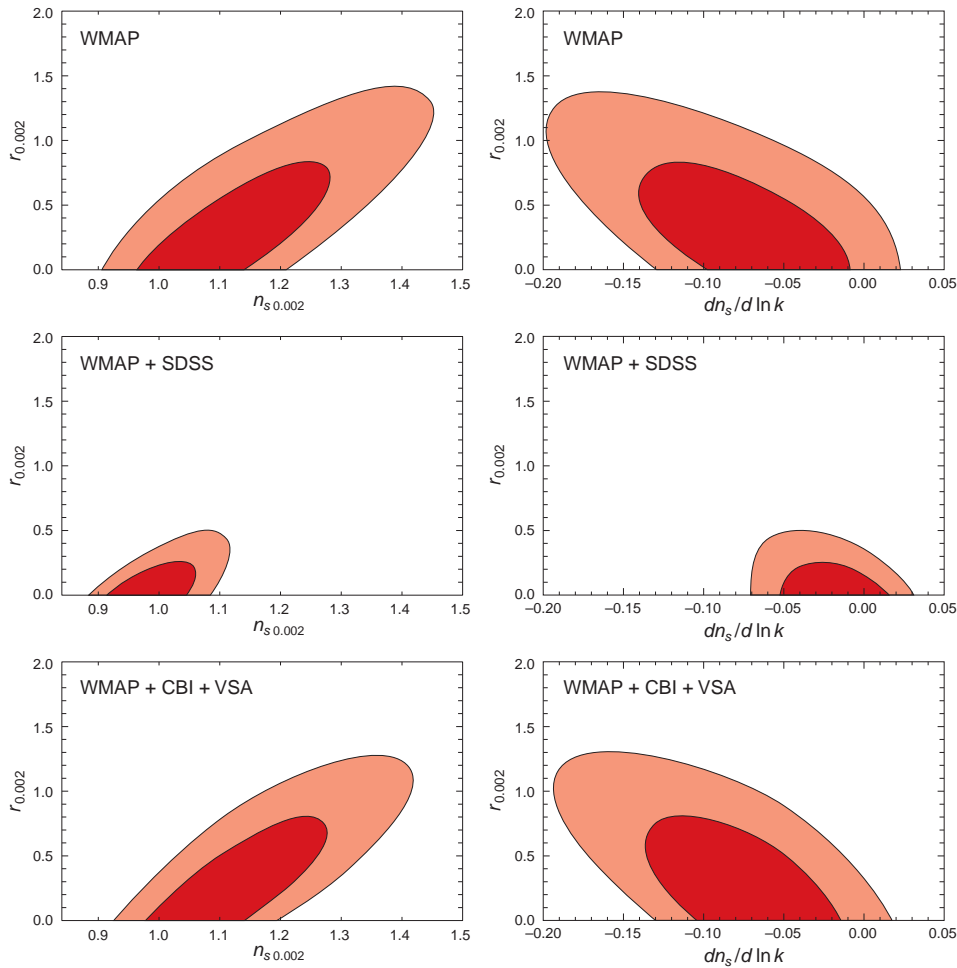}
\caption{\fg 
Joint two-dimensional marginalized 
contours (68\% and 95\%) for inflationary parameters,
$(r,n_s)$ (left panel) and $(r,dn_s/d\ln k)$ (right panel),
for Model M11 in Table~\ref{tab:how_many_parameters}, with parameters defined at $k=0.002$ Mpc$^{-1}$. 
({\it Upper}) WMAP only.
({\it Middle}) WMAP+SDSS.  
({\it Bottom}) WMAP+CBI+VSA.
Note that $n_s>1$ is favored because $r$ and $n_s$ are 
defined at $k=0.002$ Mpc$^{-1}$. At $k=0.05$ Mpc$^{-1}$ 
$n_s<1$ is favored.
The data do not require a running spectral index, $dn_s/d\ln k$, 
at more than the 95\% confidence level.
\label{fig:running_ext}}
\end{figure}

The uncertainties in the  measurements of running is slightly improved by including
the small scale experiments.
For models with only scalar fluctuations,
the marginalized value for the derivative of
the spectral index  is
\ensuremath{dn_s/d\ln{k} = -0.055^{+ 0.030}_{- 0.031}} for
WMAP only,
\ensuremath{dn_s/d\ln{k} = -0.066^{+ 0.029}_{- 0.028}} for
the WMAP+CBI+VSA data 
and \ensuremath{dn_s/d\ln{k} = -0.058\pm 0.029} for
WMAP+BOOM+ACBAR.
For models with tensors, 
\ensuremath{dn_s/d\ln{k} = -0.085\pm 0.043} for
WMAP only,
\ensuremath{dn_s/d\ln{k} = -0.090^{+ 0.038}_{- 0.039}} for
WMAP+CBI+VSA,
and \ensuremath{dn_s/d\ln{k} = -0.082\pm 0.040}
for WMAP+BOOM+ACBAR.
As Figure \ref{fig:running_ext} shows, models
with negative running allow large tensor amplitudes;
thus, when we marginalize over $r$ with a flat prior, these
models favor a more negative running.

\begin{figure}[h] 
\centering
\includegraphics[width=5in]{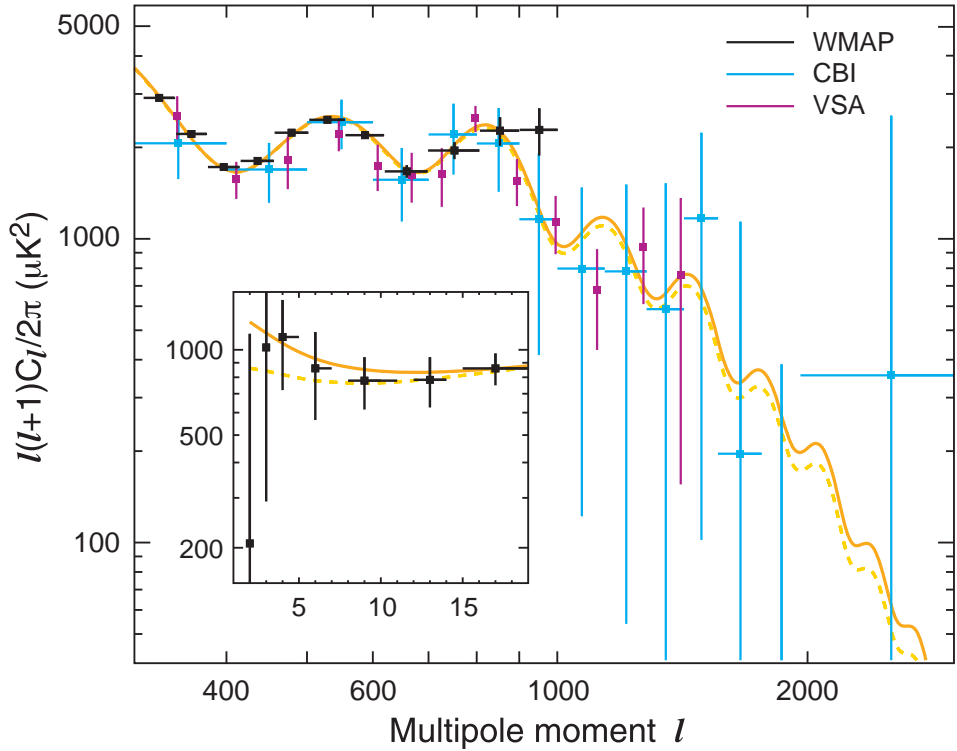}
\caption{\fg 
The running spectral index model provides
a slightly better fit to the data than the power-law spectral index model.
The solid line shows the best fit power law $\Lambda$CDM
model  and the dashed line shows the 
best fit running spectral index $\Lambda$CDM model (fit to WMAP+CBI+VSA).
The insert compares the models to the WMAP $\ell < 20$ data
and shows that the running spectral index model better fits
the decline at $\ell=2$; however, the improvement in
$\chi^2$ is only 3, not enough to strongly argue
for the addition of a new parameter.  We have also done the same analysis
for BOOMERanG and ACBAR data and found similar results: 
the current high $\ell$ data are not yet able to distinguish between the running spectral index and power law models.
\label{fig:best_fit_running}}
\end{figure}
Figure \ref{fig:best_fit_running} shows that both the power law
$\Lambda$CDM  model and the running spectral index model fit the CMB data.
At present,
the small scale data do not yet clearly distinguish the two models.

A large absolute value of running would be problematic for most inflationary models, so further testing of this suggestive trend is important for our understanding of early universe physics. Additional WMAP data and upcoming, more sensitive small-scale CMB data will further test deviations from scale invariance.  Current large scale structure data do not strengthen the case for running because these these data sets probe similar physical scales to the WMAP
experiment.  Figure \ref{fig:running_ext} shows that current data favor a large running; however, the evidence is not yet compelling.  The constraints from WMAP+lensing and WMAP+2dFGRS are similar to the WMAP+SDSS constraints shown in Figure \ref{fig:running_ext}.

\subsection{Is the Power Spectrum Featureless?}

 Since inflation magnifies fluctuations that were once on
 sub-Planckian scales to scales of the observable horizon,
 trans-Planckian physics could potentially leave its imprint on the
 CMB sky.  
Over the past few years, there has been significant interest in the
 possibility of detecting the signature of trans-Planckian 
physics in the angular power spectrum.   Several studies
 \citep{martin/brandenberger:2001,danielsson:2002, easther/etal:2002, bergstrom/danielsson:2002,
 kaloper/etal:2002, martin/brandenberger:2003, martin/ringeval:2004,
 burgess/etal:2003, schalm/etal:2004} have discussed the possible form and the expected
 amplitude  of the trans-Planckian effects which might modify the
 spectrum coming from slow roll inflation. The scalar
  power spectra resulting from power law (PL) slow roll inflation can
 be written in the terms of Hubble Flow parameters respectively as, \citep{leach/liddle:2003} 
\begin{equation}
\Delta_{\mathcal R,PL}^2(k)= A_{s} \left
(1-2(C+1)\epsilon_{1}-C\epsilon_{2}-(2\epsilon_{1}+\epsilon_{2})
\ln\left(k\over k_{0}\right) \right) 
\end{equation} 
Here, $\epsilon_1$ and $\epsilon_2$ are slow roll parameters
\citep{leach/liddle:2003}.
 After the release of the first year WMAP data, \citet{martin/ringeval:2004}
 considered a primordial power spectrum of a slightly modified form to
 account for additional trans-Planckian (TP) features, 
\begin{equation}
\begin{array}{lcl}
\Delta_{\mathcal R,TP}^2(k) &=& \Delta_{\mathcal R,PL}^2(k)
\left[1-2|x|\sigma_{0}\cos\theta(k)\right]-A_{s}|x|\sigma_{0}\pi(2\epsilon_{1}+\epsilon_{2})\sin\theta(k) \\
\mathrm{with,\ \ }\theta(k) &=& {1\over
  2\sigma_{0}}\left(1+\epsilon_{1}+\epsilon_{1}\ln\left({k\over
  k_{0}}\right)\right).
\end{array}\label{tpeq}
\end{equation}
Here $\sigma_{0} \equiv H l_{c}/ 2\pi$ is determined by the Hubble
parameter during inflation, $H$, and the characteristic length scale
for the trans-Planckian manifestation $l_{c}$, and $|x|\sigma_{0}$
characterizes the amplitude of the trans-Planckian corrections to the
fiducial spectrum. \citet{martin/ringeval:2004} report that the $\chi^{2}$ for
such a model could give an improvement of 15 over the power law
inflationary models for an additional 2 degrees of freedom with the
first year WMAP data. With three years of data, many of the glitches and bites have disappeared, and the best fit trans-Planckian model of the form in equation (11) reduce the effective $\chi^{2}$ by only 4 overall and by 5 in TT relative  to power law inflation, a far less significant effect.

The effect of the trans-Planckian corrections can be highly model
dependent (See \citet{easther/kinney/peiris:2005} and \citet{easther/kinney/peiris:2005b} for discussions). As an alternative, we consider forms that are more general
as a way to look for oscillatory signals:
\begin{eqnarray}
\Delta_{\mathcal R,TP}^{2}(k) = \Delta_{\mathcal R, PL}^{2}(k) [1+\epsilon_{TP} \cos\theta(k)] 
\end{eqnarray}
where $\theta  =\upsilon{k\over k_{0}}+\phi$  or $\theta
=\upsilon\ln\left({k\over k_{0}}\right)+\phi$ 
In these models, there are
three new parameters: the amplitude, $\epsilon_{TP}$, the frequency, $\upsilon$, and 
the phase, $\phi$.

Assuming the $\Lambda$CDM model, we fit these three parameters
to the data and
find reductions of 5 and 9.5 in the overall and TT $\chi^2_{eff}$. As in the Martin and Ringeval model, the improvements in the $\chi^2_{eff}$ are driven by improvements in the fit around $\ell\sim 30-100$ and the first peak.

\section{WMAP + Inflation \label{sec:inf}}

The inflationary paradigm \citep{guth:1981,sato:1981,linde:1982,albrecht/steinhardt:1982}
explains the homogeneity, isotropy and flatness of the universe by
positing an early epoch of accelerated expansion.  This accelerated
period of expansion also generated superhorizon fluctuations
\citep{guth/pi:1982,starobinsky:1982,mukhanov/chibisov:1981,hawking:1982,bardeen/steinhardt/turner:1983}.
In the simplest inflationary models, these fluctuations are Gaussian distributed in amplitude with 
random phase and a nearly scale invariant spectrum of
fluctuations. 

\begin{figure}[htbp!] 
\centering
\includegraphics[width=6in]{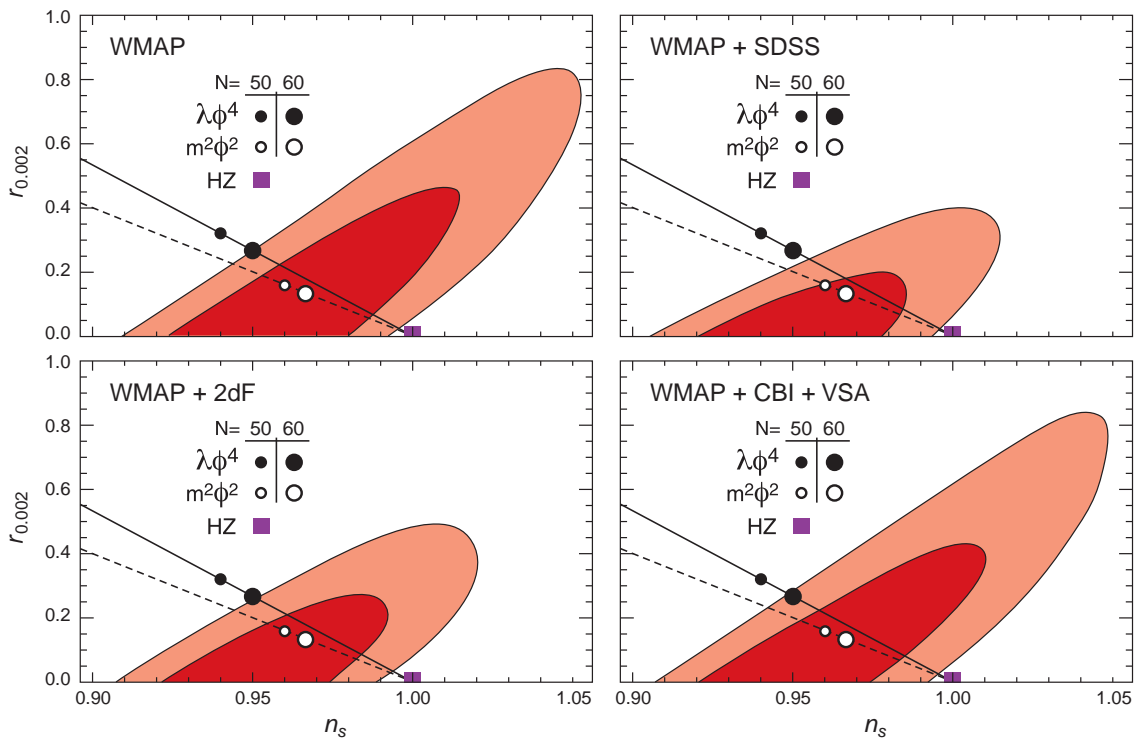}
\caption{\fg 
Joint two-dimensional marginalized 
contours (68\% and 95\% confidence levels) for inflationary parameters ($r_{0.002}$, $n_s$).
We assume a power-law primordial power spectrum, $dn_s/d\ln k=0$, as 
these models predict a negligible amount of running index, 
$dn_s/d\ln k\approx -10^{-3}$.
({\it Upper left}) WMAP only. 
({\it Upper right}) WMAP+SDSS.
({\it Lower left}) WMAP+2dFGRS.
({\it Lower right}) WMAP+CBI+VSA.
  The dashed and solid lines show the range of values predicted for monomial
inflaton models with 50 and 60 e-folds of inflation (equation
 (\ref{eq:inf_pred})), respectively.
The open and filled circles show the predictions of $m^2 \phi^2$
and $\lambda \phi^4$ models for 50 and 60 
e-folds of inflation.  The rectangle denotes the scale-invariant Harrison-Zel'dovich-Peebles (HZ) spectrum $(n_s=1,r=0)$.    Note that the current data
prefers the $m^2 \phi^2$ model over both the HZ spectrum and the
$\lambda \phi^4$ model by likelihood ratios greater than 12. ($\delta \chi^2 > 5$)
\label{fig:inf}}
\end{figure}

The detailed predictions of inflationary models depend on the
properties of the inflaton potential (see \citet{linde:2005} and
\citet{lyth/riotto:1999} for recent reviews).  Simple inflationary  models
predict that the slope of the primordial power spectrum, $n_s$,
differs from 1 and also predict the existence of a nearly scale-invariant 
spectrum of gravitational waves.  In this section, we compare the
simplest inflationary models to the WMAP three year data and to other
cosmological data sets.  Since we are constraining models with tensor modes,
we also use the WMAP constraints on the amplitude of the B mode signal in the analysis.  We characterize these models by seven basic
parameters (the  six basic parameters of the $\Lambda$CDM model plus one additional
parameter, $r$, the ratio  of the tensor to scalar power spectrum).
Figure \ref{fig:inf} shows the
likelihood contours for the slope of the scalar fluctuations and the
amplitude of the gravitational wave signal.   
\begin{table} [htbp!] 
\begin{center}
\caption{Best Fit Inflationary Parameters (WMAP data only)
\label{tab:inf}}
\begin{tabular}{|c|c|c|}
\hline
\hline
Parameter & $\Lambda$CDM + Tensor & $\Lambda$CDM + Running +Tensors \\
\hline
$\Omega_b h^2$ &
\ensuremath{0.0233\pm 0.0010} &
\ensuremath{0.0219\pm 0.0012}  \\
$\Omega_m h^2$ &
\ensuremath{0.1195^{+ 0.0094}_{- 0.0093}} &
\ensuremath{0.128\pm 0.011}  \\
$h$ &
\ensuremath{0.787\pm 0.052} &
\ensuremath{0.731\pm 0.055}  \\
$n_s$ &
\ensuremath{0.984^{+ 0.029}_{- 0.028}} &
\ensuremath{1.16\pm 0.10}  \\
$dn_s/d\ln k$ &
{\rm set\ to\ 0} &
\ensuremath{-0.085\pm 0.043}  \\
$r$ &
$<$ \ensuremath{0.65\ \mbox{(95\% CL)}} &
$<$ \ensuremath{1.1\ \mbox{(95\% CL)}}  \\
$\tau$ &
\ensuremath{0.090\pm 0.031} &
\ensuremath{0.108^{+ 0.034}_{- 0.033}}  \\
$\sigma_8$ &
\ensuremath{0.702\pm 0.062} &
\ensuremath{0.712\pm 0.056}  \\
\hline
\end{tabular}
\end{center}
\end{table}

The WMAP three year data place significant constraints on inflationary models.  The strength of
these constraints is apparent when we consider monomial models for the
inflaton potential,
$V(\phi) \propto \phi^\alpha$.  
These models \citep{lyth/riotto:1999} predict
\begin{eqnarray}
r &=& 16 \epsilon_1   \simeq \frac{4\alpha}{N}  \nonumber \\
1- n_s & =& 2\epsilon_1 + \epsilon_2  \simeq \frac{\alpha+2}{2N}
\label{eq:inf_pred}
\end{eqnarray}
where $N$ is the number of e-folds of inflation between the epoch when
the horizon scale modes left the horizon and the end of inflation.
Figure \ref{fig:inf} compares the predictions of these monomial
inflationary models to the data.  For $N=60$, $\lambda \phi^4$ predicts
$r = 4/15, n_s = 0.95$, just at the outer edge of the
$3 \sigma$ contour.  For $N=50$,
$\lambda \phi^4$ predicts $r = 0.32, n_s = 0.94$. However,
if we allow for non-minimal gravitational couplings, then the
gravity wave predictions of these models are significantly reduced
\citep{hwang/noh:1998,komatsu/futamase:1999} and the models are
consistent with the data.  Alternatively, the $m^2 \phi^2$ model
is a good fit to the observations and its predicted level of gravitational
waves, $r \simeq 0.13 - 0.16$, is within range of upcoming experiments. 

If we restrict our analysis to a specific model and value of $N$, then
the model specifies both $r$ and $n_s$.  In Table \ref{tab:inf_phi2}
we show the best fit parameters for the $m^2\phi^2$ potential  with $N=60$.
Since $n_s$ and $r$ are set to $0.9667$ and $0.1333$ for this model, there
are only five free parameters ($\Omega_m h^2,$ $\Omega_b h^2$, $h$,
$\tau$, and $A$).  The likelihood for the best fit $m^2 \phi^2$ is
$12$ times higher than the best fit ``Harrison Zel'dovich-Peebles'' model with $n_s=1$
and $r=0$.  To be conservative, this comparison uses the exact (res-4) likelihood code up to $l=30$ and the slightly lower point source correction discussed in Appendix \ref{appendix:sz_marg}.  If we cut the exact likelihood off at $l=12$, and use the original, higher point source value, the likelihood ratio is $15$.

\begin{table} [htbp!] 
\begin{center}
\caption{Best Fit Parameters for $m^2 \phi^2$ with $N=60$ (WMAP data only)
\label{tab:inf_phi2}}
\begin{tabular}{|c|c|c|}
\hline
\hline
Parameter & Mean& Best Fit\\
\hline
$\Omega_b h^2$ &
\ensuremath{0.02218\pm 0.00045} &
\ensuremath{0.0222}  \\
$\Omega_m h^2$ &
\ensuremath{0.1232\pm 0.0076} &
\ensuremath{0.125}  \\
$h$ &
\ensuremath{0.741\pm 0.029} &
\ensuremath{0.734}  \\
$\tau$ &
\ensuremath{0.081\pm 0.027} &
\ensuremath{0.0893}  \\
$\sigma_8$ &
\ensuremath{0.734\pm 0.046} &
\ensuremath{0.749}  \\
\hline
\end{tabular}
\end{center}
\end{table}

In \citet{peiris/etal:2003}, we used the inflationary flow equations
\citep{hoffman/turner:2001,kinney:2002} to explore the generic
predictions of inflationary models.  Here, we use
the slow-roll approximation to explore the
implications of the data for inflationary models.
The results of the third year
analysis are consistent with the conclusions from the first year data:
while the data rule out large regions of parameter space, there are also
wide range of possible inflationary models consistent with our current
data.  One of the most intriguing features of Figure \ref{fig:inf} is
that the data now disfavors the exact Harrison-Zel'dovich-Peebles spectrum ($n_s=1,
r = 0$).  For power law inflationary models, this suggests a detectable
level of gravity waves.
There are, however, many inflationary models that predict a much 
smaller gravity wave amplitude.  Alternative models, such as the
ekpyrotic scenario \citep{khoury/etal:2001,khoury/etal:2002} also
predict an undetectable level of gravity waves.

There are several different ways of expressing the constraints that the CMB data impose on inflationary models.
These parameters can be directly related to observable quantities: $n_s-1=-2\epsilon_1-\epsilon_2$
and $r = 16\epsilon_1$.  
For the power law models, the WMAP bound on $r$ implies that 
$\epsilon_1 < 0.03$ (95\% C.L.).     An alternative
slow roll representation (see \citet{liddle/lyth:1992,liddle/lyth:1993})  
uses
\begin{eqnarray}
\epsilon_v &\equiv& \frac{M_{Pl}^2}{2} \left(\frac{V'}{V}\right)^2  \\
\eta_v &\equiv& M_{Pl}^2 \left(\frac{V''}{V}\right)
\end{eqnarray}
These parameters can be related directly to observables:
$r = 16 \epsilon_v$ and  $n_s -1 = -6 \epsilon_v + 2 \eta_v$.  
\citet{peiris/etal:2003} discusses various classes of models
in slow roll parameter space.

Models with  a significant gravitational 
wave contributions, $r\sim 0.3$, make specific
predictions for CMB and large-scale structure observations:
(a) a modified temperature spectrum
with more power at low multipoles; and (b) a lower amplitude of density
fluctuations (for fixed CMB temperature fluctuations).  For power law models,
the strongest CMB constraints come from the shape
of the temperature spectrum and the amplitude of density
fluctuations.  In order to fit the CMB data, models with
higher $r$ values require larger values of $n_s$ 
and land a lower amplitude of scalar fluctuations, which impacts large-scale structure predictions. Thus the strongest overall constraints on the 
tensor mode contribution comes from the combination of
CMB and large-scale structure measurements (see Table \ref{tab:r_limits}).
These strong limits rely on our assumption of a power law spectral
index. If we allow for a running index, then models with 
a large tensor components are consistent with the data.

\begin{table}[htbp!] 
\label{tab:r_limits}
\begin{center}
\caption{Constraints on $r$, Ratio of Amplitude of Tensor Fluctuations
to Scalar Fluctuations (at $k=0.002$ Mpc$^{-1}$)}
\begin{tabular}{|c||c|c|}
\hline
Data Set& $r$ (no running) & $r$ (with running) \\
\hline
WMAP &  
$<$ \ensuremath{0.65\ \mbox{(95\% CL)}}  &
$<$\ensuremath{1.1\ \mbox{(95\% CL)}} \\
WMAP+BOOM+ACBAR & 
$<$\ensuremath{0.68\ \mbox{(95\% CL)}}  &
$<$\ensuremath{1.1\ \mbox{(95\% CL)}} \\
WMAP+CBI+VSA & 
$<$\ensuremath{0.62\ \mbox{(95\% CL)}}  &
$<$\ensuremath{1.1\ \mbox{(95\% CL)}} \\
WMAP+2df & 
$<$\ensuremath{0.38\ \mbox{(95\% CL)}}  &
$<$\ensuremath{0.64\ \mbox{(95\% CL)}} \\
WMAP+SDSS & 
$<$\ensuremath{0.30\ \mbox{(95\% CL)}}  &
$<$\ensuremath{0.38\ \mbox{(95\% CL)}} \\
\hline
\end{tabular}
\end{center}
\end{table}

\section{Constraining the Composition of the Universe \label{sec:constraints}}
\subsection{Dark Energy Properties\label{sec:w}}

\begin{figure}[htb] 
\centering
\includegraphics[width=5in]{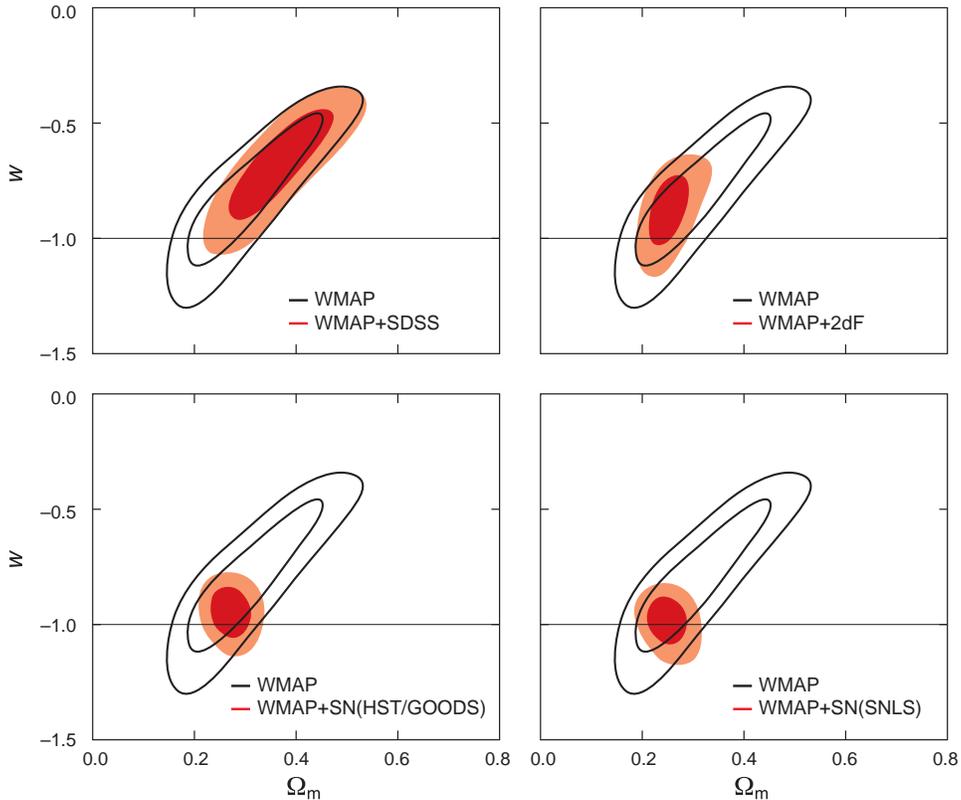}
\caption{\fg 
Constraints on $w$, the equation of state of dark energy, in
a flat universe model based on the combination of WMAP data and
other astronomical data.  
We assume that $w$ is independent of time, and ignore 
density or pressure fluctuations in dark energy.
In all of the figures, WMAP only constraints are shown in blue
and WMAP + astronomical data set in red.
The contours show the joint 2-d marginalized contours 
(68\% and 95\% confidence levels) for $\Omega_m$
and $w$. 
({\it Upper left}) WMAP only and WMAP + SDSS.  
({\it Upper right}) WMAP only and WMAP + 2dFGRS.
({\it Lower left}) WMAP only and WMAP+SN(HST/GOODS).
({\it Lower right}) WMAP only and WMAP+SN(SNLS).
In the absence of dark energy fluctuations, the excessive amount of 
ISW effect at $\l<10$  places significant constraints on models with $w<-1$.
\label{fig:omegamw}}
\end{figure}

\begin{figure}[b!] 
\centering
\includegraphics[width=5in]{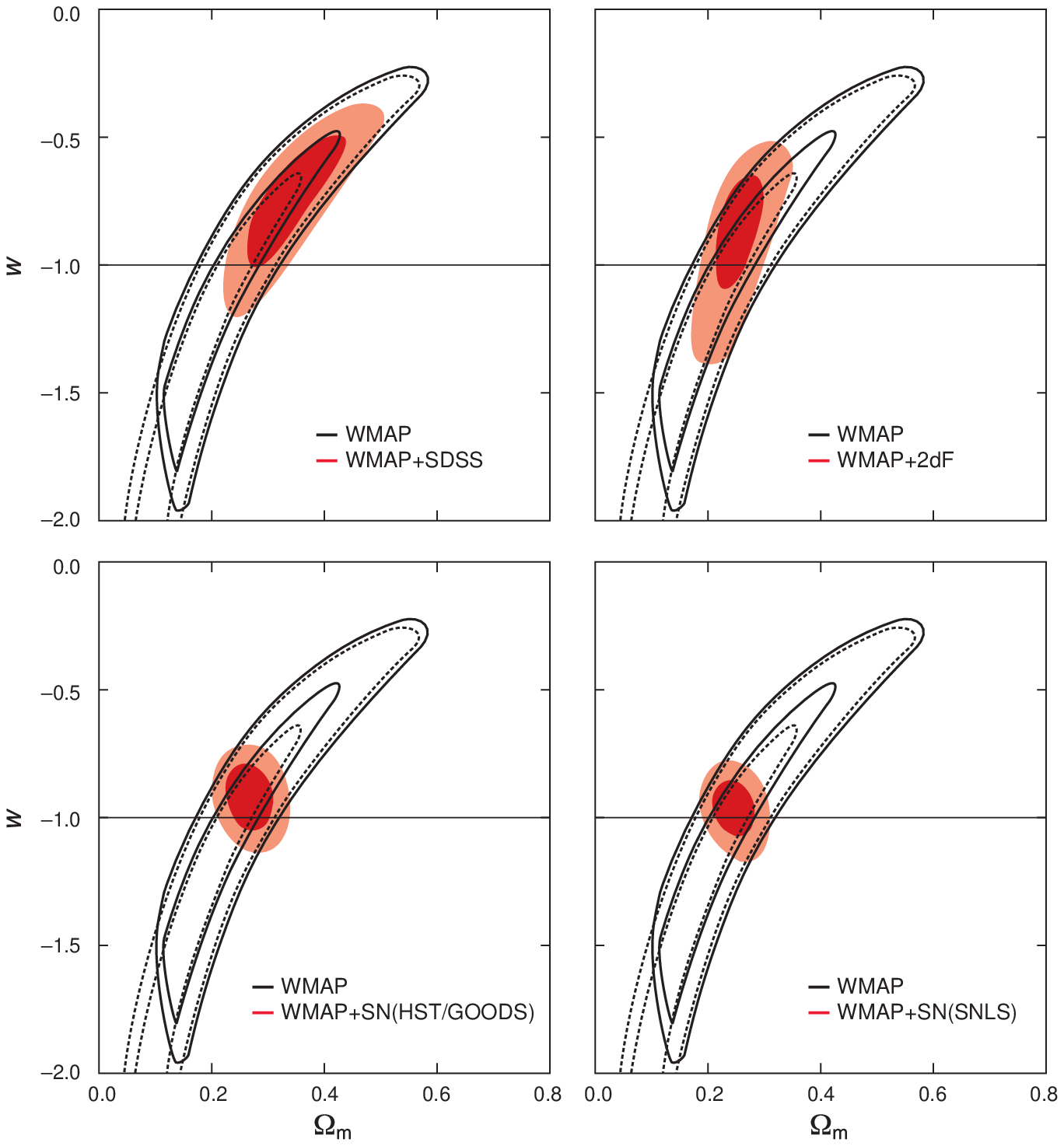}
\caption{\fg 
Constraints on $w$, the equation of state of dark energy, in
a flat universe, Model M6 in Table \ref{tab:how_many_parameters}, based on the combination of WMAP data and
other astronomical data.  
We assume that $w$ is independent of time, but include density and
pressure fluctuations in dark energy with the speed of sound in the
 comoving frame equal to the speed of light, $c_s^2=1$.
In all of the figures, WMAP data only constraints are shown in black solid lines
and WMAP + astronomical data set in red.
The contours show the joint 2-d marginalized contours 
(68\% and 95\% confidence levels) for $\Omega_m$
and $w$. 
({\it Upper left}) WMAP only  and WMAP + SDSS.  
({\it Upper right}) WMAP only and WMAP + 2dFGRS.
({\it Lower left}) WMAP only and WMAP+SNgold.
({\it Lower right}) WMAP only and WMAP+SNLS.
In the presence of dark energy fluctuations, the ISW effect at $\l<10$  
is nearly canceled by dark energy fluctuations and thus the WMAP data
 alone do not place significant constraints on $w$.  The WMAP only models are shown assuming no prior on $H_{0}$ (dashed black) and with 
 an assumed prior of $H_{0} < 100 km/s/Mpc$ (full black lines).
\label{fig:omegamw_c}}
\end{figure}
Over the past two decades, there has been growing evidence for the existence
of dark energy \citep{peebles:1984,turner/steigman/krauss:1984,ostriker/steinhardt:1995,dunlop/etal:1996,krauss/turner:1999,bahcall/etal:1999}. 
By measuring both the acceleration
\citep{riess/etal:1998,perlmutter/etal:1999} and deceleration
\citep{riess/etal:2004} of the universe, supernova observations
provide the most direct evidence for the existence of dark energy.  

The nature of this dark energy is a mystery.  
From a field theoretic perspective the most natural explanation for
this would be the presence of a residual vacuum energy density or
cosmological constant, $\Lambda$,
\citep{carroll/press/turner:1992,peebles/ratra:2003}.  However, there
are well-known fine-tuning and coincidence problems in trying to
explain the 120 orders-of-magnitude discrepancy between the expected ``natural"
Planck-scale energy density of a cosmological constant and
the observed dark energy density.  These problems motivate 
a wide range of  alternative explanations for the observations
including the presence of an extra matter candidate: for example a
dynamical, scalar ``quintessence" field
\citep{peebles/ratra:1988,wetterich:1988,zlatev/wang/steinhardt:1999}, minimally coupled
\citep{caldwell/dave/steinhardt:1998,ferreira/joyce:1998} or
non-minimally coupled to gravity \citep{amendola:1999} or other matter
\citep{bean/magueijo:2001}. In this latter case, the measured
acceleration is due to interactions in the matter bulk.
Another alternative is that modifications to gravity (e.g.,
\citet{deffayet:2001,deffayet/etal:2002}) are responsible for the observed anomalies. 

The dark energy has two distinct cosmological effects:
(1) through the Friedman equation, it alters the evolution of $H(z)$
and (2) through the perturbation equations, it alters the
evolution of $D(z)$, the growth rate of structure.
The supernova data measure only  the
luminosity distance, which depends on $H(z)$, while the large
scale structure data are sensitive to both $H(z)$ and $D(z)$.

While the presence of dark energy impacts the CMB primarily through the distance
to the surface of last scatter,
the dark energy clustering properties also affect the CMB
properties.  The dark energy response to gravitational perturbations
depends upon its isotropic and anisotropic sound-speeds \citep{hu:1998,
bucher/spergel:1999}.  This affects the CMB fluctuations through the ISW
effect.
If the dark energy can cluster, then it produces a smaller 
ISW effect and does not enhance the power spectrum at large
angular scales.  These effects are most dramatic for models
with $w < -1$, as dark energy effects in these models  turn on suddenly 
at late times and significantly enhance the quadrupole.
This can be understood in terms of the constraints imposed
by the shape of the angular power spectrum:
 if we assume that the dark energy
properties can be described by a constant value of $w$, then fixed
peak position and heights (which determine $\Omega_m
h^2$) confine our models to a narrow valley in 
the $(\Omega_m,w)$ likelihood surface 
as shown in Figure \ref{fig:omegamw} and \ref{fig:omegamw_c}.
The figures show that the 3 year data enable a more accurate
determination of $\Omega_m h^2$ which narrows the width of the
degeneracy valley. The pair of figures show that CMB data can place strong 
limits on models with $w < -1$ and non-clustering dark energy.
On the other hand, if  the dark energy is a matter component that can
cluster, even  meagerly, as is the case for scalar field theories
where $c_{s}^{2}$=1,  then this clustering counters the suppression of
perturbation growth during the accelerative epoch and the quadrupole's
magnitude is reduced. This lessens the discriminating power of the
quadrupole for measuring $w$: while CMB data rule out the $w << -1$
region in Figure \ref{fig:omegamw}, it does not constrain models in the
same region in Figure \ref{fig:omegamw_c}.

\begin{figure}[htbp!] 
\plotone{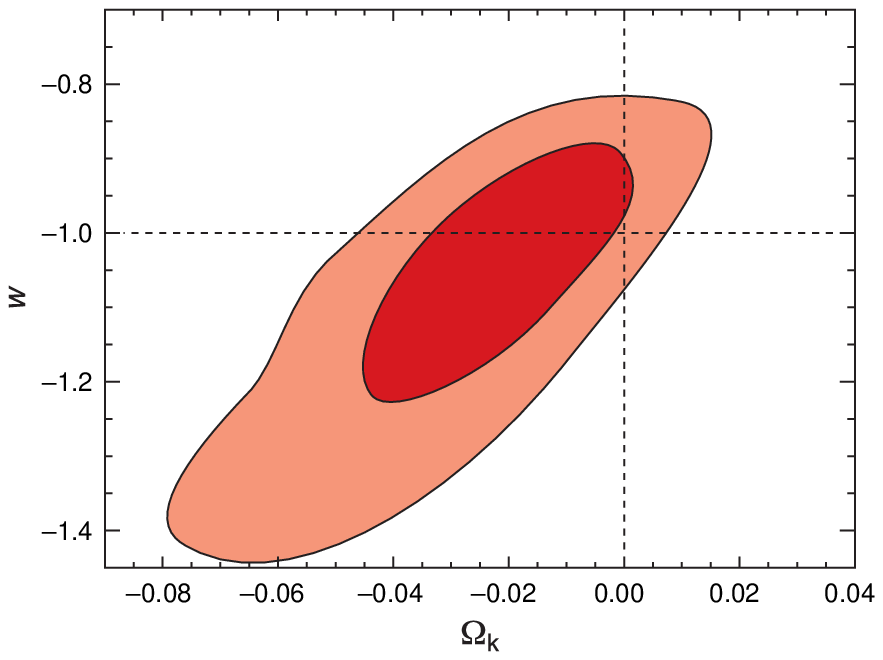}
\caption{\fg  
Constraints on a non-flat universe with quintessence-like dark energy with constant $w$ (Model M10 in
 Table~\ref{tab:how_many_parameters}).
The contours show the 2-d marginalized contours for
$w$ and $\Omega_k$ based on the 
the CMB+2dFGRS+SDSS+supernova data sets.
This figure shows that with the full combination of data sets, there
are already strong limits on $w$ without the need to assume a flat
universe prior.  The marginalized best fit values
for the equation of state  and curvature are
\ensuremath{w = -1.08\pm 0.12} and
\ensuremath{\Omega_k = -0.026^{+ 0.016}_{- 0.015}} at the 68\% confidence level.
\label{fig:omegak_w}}
\end{figure}

It's interesting to note that if we relax the assumption of spatial flatness, the data still prefers models with $w$ close to $-1$ (see  Figure  \ref{fig:omegak_w}).  In this analysis,  we assume that the dark energy clusters.
\begin{figure}[htbp!] 
\centering
\includegraphics[width=4in]{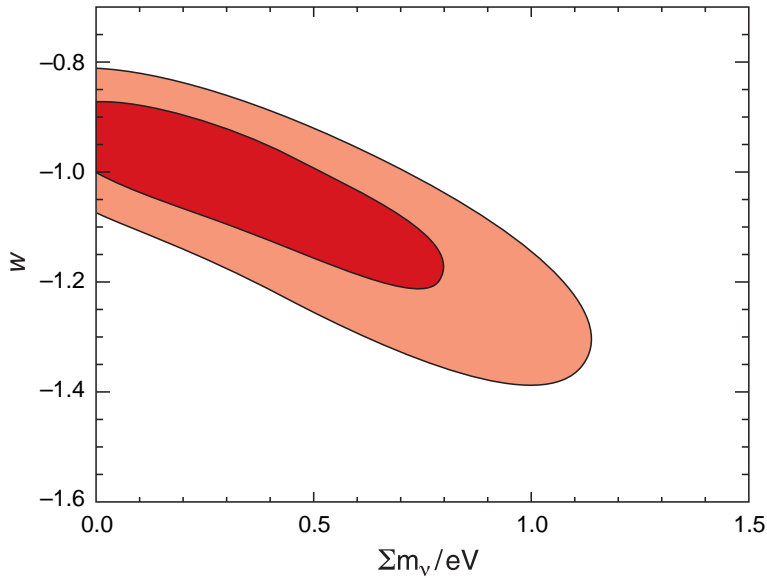}
\caption{\fg  
Constraints on a flat universe with quintessence-like dark energy and
non-relativistic neutrinos.
The contours show the 2-d marginalized constraints on r the mass of
non-relativistic neutrinos, 
$m_\nu$, and the dark energy equation of state, $w$, assumed constant, based on the 
the CMB+2dFGRS+SDSS+supernova data sets.
  In this analysis,  we assume that the dark energy clusters.
With the
combination of CMB+2dFGRS+SDSS+supernova data sets, there is not
a strong degeneracy between neutrino and dark energy properties.
Even in this more general model, we still have
an interesting constraint on the neutrino mass and
equation of state:
\ensuremath{\sum m_\nu < 1.0\ \mbox{eV}\ \mbox{(95\% CL)}} and
\ensuremath{w = -1.07\pm 0.12} (68\% CL).
This suggests that the astronomical dark energy and neutrino limits
are robust even when we start to consider more baroque models.
\label{fig:omega_nu}}
\end{figure}
\begin{table}[t] 
\caption{Constraints on $w$ in Flat Cosmologies With Different Assumption
About Dark Energy Clustering
\label{tab:wdw}}
\begin{center}
\begin{tabular}{|c||c|c|}
\hline
Data Set& with perturbations & no perturbations
\\
\hline
WMAP + SDSS &
\ensuremath{-0.75^{+ 0.17}_{- 0.16}} &
\ensuremath{-0.69\pm 0.16}  \\
WMAP + 2dFGRS &
\ensuremath{-0.89^{+ 0.16}_{- 0.14}} &
\ensuremath{-0.877^{+ 0.097}_{- 0.098}} 
\\
WMAP + SNGold &
\ensuremath{-0.919^{+ 0.081}_{- 0.080}} &
\ensuremath{-0.942^{+ 0.075}_{- 0.073}} 
\\
WMAP + SNLS &
\ensuremath{-0.967^{+ 0.073}_{- 0.072}} &
\ensuremath{-0.984^{+ 0.070}_{- 0.068}} 
\\
CMB+ LSS+ SN  & 
\ensuremath{-0.926^{+ 0.054}_{- 0.053}} &
\ensuremath{-0.915\pm 0.051} 
\\
\hline
\hline
\end{tabular}
\end{center}
\end{table}
\clearpage
\subsection{Neutrino Properties}

\subsubsection{Neutrino Mass}

Both atmospheric neutrino and solar neutrino experiments show
that neutrinos are massive and that there is significant mixing between the various neutrino interaction eigenstates (see 
\citet{mohapatra/etal:prep}
 for a recent
review).   These experiments measure the difference between the square of
the neutrino masses,
$m^2_{\nu_i} - m^2_{\nu_j}$, rather than the mass of individual neutrino mass eigenstates.
Cosmological measurements nicely complement these measurements by constraining
$\sum_i m_{\nu_i}$.
Since light massive neutrinos do not cluster as effectively as cold dark matter, the neutrino
mass has a direct impact on the amplitude and shape of the matter power spectrum
\citep{bond/efstathiou/silk:1980,bond/szalay:1983,ma:1996,
hu/eisenstein/tegmark:1998}.
The presence of a significant neutrino component  lowers the amplitude
of matter fluctuations on small scales, $\sigma$ by roughly
a factor proportional to  $(\sum m_\nu)$, where $\sum m_\nu$ is the total
mass summed over neutrino species, 
rather than the mass of individual neutrino species.
The current large-scale structure data 
restrict $\Delta \ln \sigma_8 < 0.2$, but they 
are not sensitive enough
to resolve the free-streaming scale of individual neutrino
species \citep{takada/komatsu/futamase:prep}.

Using a combination of the first year WMAP data, small-scale CMB and
large-scale structure data, \citet{spergel/etal:2003} placed an upper
limit on $\sum_i m_{\nu_i} < 0.7$ eV.  While this limit does not
depend on the Lyman $\alpha$ data, it is sensitive to galaxy bias
measurements (which normalizes the large-scale structure data) and to
the addition of small scale CMB data  (which improves the measurement
of cosmological parameters).  Over the past year, several groups
obtained comparable (but slightly different) limits
\citep{hannestad:2003, pierpaoli:2003a, elgaroy/lahav:2003}. The
differences are due to including (or removing) external data sets and
priors or to adding additional cosmological parameters.

The limits on neutrino masses from WMAP data alone is now
very close to limits based on combined CMB data sets.
\citet{ichikawa/fukugita/kawasaki:2005} used the CMB data
alone to place a limit on the neutrino mass of $\sum m_\nu < 2.0$ eV.
Using WMAP data alone, we now
find $\sum m_\nu < 1.8$  eV.

\begin{figure}[htbp!] 
\centering
\includegraphics[width=4in]{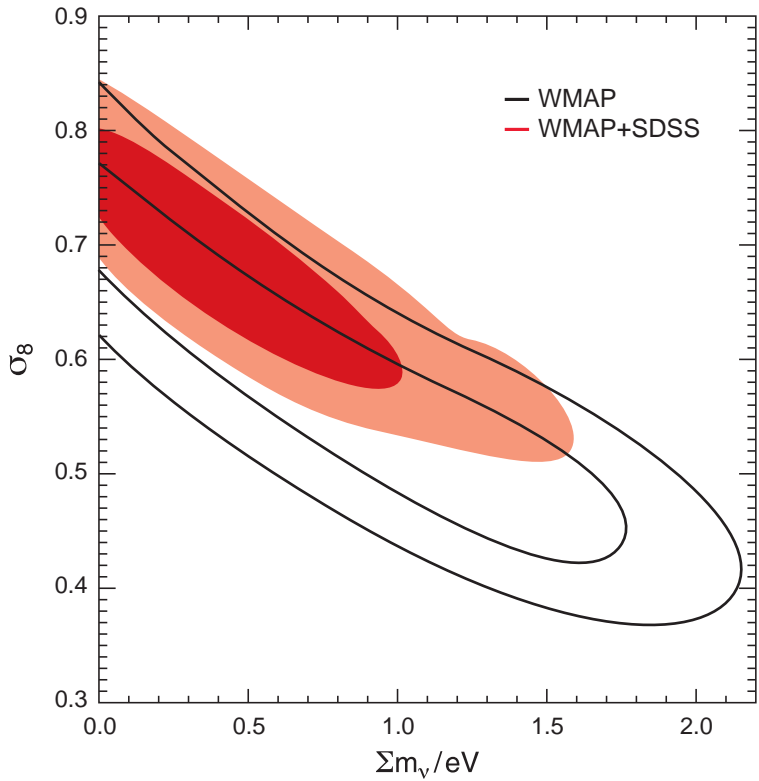}
\caption{\fg Joint two-dimensional marginalized contours 
(68\% and 95\% confidence levels) 
on $(\sigma_8,m_\nu)$ for WMAP only,  Model M7 in Table \ref{tab:how_many_parameters}, and
WMAP+SDSS.
By measuring the growth rate of structure from $z=1088$ to $z \simeq 0$,
these observations constrain the contribution of non-relativistic 
neutrinos to the energy density of the universe.
\label{fig:neutrino_mass_likelihood}}
\end{figure}

Since the presence of massive neutrinos slows the growth of
small scale structure, the combination of CMB and large-scale structure
data constrain the neutrino mass.
Figure \ref{fig:neutrino_mass_likelihood}  shows the likelihood function as a function of
neutrino mass and amplitude of mass fluctuations in the local
universe, $\sigma_8$.  
The 95\% confidence limits
on neutrino mass are given in Table \ref{tab:neutrino_limits}.
The combination of WMAP with SDSS and WMAP with 2dFGRS data
constrain $\sigma_8$ at roughly the same level,
20\% at the 95\% confidence level.  This constraint
yields comparable limits on the neutrino mass.  While the WMAP data have improved, the
error bars on $\sigma_8$ have not significantly changed from the
limits obtained from WMAPext + 2dFGRS, thus,  the
limit on neutrino mass is quite close to the  first year limit.  
Note that in 
the first year analysis, we used the \citep{verde/etal:2002} measurement of
bias for the 2dFGRS preliminary data as there had not been an equivalent
analysis done for the full 2dFGRS data set.  As discussed in 
\S \ref{sec:lss}, we now marginalize over the 2dFGRS bias and
use the bias measurements of \citep{seljak/etal:2005b} for SDSS.

If the constraints on amplitude are reliable, then
small scale matter power spectrum  structure data can significantly improve
these neutrino constraints.  \citet{goobar/etal:prep} have recently
completed a CMB + Lyman $\alpha$ study and place a limit of
$\sum m_\nu < 0.30$eV (95\% C.L.).    Similarly, cluster-based
measurements of $\sigma_8$ and lensing-based measurements of
$\sigma_8$ have the potential to tighten the constraint on $m_\nu$.
\begin{table}[htbp!] 
\begin{center}
\caption{Constraints on Neutrino Properties
\label{tab:neutrino_limits}}
\begin{tabular}{|c||c|c|}
\hline
Data Set& $\sum m_\nu$ (95\% limit for $N_\nu=3.04$) & $N_\nu$ \\
\hline
WMAP &  
\ensuremath{1.8\ \mbox{eV}\ \mbox{(95\% CL)}} &
$-$
 \\
WMAP + SDSS &
\ensuremath{1.3\ \mbox{eV}\ \mbox{(95\% CL)}} &
\ensuremath{7.1^{+ 4.1}_{- 3.5}} 
\\
WMAP + 2dFGRS &
\ensuremath{0.88\ \mbox{eV}\ \mbox{(95\% CL)}} &
\ensuremath{2.7\pm 1.4} 
\\
CMB + LSS +SN & 
\ensuremath{0.66\ \mbox{eV}\ \mbox{(95\% CL)}} &
\ensuremath{3.3\pm 1.7} 
 \\
\hline
\hline
\end{tabular}
\end{center}
\end{table}

\subsubsection{Number of Relativistic Species}

If there are other light stable neutral particles (besides the three
light neutrinos and the photon), then these particles will affect the
CMB angular power spectrum and the evolution of large-scale structure.
Because of the details of freeze-out at electron-positron annihilation 
\citep{gnedin/gnedin:1998, dolgov/hansen/semikoz:1999}, QED
corrections at finite temperature \citep{mangano/etal:2002} and
non-trivial terms in the neutrino mass matrix \citep{mangano/etal:2005}, the effective number of neutrino species
is slightly greater than 3.
Any light particle that does not couple to electrons, ions and photons
will act as an additional relativistic species.  For neutrinos,
we can compute their effects accurately as their temperature
is $(4/11)^{1/3}$ of the CMB temperature.  For other relativistic
species, the limit on $N_\nu^{eff} -3.04$ can be converted into a limit
on their abundance by scaling by the temperature.

The shape of the CMB angular power spectrum is sensitive to the epoch
of matter/radiation equality.  If we increase $N_\nu$, the effective
number of neutrino species, then we will need to also increase the
cold dark matter density, $\Omega_c h^2$, and slowly change other
parameters to remain consistent with the WMAP data
\citep{bowen/etal:2002}.
In addition,
the presence of these additional neutrino species alters the damping
tail and leaves a distinctive signature on the high $\l$ angular power
spectrum  \citep{bashinsky/seljak:2004} 
and on the small scale matter power spectrum.

The high matter density also alters the growth rate of structure,
thus, the combination of large-scale structure and CMB data constrains
the existence of any new light relativistic species.  These limits
constrain both the existence of new particles and the interaction
properties of the neutrino \citep{bowen/etal:2002,hall/oliver:2004}.
\citet{hannestad:2001} 
used the  pre-WMAP CMB and large-scale structure data to place an
upper limit of $N_\nu < 17$.  After the release of the first year WMAP
data, several authors  \citep{hannestad:2003,
pierpaoli:2003a, barger/etal:2003,
  crotty/lesgourgues/pastor:2003, elgaroy/lahav:2003, 
  barger/marfatia/tregre:2004, 
hannestad:2005} 
used the combination of WMAP, 2dFGRS and various external data to reduce this limit by a factor of 2-3.
Table \ref{tab:neutrino_limits}
shows the maximum likelihood estimate of the number of neutrino species  for
different data set combinations using the new WMAP data.  
The SDSS and 2dFGRS data differ in the shapes of the two measured power
spectra: this difference leads to the disagreement  in their
best fit values for $N_\nu^{eff}$.
\clearpage
\subsection{Non-Flat Universe
\label{sec:geom}}
\begin{figure} 
\centering
\includegraphics[width=5in]{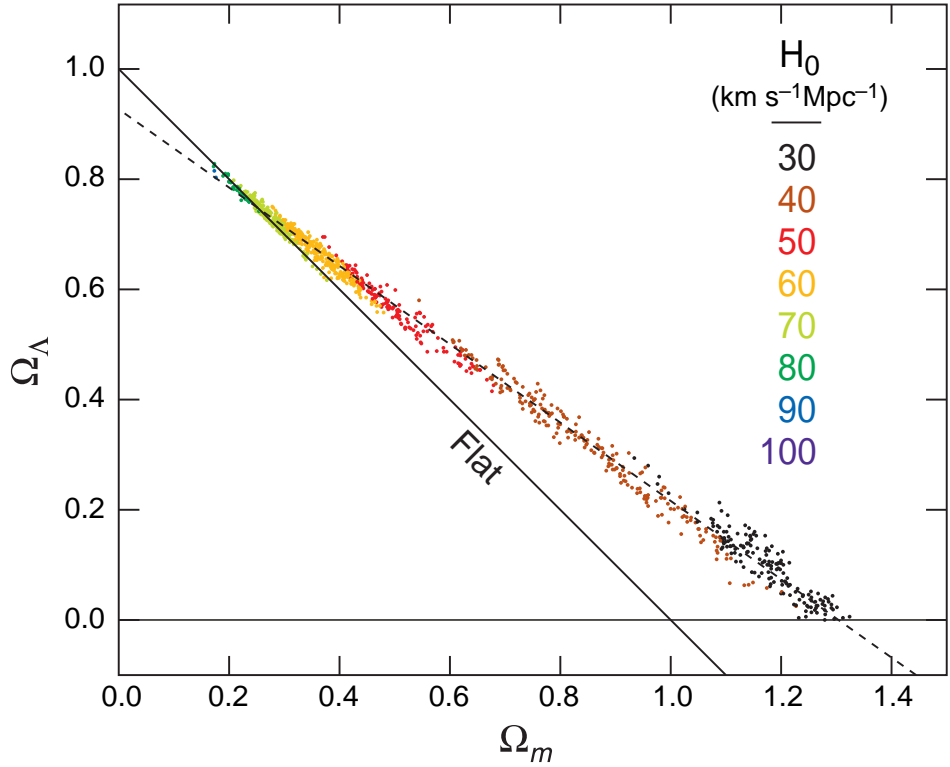}
\caption{\fg Range of non-flat cosmological models consistent
with the WMAP data only.  The models in the figure are all power-law CDM
models with dark energy and dark matter, but without the constraint that
$\Omega_m + \Omega_\Lambda = 1$ (model M10 in Table 
\ref{tab:how_many_parameters}).
The different colors
correspond to values of the Hubble constant as indicated in the figure.  
While models with $\Omega_\Lambda=0$ are not disfavored by the WMAP data
 only ($\Delta\chi^2_{eff}=0$; Model M4 in 
Table~\ref{tab:how_many_parameters}),
the combination of WMAP data plus measurements of the Hubble constant strongly constrain
the geometry and composition of the universe within the framework of
 these models.
The dashed line shows an approximation to the degeneracy track:
$\Omega_{K} =   -0.3040 +    0.4067 \Omega_{\Lambda}$.  Note that
for these open universe models, we assume a flat prior on $\Omega_\Lambda$.
\label{fig:Wv-Wm}}
\end{figure}

\begin{figure}[htbp!] 
\centering
\includegraphics[width=5in]{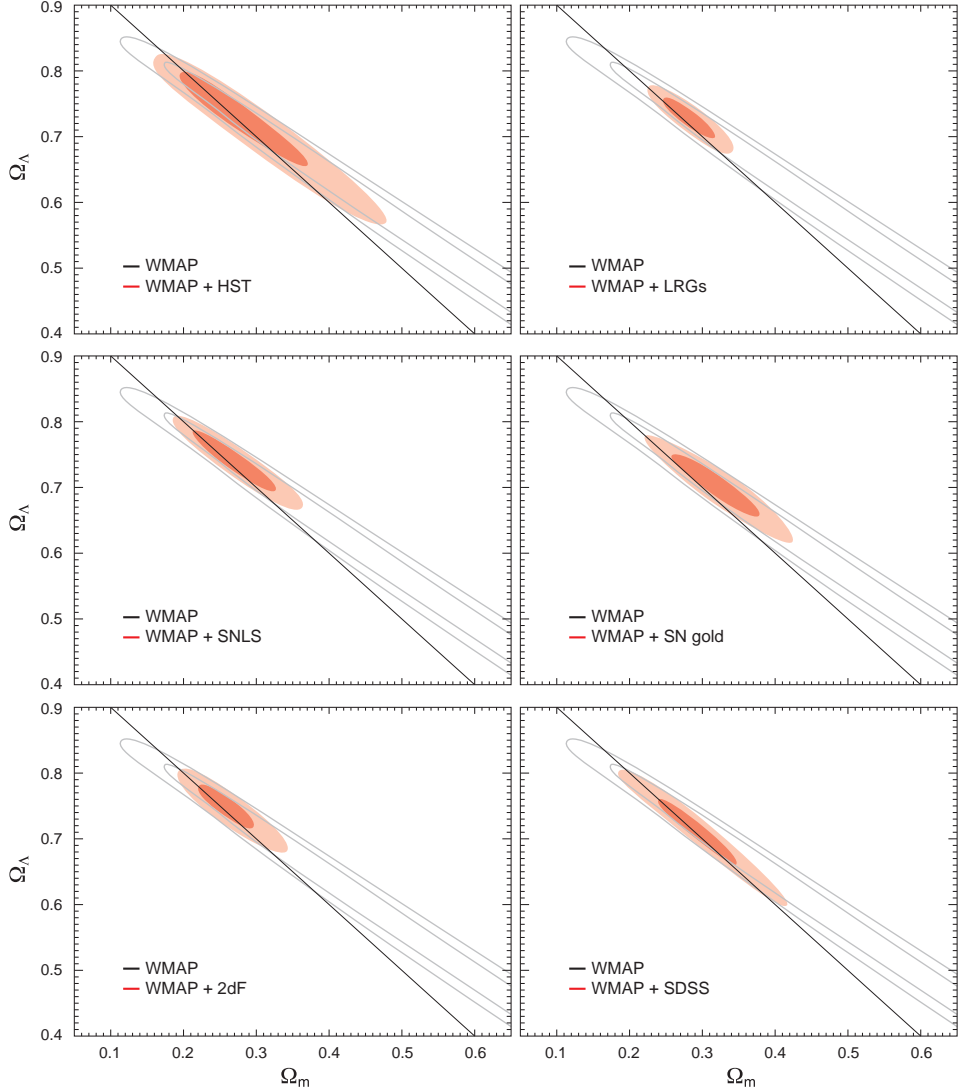}
\caption{\fg Joint two-dimensional marginalized contours
(68\% and 95\%) for matter density,
$\Omega_m$, and vacuum energy density, $\Omega_{\Lambda}$ for power-law CDM
models with dark energy and dark matter, but without the constraint that
$\Omega_m + \Omega_\Lambda = 1$ (model M10 in Table \ref{tab:how_many_parameters}).  The panels
show various combinations of WMAP and other data sets.
While 
models with $\Omega_m = 0.415$ and $\Omega_\Lambda = 0.630$ are 
a better fit to the WMAP three year data alone than the flat model,
the combination of WMAP three year data and other astronomical data
favors nearly flat cosmologies.
({\it Upper left}) 
WMAP+HST key project measurement of $H_0$.
({\it Upper right}) 
WMAP+SDSS LRG measurement of the angular diameter distance to $z=0.35$. 
({\it Middle left}) 
WMAP+SNLS data.
({\it Middle right}) 
WMAP+SNGold. 
({\it Lower left}) 
WMAP+2dFGRS.
({\it Lower right}) 
WMAP+SDSS.
\label{fig:omegam_omegav}}
\end{figure}
The WMAP observations place significant constraints on the geometry
of the universe through the positions of the acoustic peaks.
The sound horizon size, $r_s$, serves as a very useful ruler for measuring
the distance to the surface of last scatter.  For power law open universe models,
$\ensuremath{\theta_A = 0.535^{+ 0.030}_{- 0.027}\ \mbox{\arcdeg}}$.
Figure \ref{fig:omegam_omegav} shows that this constraint confines the likelihood function
to a narrow degeneracy surface in $(\Omega_m,\Omega_\Lambda)$.
This degeneracy line is well fit by $\Omega_{K} =   -0.3040 +    0.4067 \Omega_{\Lambda}$.
However, the CMB data alone does not distinguish between models along the valley: it is consistent
with both flat models and models with $\Omega_\Lambda = 0$.  If we allow for a large SZ signal,
then the WMAP data alone favors a model with $\Omega_K = -0.04$; however, this model
is not consistent with other astronomical data.

The combination of WMAP data and other astronomical data places strong constraints on the geometry of the
universe (see Table \ref{tab:omegam_omegav}):  
\begin{itemize}
\item The angular scale of the baryon acoustic oscillation (BAO) 
peak in the SDSS LRG sample \citep{eisenstein/etal:2005} measures 
the distance to  $z = 0.35$.  The combination of the BAO and CMB observations strongly constrain
the geometry of the universe. The position of the peak in the galaxy spectrum in the SDSS
and 2dFGRS surveys provide local measurements of the angular diameter distance.
\item Figure \ref{fig:Wv-Wm} shows that the Hubble constant varies along
this line, so that the HST key project 
constraint on the Hubble
constant leads to a strong bound on the curvature.
\item SNe observations measure the luminosity distance to $z \sim 1$.  The combination of
SNe data and CMB data also favors a nearly flat universe.
\end{itemize}
 
The strong limits quoted in 
Table \ref{tab:omegam_omegav} rely on our assumption that the dark energy
has the equation of state, $w=-1$.  In section \ref{sec:w}, we discussed
relaxing this assumption and assuming that $w$ is a constant.  Figure
\ref{fig:omegamw} shows that by using the combination of CMB, large-scale
structure and supernova data, we can simultaneously constrain
both $\Omega_k$ and $w$.  This figure confirms that our minimal
model, $\Omega_k = 0$ and $w=-1$ is consistent with the current data.

\begin{table}[htbp!] 
\begin{center}
\caption{Joint Data Set Constraints on Geometry and Vacuum Energy
\label{tab:omegam_omegav}}
\begin{tabular}{|c||c|c|}
\hline
Data Set& $\Omega_K$ & $\Omega_\Lambda$ \\
\hline
\hline
WMAP + $h = 0.72 \pm 0.08$ &
\ensuremath{-0.014\pm 0.017} &
\ensuremath{0.716\pm 0.055}  \\
WMAP + SDSS &
\ensuremath{-0.0053^{+ 0.0068}_{- 0.0060}} &
\ensuremath{0.707\pm 0.041}  \\
WMAP + 2dFGRS &
\ensuremath{-0.0093^{+ 0.0098}_{- 0.0092}} &
\ensuremath{0.745^{+ 0.025}_{- 0.024}}  \\
WMAP + SDSS LRG &
\ensuremath{-0.012\pm 0.010} &
\ensuremath{0.728\pm 0.021}  \\
WMAP + SNLS&
\ensuremath{-0.011\pm 0.012} &
\ensuremath{0.738\pm 0.030}  \\
WMAP + SNGold &
\ensuremath{-0.023\pm 0.014} &
\ensuremath{0.700\pm 0.031}  \\
\hline
\hline
\end{tabular}
\end{center}
\end{table}

\section{Are CMB\ Fluctuations Gaussian?\label{sec:ng}}

The detection of primordial non-Gaussian fluctuations in the CMB
would have a profound impact on our understanding of the physics of
the early universe.  While the simplest inflationary models predict
only mild non-Gaussianities that should be undetectable in the WMAP
data,
there are a wide range of plausible mechanisms for generating significant 
and detectable non-Gaussian fluctuations (\citet{bartolo/etal:2004}
for a recent review).  There are a number of plausible extensions of
the standard inflationary model
\citep{lyth/ungarelli/wands:2003,dvali/gruzinov/zaldarriaga:2004b,
  bartolo/matarrese/riotto:2004} or alternative early universe models
\citep{arkani-hamed/etal:2004,alishahiha/silverstein/tong:2004}  that predict skewed primordial
fluctuations at a level detectable by WMAP. 

There are other cosmological mechanisms for generating non-Gaussianity.
The smallness of the CMB quadrupole seen by both WMAP and COBE has
stimulated interest in the possibility that the universe may be finite
\citep{luminet/etal:2003,aurich/etal:2005}.  If the universe were
finite and had a size comparable to horizon size today, then the
CMB fluctuations would be non-Gaussian
\citep{cornish/spergel/starkman:1996,levin/etal:1997,bond/pogosyan/souradeep:2000,inoue/tomita/sugiyama:2000}.
While analysis of the first year data did not find any evidence for a
finite universe \citep{phillips/kogut:2006, cornish/etal:2004}, these
searches were non-exhaustive so the data rule out most but not all
small universes.   

Using an analysis of Minkowski functionals,
\citet{komatsu/etal:2003} did not find evidence for
statistically isotropic but non-Gaussian fluctuations
in the first year sky maps .  
The \citet{colley/gott:2003} reanalysis of the
maps confirmed the conclusion that there was no evidence of
non-Gaussianity.   \citet{eriksen/etal:2004a} measured the Minkowski functionals
and the Length of the Skeleton for the first year maps on 11 different smoothing scales.  While they found no evidence for deviations from non-Gaussianity using the Minkowski area, Minkowski length
and the length of the skelton, they did find a intriguingly high $\chi^2$ for the genus statistic.

For a broad class of theories, we can parameterize the effects of
non-linear physics by a simple coupling term that couples a Gaussian
random field, $\psi$, to the Bardeen curvature potential, $\Phi$: 
\begin{equation}
\Phi(\vec x) = \psi(\vec x) + f_{NL} \psi^2(\vec x)
\label{eq:fnl}
\end{equation}
Simple inflationary models based on a single slowly-rolling scalar field
with
the canonical kinetic Lagrangian predict $|f_{NL}| < 1$
\citep{maldacena:2003, bartolo/etal:2004}; however, curvaton inflation
\citep{lyth/ungarelli/wands:2003},  ghost inflation 
\citep{arkani-hamed/etal:2004}, and Dirac-Born-Infeld (DBI) inflation models
\citep{alishahiha/silverstein/tong:2004} can generate much larger
non-Gaussianity, $|f_{NL}| \sim 100$. Using the
WMAP first year data, \citet{komatsu/etal:2003} constrained $-54 <
f_{NL} < 134$ at the 95\% confidence level. Several different groups
\citep{gaztanaga/wagg:2003,   mukherjee/wang:2003, cabella/etal:2004,
  phillips/kogut:2006,creminelli/etal:prep} have applied  alternative techniques to measure
$f_{NL}$ from the maps and  have similar limits on $f_{NL}$.
\citet{babich/creminelli/zaldarriaga:2004} note that these limits 
are sensitive to the physics that generated  the non-Gaussianity as different mechanisms
predict  different forms for the bispectrum.

Since the release of the WMAP data, several groups have  claimed detections of significant non-Gaussianities
\citep{tegmark/deoliveira-costa/hamilton:2003,eriksen/etal:2004,copi/huterer/starkman:2003,vielva/etal:2004,hansen/etal:2004,park:2004,larson/wandelt:2004,
cruz/etal:2005}.  Almost all of these
claims imply that the CMB fluctuations are not stationary
and claim
a preferred direction or orientation in the data.
\citet{hajian/souradeep/cornish:2005} argue that these detections
are based on {\it a posteriori} selection of preferred directions
and do not find evidence for preferred axes or directions.
Because of the potential
revolutionary significance of these detections, they must be treated
with some caution.  Galactic foregrounds are non-Gaussian and
anisotropic,
and even low level contamination in the maps can produce detectable
non-Gaussianities \citep{chiang/etal:2003, naselsky/etal:2005},
but have only minimal effects on the angular
power spectrum \citep{hinshaw/etal:2003}.  Similarly, point sources can be a source of
non-Gaussianity at small angular scales \citep{eriksen/etal:2004b}.
Because of the WMAP scan pattern, the variance in the noise in the
maps is spatially variable.  There is significant $1/f$ noise in several
of the Difference Assemblies (DAs) (particularly W4)--- which leads to anisotropies in the
two-point function of the noise.  Finally, most of the claimed
detections of significant non-Gaussianities are based on {\it a
  posteriori} statistics.  Many of the claimed detections of
non-Gaussianity can be tested with the three year WMAP data (available at
lambda.gsfc.nasa.gov).  Future tests should use the simulated noise
maps, Monte Carlo simulations and the  difference maps (year 1 $-$ year 2, year 2 $-$ year 3, etc.) 
to confirm that the tests
are not sensitive to noise statistics and the multi-frequency data to
confirm that any claimed non-Gaussianity has a  thermal spectrum.
Claims of non-Gaussianity incorporating data close to
the galactic plane (within the Kp2 cut) should be treated with caution,
as the foreground corrections near the plane are large and uncertain.

The following subsections describe a number of statistical tests designed to search for
non-Gaussianities in the microwave sky.  
All of these analyses use three year maps cleaned with the
KKaHaDust templates \citep{hinshaw/etal:prep}. We refer
to these maps as the ``template-cleaned maps".
In the first subsection, we
show that the probability distribution function of the cleaned
CMB  maps is consistent with Gaussianity.  In the
second subsection, we show that the Minkowski functionals are
consistent with expectations for Gaussian fluctuations. Next, we
compute the bispectrum of the cleaned maps.  
The final subsection describes measurements of the four point function. 

\subsection{One Point Distribution Function}

One of the simplest tests of non-Gaussianity is a measurement of the one point probability function.
However, because the detector noise in the map is inhomogeneous
(higher in the ecliptic plane and lower near the poles), this test is
non-trivial.  We account for the spatial variations in noise by
computing a variance-normalized temperature for each pixel in a given
map:  
\begin{equation}
u_i = \frac{T_i}{\sqrt{\sigma^2_{noise}/N_{obs}+ \sigma_{CMB}^2}}
\label{eq:pdf}
\end{equation}
where $T_i$ is the measured temperature signal, the detector noise depends on the 
number of observations of a given pixel, $N_{obs}$. Here,
we apply the analysis to template-cleaned maps outside the Kp2 skycut.
For this analysis, we compute $\sigma_{noise}$, the noise
per observation, from the year 1 $-$ year
2 difference maps and fit $\sigma_{CMB}$, the CMB signal,  to the sum of
the year one and year two maps. 
With $N_{side} = 1024$, the computed
$\sigma_{noise}$ value is within 0.5\% of the value of $\sigma_0$
estimated from the time series \citep{jarosik/etal:prep}.  As we lower
the resolution, the value of $\sigma_{noise}$ slowly increases with increasing pixel
size.  For W4, the channel with the large $1/f$ noise,
this change is most dramatic; the value of $\sigma_0$
at resolution $N_{side} = 32$ is 6\% higher
than the value computed for $N_{side} = 1024$. 

\begin{figure}[htbp!] 
\includegraphics[width=2in]{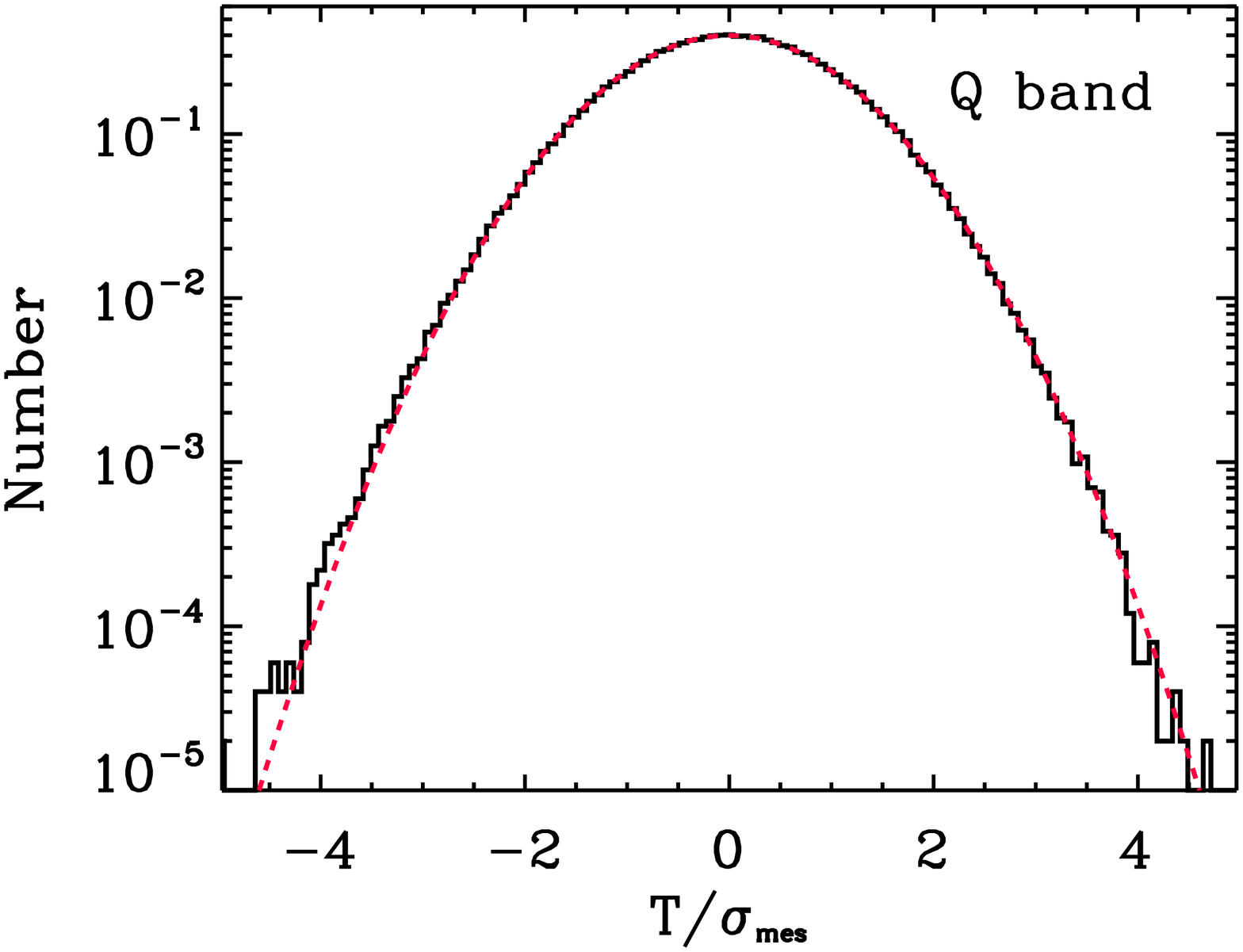}
\includegraphics[width=2in]{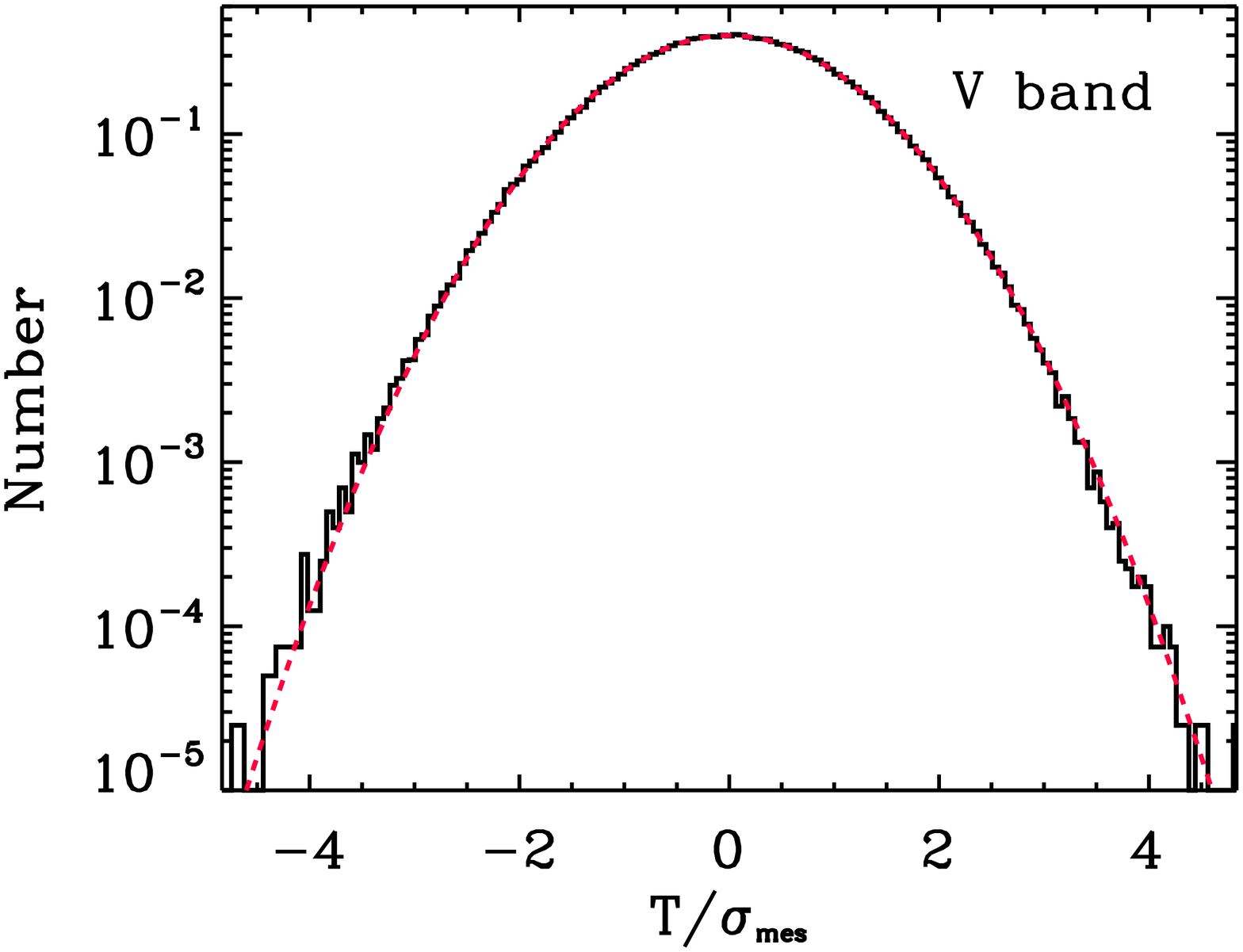}
\includegraphics[width=2in]{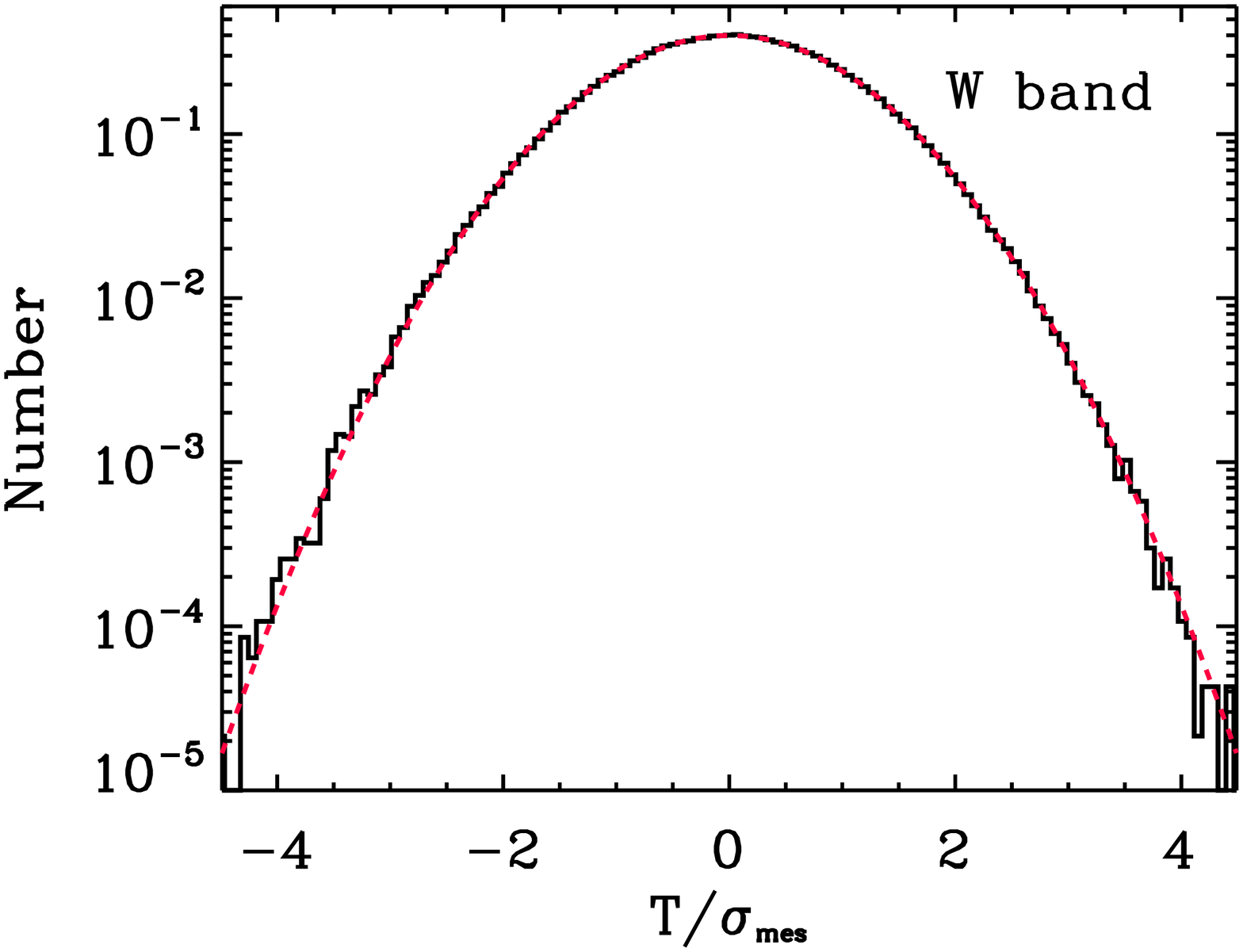}
\caption{\fg Normalized one point distribution function
of temperature anisotropy, defined in equation (\ref{eq:pdf}),
for the  
template-cleaned  Q (left), 
V (middle) and W (right) band maps outside
the Kp2 cut.  The sky maps
have been degraded to $N_{side} = 256$ for this figure.  
The red line
shows the Gaussian distribution, which is an excellent fit
to the one point distribution function.
\label{fig:pdf_freq}}
\end{figure}

\begin{figure}[htbp!] 
\includegraphics[width=2in]{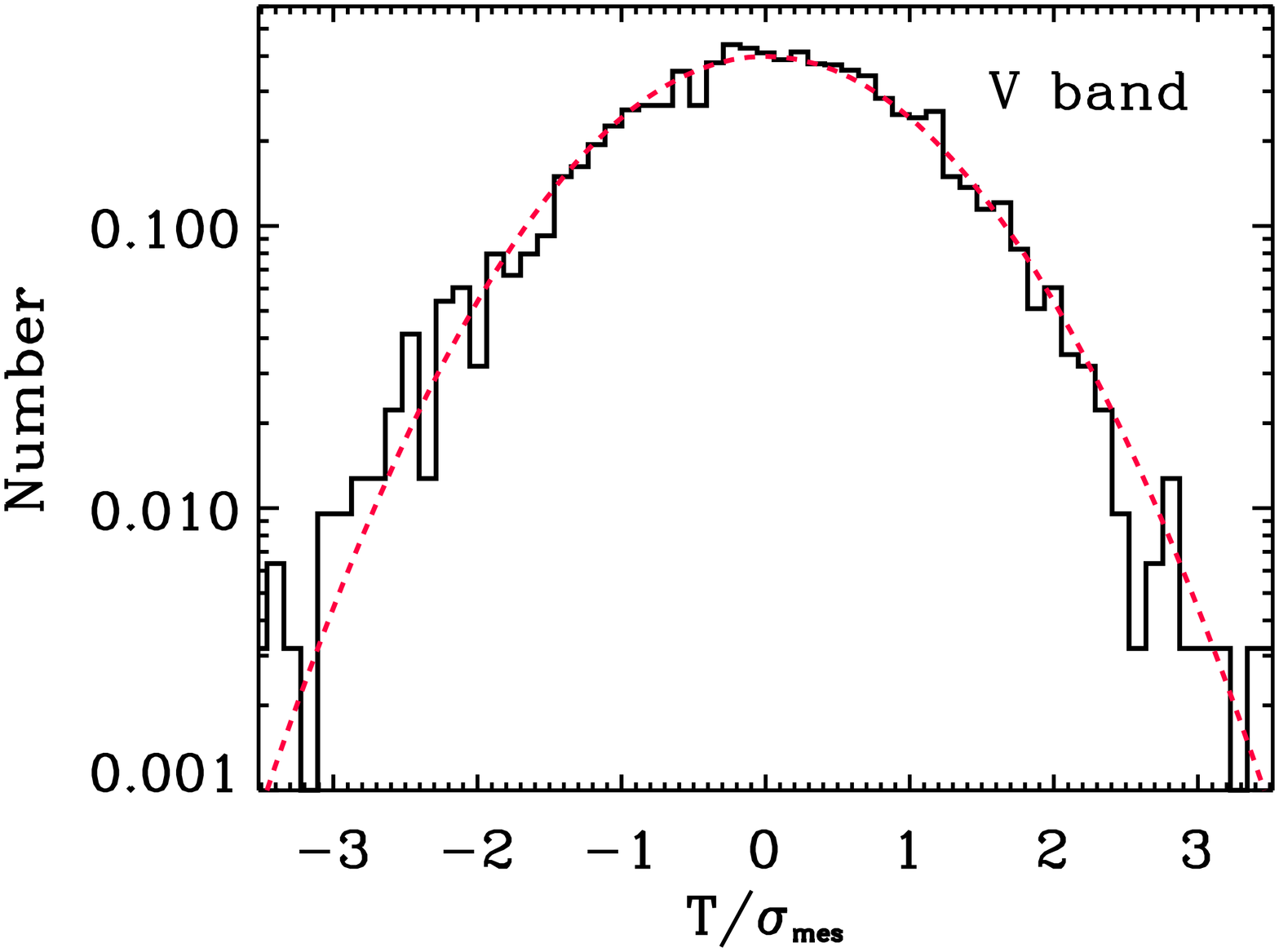}
\includegraphics[width=2in]{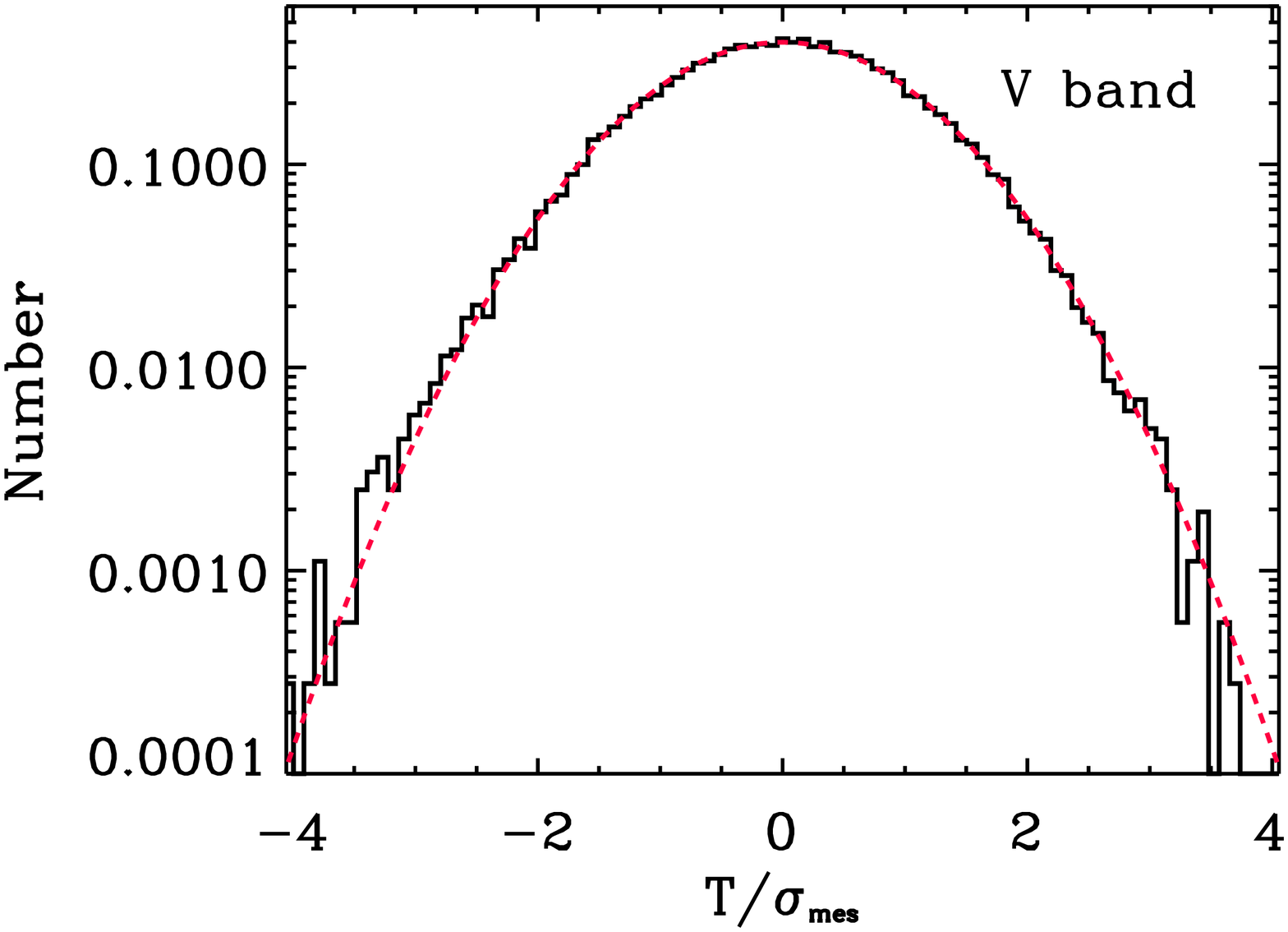}
\includegraphics[width=2in]{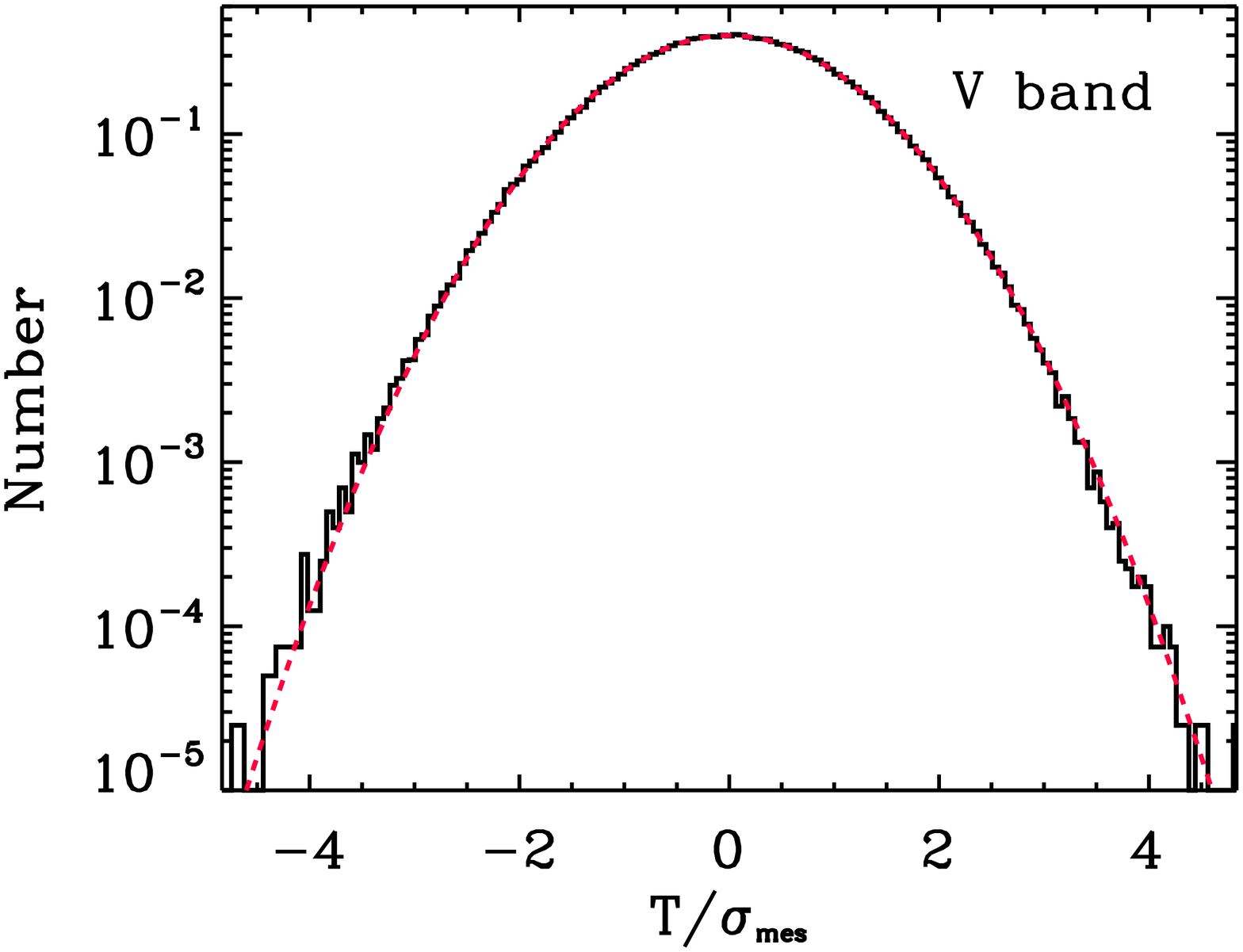}
\caption{\fg Normalized one point distribution function of temperature
 anisotropy, defined in equation (\ref{eq:pdf}),
for the template-corrected V band data maps 
outside the Kp0 cut.  The sky maps
have been degraded to $N_{side} = 16$(left), $64$(middle) and $256$(right) for this figure.  The red line shows the best fit Gaussian, which is an
excellent fit to the one point distribution function.
\label{fig:pdf_scale}}
\end{figure}

Figures \ref{fig:pdf_freq} and \ref{fig:pdf_scale} shows the one-point
distribution function of the cleaned sky maps as a function of
resolution.  Most of the distributions appear to be rather well fit
by a Gaussian.   This result is consistent with that from
the area of hot and cold spots (one of the Minkowski functionals),
which measures the cumulative one point probability function.
However,  with $N_{side} = 16$, there is a slight excess at negative $T$ in the first panel of
Figure \ref{fig:pdf_scale}.  This is due to the cold spot near $l =  209$ and $b= -57$
discussed by \citep{vielva/etal:2004} in the context of the first year data.  \citet{cruz/etal:2006} argue that there is a 1.85\% probability of seeing the level of skewness and kurtosis associated with this cold spot.
While this probability would increase if they allowed for variable smoothing scales, the results are intriguing and merit further investigation.

\clearpage 

\clearpage

\subsection{Size and Shape of Hot and Cold Spots}

Minkowski functionals \citep{minkowski:1903, gott/etal:1990, mecke/buchert/wagner:1994,
schmalzing/buchert:1997, schmalzing/gorski:1998, winitzki/kosowsky:1998} 
measure the contour length, area, and number of hot and cold spots.   Following the approach used in the first year analysis,
we compute the Minkowski functionals as a function of temperature threshold, $\nu = \Delta T/\sigma$, where
$\sigma$ is the standard deviation of the map.
For a two dimensional map, we measure three
Minkowski functionals, the genus, $G(\nu)$, of the maps, the contour length, $C(\nu)$ and the area within
the contours, $A(\nu)$.

\begin{figure}[htbp] 
\centering
\includegraphics[width=4.7in]{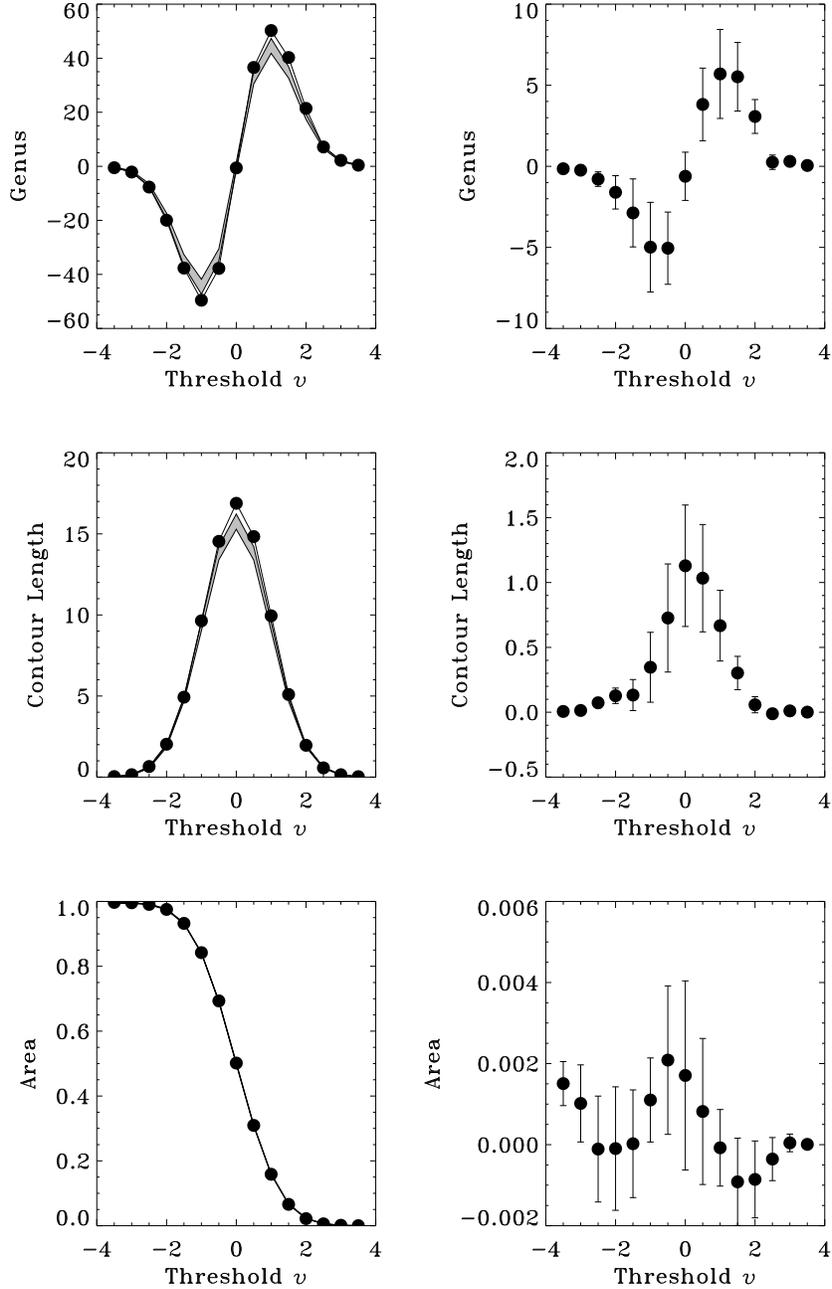}
\caption{
\fg
 The {\sl WMAP} data are in excellent agreement with the Gaussian simulations
based on the analysis of
 the Minkowski functionals for the
three year  {\sl WMAP} data  outside the Kp0 cut.
The filled circles in the left panel shows the  values for the data
 at $N_{side}=128$ (28$\arcmin$ pixels).
 The gray band shows the 68\% confidence interval for the Gaussian 
 Monte Carlo simulations.
 The right panels show the residuals between the mean of the Gaussian
 simulations and the {\sl WMAP} data.  Note that the residuals
are highly correlated from bin to bin, so the $\chi^2$ are
consistent with the Gaussian simulations.
 \label{fig:minkowski}
 }
\end{figure}

We compare the measured values of the Minkowski functionals to  their expected amplitude for a Gaussian sky.  We
simulate a series of maps based on the best fit parameters for $\Lambda$CDM  and the WMAP noise patterns.   For the analysis, we use the template-cleaned
V+W maps outside the Kp0 sky mask region. 
Following the approach used in \citet{komatsu/etal:2003}, we compute the Minkowski
functionals at 15 thresholds from $-3.5 \sigma$ to $+3.5 \sigma$ and compare each functional to the simulations using
a goodness of fit statistic,
\begin{equation}
\chi^2 = \sum_{{\nu_1}\ {\nu_2}} \left[F_{WMAP}^i - \left< F_{sim}^i\right>\right]_{\nu_1} \Sigma_{{\nu_1}\ {\nu_2}}^{-1}
                                 \left[F_{WMAP}^i - \left< F_{sim}^i\right>\right]_{\nu_2}
\end{equation} 
where $F_{WMAP}^i$ is the Minkowski functional computed from the WMAP data, $F_{sim}^i$ is the Minkowski functional 
computed from the simulated data, and $\Sigma_{\nu_1 \nu_2}$ is the bin-to-bin covariance from the simulations.
Figure \ref{fig:minkowski} shows the Minkowski functionals as a function of threshold for a map with $N_{side} = 128$
(28 $\arcmin$ pixels).  These pixels are small enough to resolve the
acoustic spots, but not so small as to be noise dominated. The figure
shows that the contour length, area, and number of spots is consistent with
expectations for a Gaussian theory. Table \ref{tab:minkowski} lists
the probability of measuring the observed values of the Minkowski
functionals as a function of pixel size.  At all resolutions, the maps
are consistent with Gaussian simulations. 

We have also simulated non-Gaussian sky with 
non-Gaussian
signals generated according to equation (\ref{eq:fnl}).  By comparing
these simulations to the data, we can constrain $f_{NL} = 7 \pm 66$
at the 68\% confidence level, consistent with the bispectrum measurement
(\S \ref{section:bispectrum}).

\begin{table}[h] 
 \begin{center}
 \caption{
  $\chi^2$ for Minkowski Functionals
for 15 thresholds for the template-cleaned VW
\label{tab:minkowski}
  }
  \begin{tabular}{|c| c| c c c c |}
\hline
Pixels &   Minkowski    & $\chi^2$  &  DOF     & $<Sim>$  &   F$>$WMAP \\
\hline
128 &   Genus       & 20.9  &  15    &  15.4 &  0.17 \\
128 &    Contour    &  19.2 &   15   &   15.1&   0.19 \\
128 &    Spot Area  &  14.0 &   15   &   15.3&   0.54 \\
128 &    Combined   &  51.6 &   45   &   47.2&   0.31 \\
\hline
64 &    Genus       & 18.3  &  15    &  14.9 &  0.24 \\
64 &    Contour     & 19.3  &  15    &  14.9 &  0.19 \\
64 &    Spot Area   &  8.4  &  15    &  15.5 &  0.93 \\
64 &    Combined    & 50.0  &  45    &  47.2 &  0.36 \\
\hline
32 &    Genus       & 17.3  &  15    &  15.4 &  0.31 \\
32 &    Contour     & 27.8  &  15    &  15.8 &  0.04 \\
32 &    Spot Area   &  8.5  &  15    &  15.8 &  0.89 \\
32 &    Combined    & 43.8  &  45    &  49.1 &  0.61 \\
\hline
16 &   Genus        &28.2   & 15     & 15.8  & 0.05 \\
16 &   Contour      &19.0   & 15     & 15.7  & 0.29 \\
16 &   Spot Area    &14.1   & 15     & 15.6  & 0.47 \\
16 &   Combined     &84.6   & 45     & 49.4  & 0.03 \\
\hline
8 &   Genus        &  10.8   & 15    &  15.5 &  0.62 \\
8 &    Contour      &  24.3  &  15   &   16.0&   0.09 \\
8 &    Spot Area  &    28.8  &  15   &   15.0&   0.05 \\
8 &    Combined     & 100.5  &  45  &    49.0 &  0.03 \\
  \hline
  \end{tabular}
\end{center}
\end{table}

\subsection{Bispectrum
\label{section:bispectrum}}

The bispectrum is sensitive to both primordial non-Gaussianity and
various sources of secondary anisotropy \citep{spergel/goldberg:1999,goldberg/spergel:1999,komatsu/spergel:2001}.
Here, we use the WMAP 3 year
data to constrain the amplitude of primordial non-Gaussianity and to
detect the amplitude of the point source signal in
the cleaned Q, V and W band maps. 

The amplitude of the primordial non-Gaussian signal can be found by
computing a cubic statistic on the map \citep{komatsu/spergel/wandelt:2005}: 
\begin{equation}
S_{primordial} = \frac{1}{f_{sky} }\int 4\pi r^2 dr \int \frac{d^2\hat n}{4 \pi} A(r,\hat n) B^2(r,\hat n)
\label{eq:KSW}
\end{equation}
where $f_{sky}$ is the fraction of the sky used in the
analysis, $B(r,\hat  n)$ is a Weiner filtered map of the primordial
fluctuations and $A$ is optimized to detect the form of the
non-linearities.  The amplitude of $S_{primordial}$ can be related
directly to $f_{NL}$. Here, we use $A$ and $B$ as defined in
\citet{komatsu/spergel/wandelt:2005}. While we used $\l_{max}=265$ for
the first-year analysis, we use $\l_{max}=350$ for the present analysis,
as noise is significantly lower with three years of data.
The error on $f_{NL}$ begins to increase at $\l_{lmax}>350$ due to
the presence of inhomogeneous noise.
Note that \citet{creminelli/etal:prep} argue that the optimal estimator
for $S_{primordial}$ should include a term that is linear in
temperature anisotropy as well as a cubic term that we already have in 
equation~(\ref{eq:KSW}).
They claim that their estimator could reduce the error on $f_{NL}$ by
about 20\%. While their result is attractive, 
we shall not include the linear term
in our analysis, as their estimator has not been tested against
non-Gaussian simulations and thus it is not yet clear if it is unbiased.

Point sources are an expected cause of non-Gaussianity.  Because
point sources are not very correlated on the angular scales probed by
WMAP, the point sources make a constant contribution to the
bispectrum, $b_{src}$.  \citet{komatsu/spergel/wandelt:2005} develops a cubic
statistic approach for computing $b_{src}$: 
\begin{equation}
S_{ps} = \frac1{m_3} \int \frac{d^2\hat n}{4\pi} D^3(\hat n)
\end{equation}
where 
$m_3=(4\pi)^{-1}\int d^2\hat{n}M^3(\hat{n})$,
$M(\hat{n})=[\sigma^2_{\rm CMB}+N(\hat{n})]^{-1}$, and
$D(\hat n)$ is a match filter optimized for point source detection:
\begin{equation}
D(\hat n) = \sum_{\l,m} \frac{b_\l}{\tilde C_\l} a_{lm} Y_{lm}(\hat n) 
\end{equation}
where $b_\l$ is a beam transfer function and $\tilde{C}_\l=C_\l^{cmb}b_\l^2 + N_l$.
We weight the temperature maps by $M(\hat{n})$ before we calculate
$a_{lm}$. We use $\l_{max}=1024$ for calculating $D(\hat{n})$.
(See \S~3.2 of \citet{komatsu/etal:2003} for details of the 
weighting method.) 
Given the uncertainties in the source cut-off and the luminosity
function, the values for $b_{src}$ in Table \ref{tab:bispectrum}
are consistent with the values of $A_{ps}$ in \citet{hinshaw/etal:prep}.

Table \ref{tab:bispectrum} lists the measured amplitude of the
non-Gaussian signals in the 3 year maps.  The values are computed for
template-cleaned Q, V and W band maps.  With three years of data, the
limits on primordial non-Gaussianity have improved from
$-58<f_{NL}<137$ to $-54<f_{NL}<114$ at the 95\% confidence level.
The improvement in limit on $f_{NL}$ is roughly consistent with
the expectation that in the signal-dominated regime, 
$\Delta f_{NL}\propto l_{max}^{-1}$ \citep{komatsu/spergel:2001}. 
The level of
point source non-Gaussianity in the 3 year maps is lower than in the
first year maps.  This drop is due to the more sensitive 3 year masks
removing additional sources.  

\begin{table} 

\caption{Amplitude of Non-Gaussianity
\label{tab:bispectrum}
  }
\begin{center}
  \begin{tabular}{|c|c|c|}
   \hline\hline
   & $f_{\rm NL}$ & $b_{\rm src}$ \\
   & & [$10^{-5}~{\rm \mu K^3~sr^2}$]  \\
\hline
   Q
   & $41\pm 55$ & $4.8\pm 2.0$\\
   V 
   & $25\pm 50$ & $0.12\pm 0.52$\\
   W 
   & $11\pm 50$ & $-0.21\pm 0.34$\\
   V$+$W
   & $18 \pm 46$ & $0.25\pm 0.26$ \\
  Q$+$V$+$W
   & $30 \pm 42$ & $0.73\pm 0.36$\\
\hline
\hline
  \end{tabular}
 \end{center}
\end{table}

\subsection{Trispectrum}
\begin{figure}[h] 
\includegraphics[width=2in]{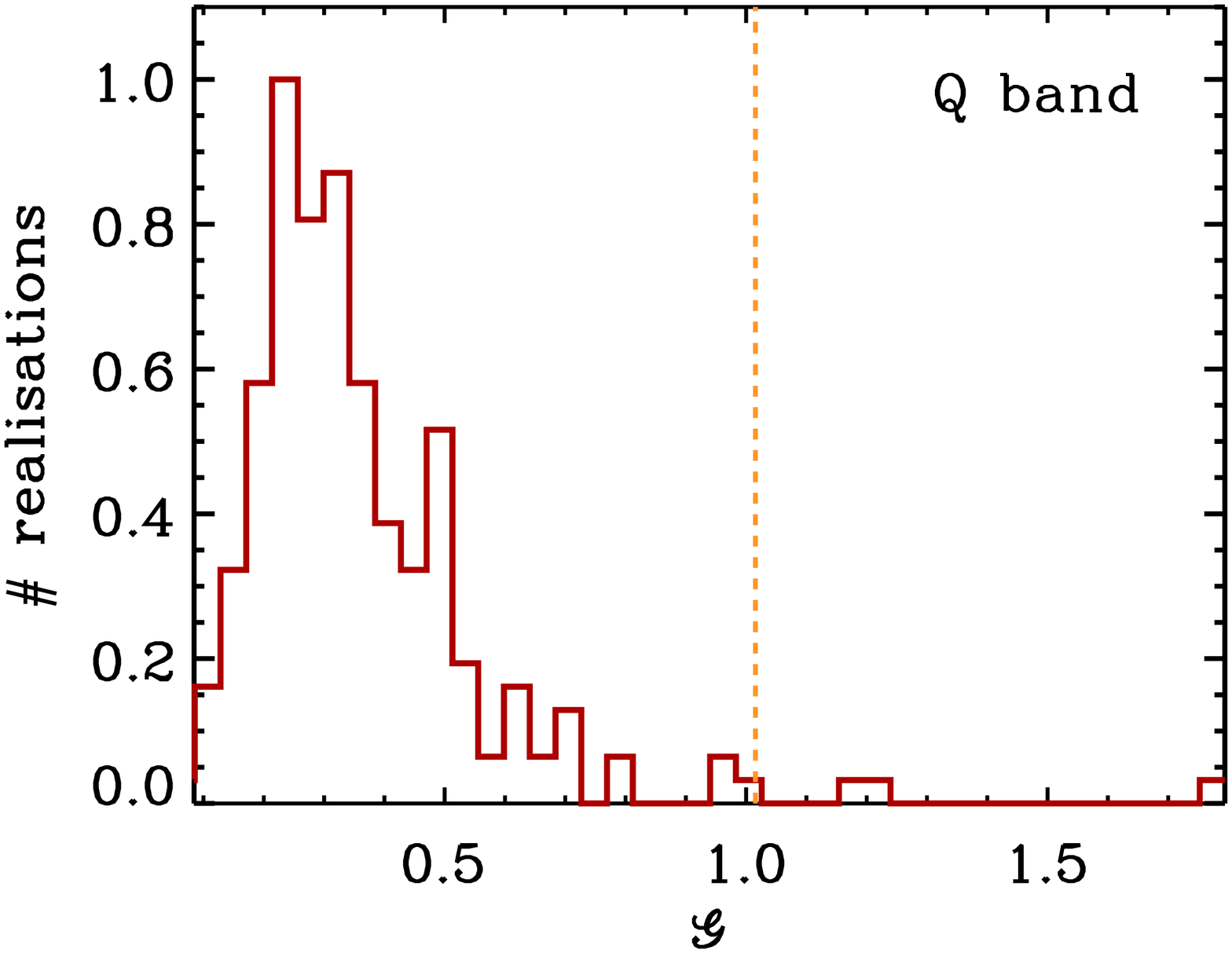}
\includegraphics[width=2in]{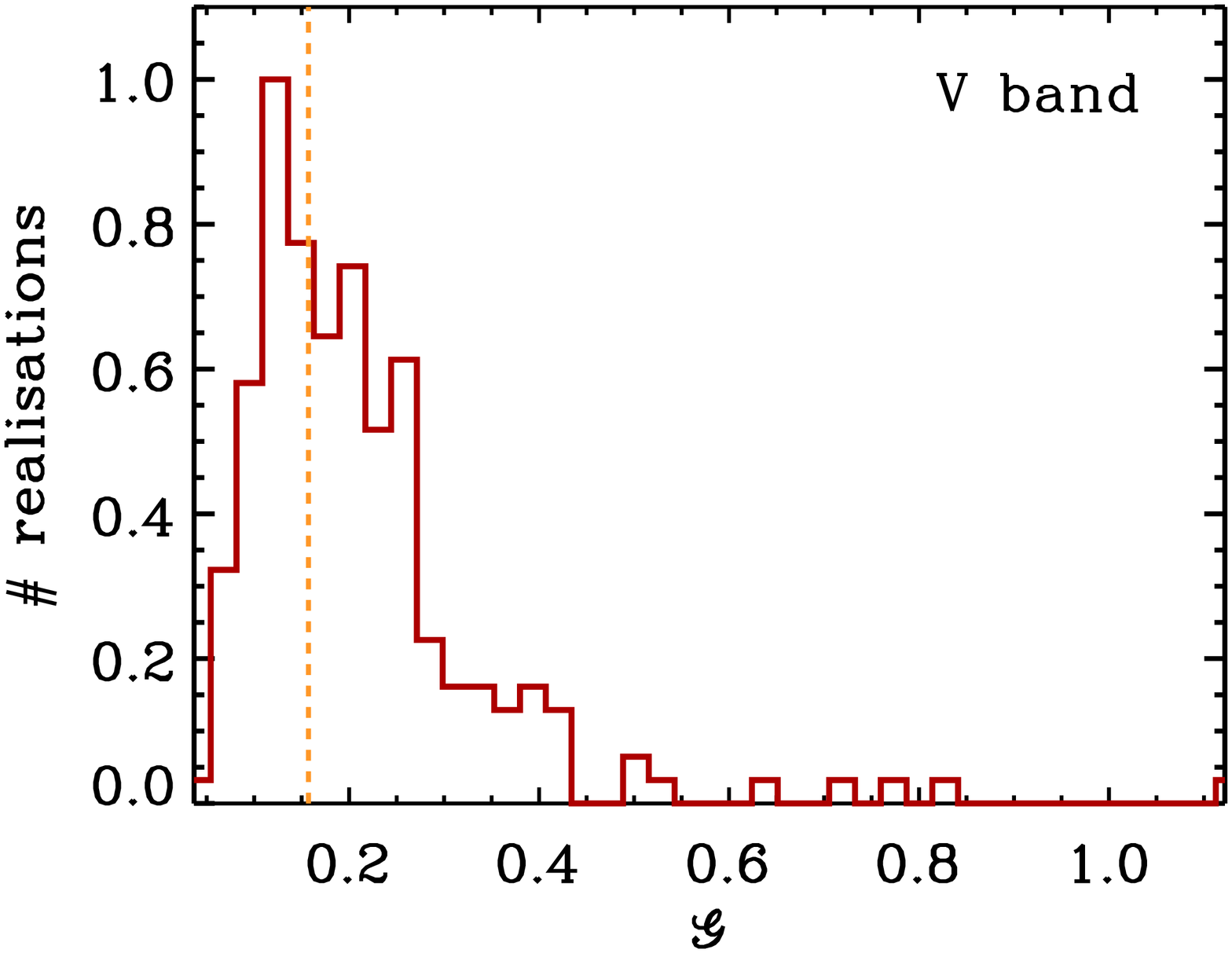}
\includegraphics[width=2in]{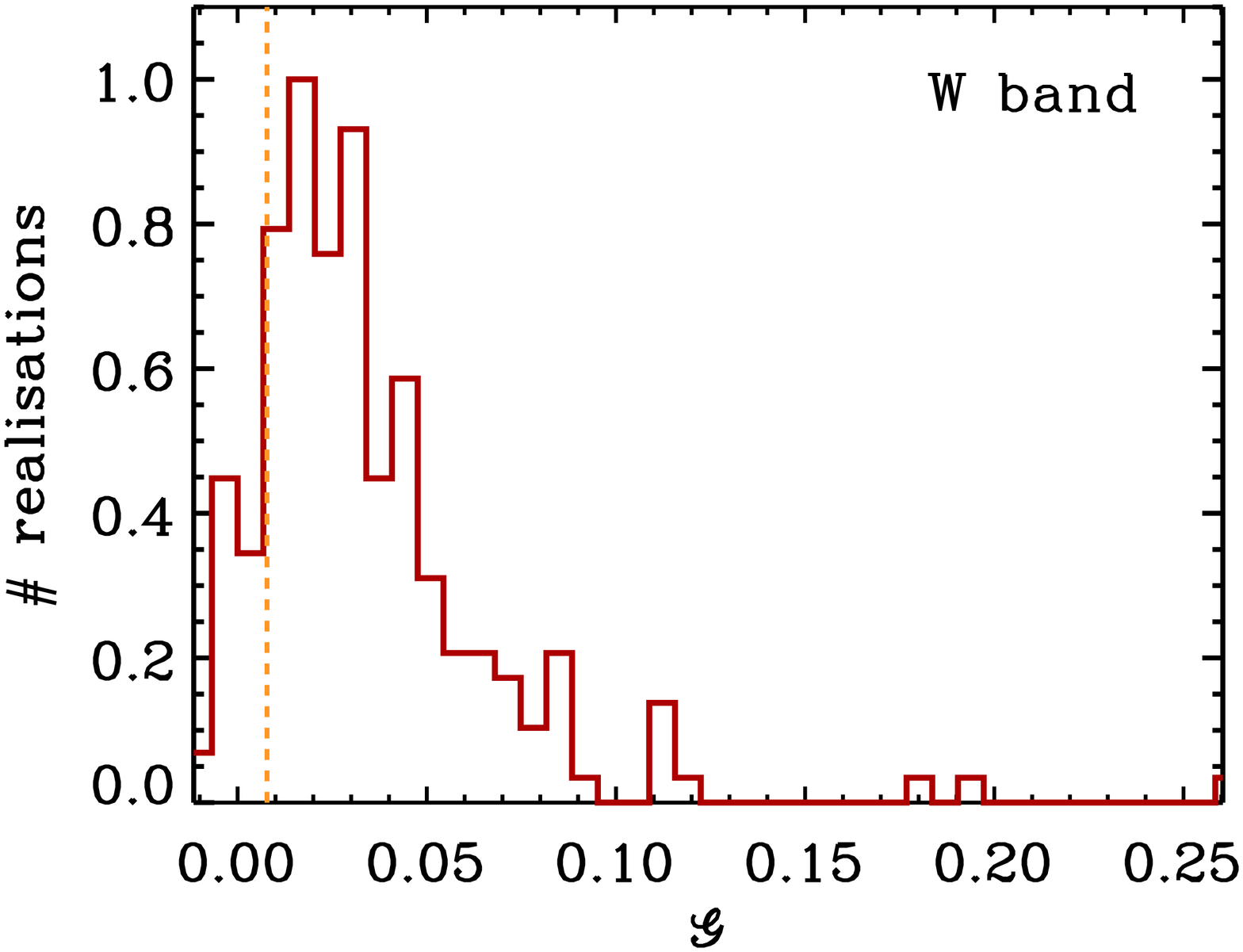}
\caption{\fg Constraints on the amplitude of the four point function.
The measured amplitude 
of the four point function (expressed in terms 
of a non-Gaussian amplitude defined in equation (\ref{eq:G})) is compared
to the same statistic computed for simulated Gaussian random fields.  The
yellow line
shows the results for Q, V and W bands and the red histogram
shows the distribution of the results of the Monte-Carlo realizations.
Note that in both the simulations and the data $A$ is greater than 0
due to the inhomogeneous noise.  The excess in Q is may be due to 
point source contamination.
\label{fig:trispectrum}}
\end{figure}

Motivated by claims that there are large scale variations in the amplitude of fluctuations, we consider a non-Gaussian model
that generates a non-trivial four point function for the curvature (and temperature) fluctuations, but does not
produce a three-point function.  This model describes a cosmology where the value of one field modulates the amplitude of 
fluctuations in a second field:
\begin{equation}
\Phi(\vec x) = 
 \phi(\vec x) [ 1 + g_{NL} \psi (\vec x)]
\end{equation}
where $\phi$ and $\psi$ are Gaussian random fields and $\Phi$ is the Bardeen curvature potential.
The presence of such a term would generate variations in the amplitude of fluctuations across the sky.

Appendix \ref{appendix:trispectrum} derives an estimator for the amplitude of the non-Gaussian term, $g_{NL}^2 |\psi|^2$.
This estimator is based on approximating the CMB fluctuations as arising from an infinitely thin surface of last scatter.
We measure the amplitude of the four point function by computing
\begin{equation}
\G = \sum_i (T^f_i \nabla^2 T_i^f - N_i^2)^2,
\label{eq:G}
\end{equation}
where $T^f$ is a smoothed map (e.g., an $N_{side}=128$ map), $T_i$ is an unsmoothed map, and $N_i$ is the expected
value of $T_f \nabla^2 T_i$ for a map without any signal.

Figure \ref{fig:trispectrum} shows  measurements of $\G$ from
the Q, V and W band data.  The V and W band data  show any evidence
for a non-trivial four point function, while Q band data may show
the contamination from point sources.
At the S/N level of the 3 year data, 
there are no significant cosmological or systematic effects
modulating the amplitude of the fluctuations as a function of scale.
\section{Conclusions \label{sec:conclusions}}

The standard model of cosmology has survived another rigorous
set of tests.  
The errors on the WMAP data at large $\ell$ are now three times smaller
and there has been significant improvements
in other cosmological measurements.  Despite the overwhelming force of the data, the model continues to thrive.  This was the basic result of \citep{spergel/etal:2003} and
was reinforced by subsequent analyses of the first year WMAP data with
the SDSS  \citep{tegmark/etal:2004b}, and 
analysis of first year WMAP plus the final 2dFGRS survey \citep{sanchez/etal:2006}.
After the analyses in this paper were completed, a larger SDSS LRG sample was released.
An analysis of the three year WMAP data combined with the new larger sample reached the same
basic conclusions as this paper: the $\Lambda$CDM model is consistent with both CMB and large-scale structure measurements and its basic parameters are now reasonably well constrained
\citep{tegmark/etal:2006}.

The data are so constraining that there is little room for significant
modifications of the basic $\Lambda$CDM model.   The combination of WMAP measurements
and other astronomical measurements place significant limits
on the geometry of the universe, the nature of dark energy, and
even neutrino properties. While allowing for a running spectral
index slightly improves the fit to the WMAP data, the improvement in the
fit is not significant enough to require a new parameter.

Cosmology requires new physics beyond the standard model of particle
physics: dark matter, dark energy and a mechanism to generate
primordial fluctuations.  The WMAP data provides insights into
all three of these fundamental problems: 
\begin{itemize}
\item The clear detection of
the predicted acoustic peak structure implies that the dark matter
is non-baryonic.  
\item
The WMAP data are consistent with a
nearly flat universe in which the dark energy
has an equation of state close to that of 
a  cosmological constant, $w=-1$.
The combination of WMAP data with
measurements
of the Hubble Constant, baryon oscillations, supernova data and large-scale
structure observations all reinforces the evidence for 
dark energy.
\item The simplest
model for structure formation, a scale-invariant spectrum of fluctuations,
is not a good fit to the WMAP data.  The WMAP data requires either
tensor modes or a spectral index with $n_s < 1$ to fit the angular
power spectrum.  These observations match the basic inflationary predictions
and are well fit by the predictions of the simple $m^2 \phi^2$ model.
\end{itemize}

Further WMAP observations and future analyses will test the inflationary paradigm.  While we do not find convincing evidence for significant non-Gaussianities,
an alternative model that better fits the low $\ell$ data would be an exciting
development.  Within the context of the inflationary models,  measurements
of the spectral index as a function of scale and measurements of tensor
modes directly will provide a direct probe into the physics of the first
moments of the big bang.

\section{Acknowledgments}

The {\WMAP} mission is made possible by the support of the Office of Space
Sciences at NASA Headquarters and by the hard and capable work of scores of
scientists, engineers, technicians, machinists, data analysts, 
budget analysts, managers, administrative staff, and reviewers.
We thank the referee for helpful comments,
Henk Hoekstra, Yannick Mellier and Ludovic Van Waerbeke for providing
the CFHTLS data and  discussions of the lensing data, 
John Sievers for discussions of small-scale CMB experiments,
 Adam Riess and Kevin Kriscinius
for discussion of supernova data, Daniel Eisenstein for discussion
of the SDSS LRG data, Max Tegmark for discussions of SDSS P(k) data, and John Peacock for discussions about the 2dFGRS data.
EK acknowledges support from an Alfred P. Sloan Research Fellowship. HVP is supported by NASA through Hubble Fellowship grant \#HF-01177.01-A
awarded by the Space Telescope Science Institute which is operated
by the Association of Universities for Research in Astronomy, Inc., for
NASA under contract NAS 5-26555.
This research has made use of NASA's Astrophysics Data System Bibliographic
Services, the HEALPix software, CAMB software, 
and the CMBFAST software.  
CosmoMC \citep{lewis/bridle:2002} 
 was used to produce Figures \ref{fig:nstauwmaponly} and
\ref{fig:sz_marg}.
We also used  CMBWARP software
\citep{jimenez/etal:2004}
for initial investigations of the parameter space.
This research was additionally supported by NASA LTSA03-000-0090,
NASA ADP04-000-0093,
NASA ATPNNG04GK55G, and NASA ADP05-0000-092 awards. 

\appendix
\section{SZ Marginalization, Priors, and Likelihood Approximations
\label{appendix:sz_marg}}

In this Appendix we discuss the affect that various analysis choices have on the estimated cosmological parameters, focusing on the six parameter $\Lambda$CDM model.

The three-year analysis now accounts for the SZ effect by marginalizing over the amplitude of the SZ contribution, parameterized by the model of Komatsu \& Seljak (2002).  Specifically, we evaluate their predicted spectrum using $\Omega_{m}=0.26, \Omega_{b}=0.044, h=0.72, n_{s}=0.97$ and $\sigma_{8}=0.8$  We define the amplitude of the SZ signal (relative to this model) by $A_{SZ}$ and marginalize over this parameter with a flat prior, $0 < A_{SZ} < 2$. This range is based on the assumption that the \citet{komatsu/seljak:2002}(KS) model estimates the true SZ signal with an uncertainty of order one.  Numerical simulations \citep{nagai:prep} and analytical studies \citep{reid/spergel:prep} find a tight correlation between mass and SZ signal, with uncertainties that are dominated by the cluster gas fraction.   These results support the \citet{komatsu/seljak:2002} approach and suggest that the range of this prior is generous.  \citet{afshordi/lin/sanderson:2005} analyze of the SZ signal from 116 nearby clusters in the first-year WMAP data and find that the signal from nearby clusters is 30-40\% weaker than expected.  Since these nearby clusters are the dominant source of fluctuations in the WMAP angular power spectrum, this implies that $A_{SZ} < 1$.

We now use the amplitude of the angular power spectrum peak, $C_{220}$, rather than $A$ as the amplitude parameter in the Monte Carlo Markov Chains.  This choice
of prior leads to a slightly lower best fit amplitude.

\begin{figure} 
\includegraphics[width=5in]{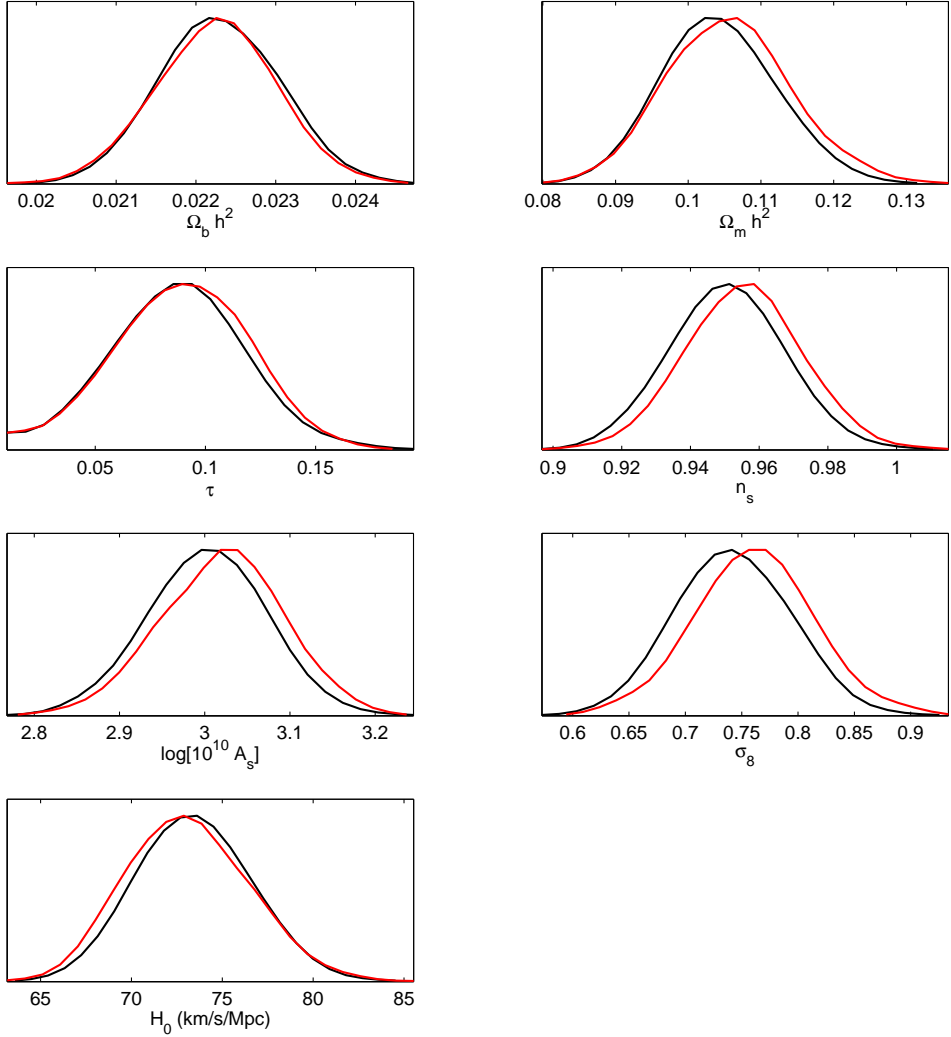}
\caption{\fg The effect of SZ marginalization on
the likelihood function.  The red curve is the likelihood
surface for the three-year WMAP data for the power-law $\Lambda$CDM 
model with $A_{SZ} = 0$.  The black curve is the likelihood
surface after marginalizing over the amplitude of the SZ contribution.
\label{fig:sz_marg}}
\end{figure}

Figure \ref{fig:sz_marg} shows how the SZ marginalization and the change in amplitude prior affect our estimates of cosmological parameters.  Except
for $\sigma_8$, the effects are all relatively small.  For $\sigma_8$ we estimate that roughly half of this change is due to the SZ marginalization, and half is due to the change in amplitude prior.  \citet{lewis:2006} provides a detailed study
of the effects of gravitational lensing on the analysis and show that it is smaller than the SZ effect.

Since the first draft of this paper was circulated, it has been pointed out \citep{eriksen/etal:2006} that the pseudo-$C_l$-based power spectrum is biased slightly high in the range $12<l<30$, which in turn biases the spectral index, $n_s$, slightly low.  As a result, we have increased the sky map resolution used in our exact TT likelihood module (to $N_{side}=16$), which allows us to use this form up to l=30.  We have established that this approach agrees well with an independent exact form based on Gibbs sampling methodology \citep{eriksen/etal:2006}.

\citet{huffenberger/etal:2006} argue that the point source amplitude quoted in the original version of \citet{hinshaw/etal:prep} was too high: they find $A_{ps} = 0.011 \pm 0.001$, while \citet{hinshaw/etal:prep} originally quoted $A_{ps} = 0.017 \pm 0.002$.  Subsequently, \citet{hinshaw/etal:prep} have re-analyzed the source contribution and now favor an intermediate value with a more conservative error, $A_{ps} = 0.014 \pm 0.003$.  The original source correction also had the effect of biasing the spectral index slightly low.

Table \ref{tab:ns_sens} shows the sensitivity of the spectral index, $n_s$, to
the various treatments discussed above.  The first row (Fiducial) gives the result originally quoted in this paper.  The second row (No SZ Correction) gives the result if we assume $A_{SZ}=0$.  The third row (No Beam Error) shows the effect of assuming the beam window function has no error.  The fourth row (Lower Point Source Amplitude) shows the change induced by changing the source amplitude (and its uncertainty) from $0.017 \pm 0.002$ to $0.014 \pm 0.003$.  The sixth row (Foreground Marginalization) shows the effect of disabling marginalization over the foreground emission template in the exact likelihood module (up to l=12).  The seventh row (Res 4 Likelihood) shows the effect of using the exact likelihood module to l=30, and the last row shows our current best estimate of $n_s$.

The detailed form of the likelihood function and the treatment of
point sources and the SZ effect has a $\sim 0.5 \sigma$ effect on the
best fit slope.  
We have also used simulations where we know the input model to check the effects
of including various secondary effects.  \citet{lewis:2006} provides a detailed study
of the effects of gravitational lensing on the analysis and show that it is smaller than the SZ effect.

\begin{table}[h!] 
\begin{center}
\caption{Sensitivity of Slope to Likelihood Model and Prior\label{tab:ns_sens}}
\begin{tabular} {|c|c|}
\hline
Fiducial  ($N_{side}=8$ Likelihood, $A_{PS} = 0.017$) &   $0.951 \pm 0.016$ \\
No SZ Correction &   $0.954 \pm 0.016$ \\ 
No Beam Error &   $0.953 \pm 0.016$ \\
Reduced Point Source Amplitude ($A_{PS} = 0.014$) & $0.955 \pm 0.016$ \\
\citet{huffenberger/etal:2006} Point Source Amplitude ($A_{PS}=0.011$) &  $0.957 \pm 0.016$ \\
Foreground Marginalization &
 $0.953 \pm 0.016$ \\
 $N_{side}=16$ Likelihood & $0.957 \pm 0.016$ \\
 $N_{side}=16$ Likelihood + $A_{PS}= 0.014$ + no SZ correction & $0.960 \pm 0.016$ \\
 $N_{side}=16$ Likelihood + $A_{PS}= 0.014$ & $0.958 \pm 0.016$ \\
 \hline
\end{tabular}
\end{center}
\end{table}

\section{Trispectrum methodology
\label{appendix:trispectrum}}

\subsection{Predicted Trispectrum Signal}

We consider here a model that generates a non-trivial trispectrum, but no bispectrum signal.  We assume that 
the gravitational potential 
$\Phi$,  is a product of two independent Gaussian fields, $\phi$
and $\psi$:
\begin{equation}
\Phi(\vec x) = \phi(\vec x) [ 1 + g_{NL} \psi (\vec x)]
\end{equation}
where $g_{NL}$ characterizes the strength of the non-linear term.

Following \citet{komatsu/spergel/wandelt:2005} approach for the
bispectrum, extended recently to the trispectrum in \citet{kogo/komatsu:prep}, the observed temperature multipoles are:
\begin{equation}
a_{\ell m} = b_\ell \int r^2 dr \Phi_{\ell m}(r) \alpha_\ell(r) + n_{\ell m}
\end{equation}
where $b_\ell$ is the beam, $n_{\ell m}$ is the noise and $\alpha_\ell(r)$  is the radiation transfer function:
\begin{equation}
\alpha_\ell(r) = \frac{2}{\pi} \int k^2 dk g_{T\ell}(k) j_\ell (kr)
\end{equation}

The non-linear coupling term generates a second order term:
\begin{equation}
a_{\ell m} = n_{\ell m} + b_\ell \int r^2 dr\alpha_\ell(r)  \left[ \phi_{\ell m}(r)  +
  \phi_{\ell'm'}(r) \psi_{\ell''m''}(r) C_{\ell m}^{\ell'm'\ell''m''}\right] 
\end{equation}
where
\begin{equation}
C_{\ell m}^{\ell'm'\ell''m''} = \sqrt{\frac{4\pi}{(2\ell+1)(2\ell'+1)(2\ell''+1)}}
                        \left(\begin{array}{ccc}
                                 \ell & \ell' & \ell''   \\
                                0&  0 & 0
                                \end{array}\right)
                                \left(\begin{array}{ccc}
                                 \ell&\ell' &\ell''    \\
                                m&  m' & m''
                                \end{array}\right).
\end{equation}
This term does not have any effect on the bispectrum as  $<\phi^3> = 0$ and $<\psi^3> = 0. $ However,
it does have a non-trivial effect on the trispectrum.

As with gravitational lensing (see \citet{hu:2001}), the largest trispectrum term is the diagonal term,
$T_{\ell\ell}^{\ell\ell}(0) = <C_\ell C_\ell> - 3<C_\ell>^2$.
This term would generate an excess in the $\chi^2$ of the fit of the model to the data:
\begin{eqnarray}
  T_{\ell\ell}^{\ell\ell}(0) &=&  g_{NL}^2
  C_\ell \int r^2 dr \int \tilde r^2 d\tilde r \alpha_\ell(r) \alpha_\ell(\tilde r)  \nonumber \\
  &&\sum_{m\ell'm'\ell''m''}\sum_{\tilde \ell' \tilde m' \tilde \ell'' \tilde m''}
  <\phi_{\ell'm'}(r) \phi_{\tilde \ell' \tilde m'} (\tilde r)>
  <\psi_{\ell''m''}(r) \psi_{\tilde \ell'' \tilde m''} (\tilde r)>
  C_{\ell m}^{\ell'm'\ell''m''} C_{\ell m}^{\tilde \ell' \tilde m'\tilde \ell'' \tilde m''}.
\end{eqnarray}
We can then use
\begin{equation}
<\phi_{\ell'm'}^{}(r) \phi_{\tilde \ell' \tilde m'}^{} (\tilde r)> =
        \delta_{\ell'\tilde \ell'}^{} \delta_{m'\tilde m'}^{}
        \int k^2 dk P_{\phi}(k)  j_{\ell'} (kr) j_{\ell'}(k\tilde r)
\end{equation}
and the equivalent relationship for $\psi$ to rewrite the trispectrum as
\begin{equation}
T_{\ell\ell}^{\ell\ell}(0) =  g_{NL}^2 \xi_\ell C_\ell^2
\end{equation}
where
\begin{eqnarray}
\xi_\ell &=& \frac{4 \pi}{C_\ell} \int k^2 dk P_{\phi}(k)\int (k')^2 dk' P_{\psi}(k') \nonumber \\
&&\left(\begin{array}{ccc}
  \ell & \ell' & \ell''   \\
  0&  0 & 0
\end{array}\right)^2    
\int r^2 dr \int \tilde r^2 d\tilde r \alpha_\ell(r) \alpha_\ell(\tilde r)
j_{\ell'} (kr) j_{\ell''} (k'r)  j_{\ell'} (k\tilde r) j_{\ell''} (k' \tilde r).
\end{eqnarray}
While this full integral is numerically intractable, we approximate the surface
of last scatter as a thin screen so that
\begin{equation}
a_{\ell m} = b_\ell \Phi_{\ell m}(r_*) {\bf \alpha_\ell} + n_{\ell m}
\end{equation}
then, the trispectrum coupling term reduces to
\begin{equation}
  \xi_\ell =  \frac{4 \pi {\bf \alpha_\ell}^2}{C_\ell} \int k^2 dk P_{\phi}(k) j_{\ell'}^2(kr_*)
  \int k'^2 dk' P_{\psi}(k') j_{\ell''}^2(k'r_*)
  \left(\begin{array}{ccc}
    \ell & \ell' & \ell''   \\
    0&  0 & 0
  \end{array}\right)^2.
\end{equation}
Recall that in this limit,
\begin{equation}
C_\ell = {\bf \alpha_\ell}^2 \int k^2 dk P_{\phi}(k) j_{\ell}^2(kr_*)
\end{equation}
Thus,
\begin{equation}
\xi_\ell =  \frac{4 \pi {\bf \alpha_\ell}^2 C_{\ell'}}{\bar \alpha_{\ell'}^2 C_\ell} 
\int k'^2 dk' P_{\psi}(k') j_{\ell''}^2(k'r_*)
\left(\begin{array}{ccc}
  \ell & \ell' & \ell''   \\
  0&  0 & 0
\end{array}\right)^2
\end{equation}
The amplitude of $\xi_\ell$ is, thus, roughly the variance in the $\psi$
field on the scale $r_*/\ell$. Note that $\xi_\ell$ is a positive definite quantity so that $T_{\ell\ell}^{\ell\ell}(0)$ should be always positive.

\subsection{Detecting the Non-Gaussian Signal}

If we assume that $\xi_\ell$ is constant, then we can follow \citet{hu:2001} and compute
an optimal quadratic statistic. We approximate the optimal statistic
as $\sum_i (T^f_i \nabla^2 T_i^f - N_i^2)^2$, where $T^f$ is a
smoothed map (\eg a res 7 map) and we use the approximation that $C_\ell
= A/\ell(\ell+1)$.  This has the advantage that we can easily compute it
and has well-defined noise properties.    

\subsubsection{Practical implementations}

We define for this purpose the dimensionless $\G$ statistic as
\bea
\G & = & \sum_{p,b_1,b_2,b_3,b_4}
  w_{p,b_1}^{}\hat T_{p,b_1}^{}w_{p,b_2}^{}\hat T_{p,b_2}^{}w_{p,b_3}^{}\hat
  T_{p,b_3}^{}w_{p,b_4}^{}\hat T_{p,b_4}^{}\nonumber \\
& - & \sum_{b_1,b_2,b_3,b_4}
\Bigl(\sum_{p_1} w_{p_1,b_1}^{}\hat T_{p_1,b_1}^{}w_{p_1,b_2}^{}\hat T_{p_1,b_2}^{}
\Bigr)
\Bigl(\sum_{p_2} w_{p_2,b_1}^{}\hat T_{p_2,b_1}^{}w_{p_2,b_2}^{}\hat T_{p_2,b_2}^{}
\Bigr) \nonumber\\
\eea
where $b_i$ refers to various bands (Q, V, and W for yr1, yr2 and yr3) that are
all distinct for a single term so that the noise bias is null for this
statistic, $w_{p,b_i}$ is a particular pixel weight (we will consider
it equal unity first) and $\hat T^{}_b$ is a filtered map defined as 
\bea
\hat T^{}_{pb} & = & {T^{}_{pb}\over\sqrt{\sum_q T^{2}_{qb}}}\\
T^{}_{pb} & = & \sum_{\ell m} f_{\ell b}^{} \tilde a_{\ell mb}^{M} Y_{\ell
  m}^{}(\hatn_p)
\eea
where $\tilde a_{\ell m}^{M}$'s are the spherical harmonic coefficients of
the masked sky. We use at this level the Kp12 mask to hide the brighter part
of the galaxy (and potentially the brighter point sources) and
ignore cut sky effects in considering those pseudo-$a_{\ell m}^{}$s. But
when computing the sum over pixels in $\G$, we consider only 
pixels outside the Kp2 area. The obvious advantage of this simple real
space statistic is its ability to handle inhomogeneous noise and to
localize its various contributions in real space. The second term in
the definition of $\G$ aims at subtracting off the Gaussian
unconnected part, so that if the $\hat T$ fields are homogeneous
Gaussian fields, we obtain $\langle\G^{ps}\rangle = 0$. 

The exact nature of $f_\ell^{}$ will depend on the source of the signal.
For example, point sources do contribute to all $n$-points
functions in real or harmonic space and are as such visible in the power
spectrum, bispectrum and trispectrum. The first two have been used to
set limits and corrections. 

Should we want to isolate the point sources contribution with the $\G$
statistic, we would proceed in the following manner. The point sources
power spectrum is well approximated by a constant, white noise like,
power spectrum $C^{ps}_\ell$ (see \citet{komatsu/etal:2003}). Given the
measured power spectrum, $\tilde C_\ell^{} = C_\ell^{}b_\ell^{2} + C^{ps}_\ell
b_\ell^{2} + N_\ell^{} $, the Wiener like filter to reconstruct point
sources is to $f^{ps}_\ell = b_\ell^{2}/\tilde C_\ell^{}$.  Note that this
filter would be optimal  only if point sources were drawn from a
Gaussian distribution, which is not true. We can however expect it to
be close to optimal. 

In order to constrain the CMB contribution to the trispectrum and constrain
$g_{NL}^{}$ close to optimality, we will set $f_{\ell1}^{}$ and
$f_{\ell2}^{}$ to the Wiener filter for the CMB field, $f_{\ell1}^{} =$
$f_{\ell3}^{} =$ $C_\ell^{theory}b_\ell^2/C_\ell^{measured}$ and $f_{\ell2}^{} =
f_{\ell4}^{}=\ell(\ell+1)f_{\ell1}^{}$, where $C_\ell^{theory}$ is the best fit
model angular power spectrum and $C_\ell^{measured}$ is the measured raw
power spectrum including the signal and the noise and not corrected
for the beam window function.

We restrict ourselves to a unit weighting which is nearly optimal in the signal
dominated regime where we draw our conclusions from, \ie at resolution
lower than $N_{side}=64$

\subsubsection{Explicit relation to the trispectrum}

Ignoring the weights, it is easy to show using the relation recalled
in the next section that 
\bea
\lefteqn{\langle T_1(\hatn)T_2(\hatn)T_3(\hatn)T_4(\hatn) \rangle_c^{}
  =  \sum_p T_{1p}T_{2p}T_{3p}T_{4p}              }&&\\ 
& = &  \sum_{\ell_1m_1\ell_2m_2\ell_3m_3\ell_4m_4} \langle
t_{\ell_1m_1}^{}t_{\ell_2m_2}^{}t_{\ell_3m_3}^*t_{\ell_4m_4}^* \rangle_c^{} \int
d\Omega(\hatn)
Y_{\ell_1m_1}^{}(\hatn)Y_{\ell_2m_2}^{}(\hatn)Y_{\ell_3m_3}^{*}(\hatn)Y_{\ell_4m_4}^{*}(\hatn)\nonumber\\
&&\\
& = &
\sum_{\ell_1m_1\ell_2m_2\ell_3m_3\ell_4m_4LM}\sqrt{{(2\ell_1+1)(2\ell_2+1)\over4\pi(2L+1)}}\sqrt{{(2\ell_3+1)(2\ell_4+1)\over4\pi(2L+1)}}C_{\ell_10\ell_20}^{L0}C_{\ell_30\ell_40}^{L0}C_{\ell_1m_1\ell_2m_2}^{LM}  
C_{\ell_3m_3\ell_4m_4}^{LM}\nonumber\\ 
&\times& \langle 
t_{\ell_1m_1}^{}t_{\ell_2m_2}^{}t_{\ell_3m_3}^*t_{\ell_4m_4}^* \rangle_c^{}\; .
\eea
It is then easy to relate to standard expression for the connected
part of the trispectrum as in \citet{hu:2001} and \citet{komatsu:prep}.

\end{document}